\documentclass[prd,aps,nofootinbib,twocolumn,superscriptaddress,preprintnumbers,balancelastpage,longbibliography]{revtex4-1}
\usepackage{ae,aecompl}
\usepackage{graphicx}	
\usepackage{amsmath}	
\usepackage{graphicx}
\usepackage{dcolumn}
\usepackage{bm}
\usepackage{graphics}
\usepackage{afterpage}
\usepackage{float}
\usepackage{subfigure}
\usepackage{rotating}
\usepackage{multirow}
\usepackage{tabularx}
\usepackage{booktabs}
\usepackage{multirow}
\usepackage{fancyheadings}
\usepackage{fancyhdr}

\usepackage{theorem}
\usepackage{moreverb}
\usepackage{euscript}
\usepackage{psfrag}
\usepackage{slashed}
\usepackage{mathtools}
\usepackage{makecell}
\usepackage[flushleft]{threeparttable}
\usepackage{adjustbox}
 
\usepackage[utf8]{inputenc}
\usepackage[english]{babel}

\usepackage{dcolumn}
\usepackage{bm}
\usepackage[mathlines]{lineno}

\usepackage[colorlinks=true
,urlcolor=blue
,anchorcolor=blue
,citecolor=blue
,filecolor=blue
,linkcolor=black
,menucolor=blue
,linktocpage=true
,pdfproducer=medialab
,pdfa=true
]{hyperref}
\usepackage{listings}
\usepackage{color,xcolor}

\renewcommand{\arraystretch}{1.8}
\newcolumntype{C}[1]{>{\centering\let\newline\\\arraybackslash\hspace{0pt}}m{#1}}

\lstdefinestyle{python}{
  belowcaptionskip=1\baselineskip,
  breaklines=true,
  frame=L,
  xleftmargin=\parindent,
  language=Python,
  showstringspaces=false,
  basicstyle=\small\ttfamily,
  morekeywords={models, lambda, forms,True,False,None},
  keywordstyle=\bfseries\color{deepgreen!40!black},
  commentstyle=\itshape\color{gray},
  identifierstyle=\color{black},
  stringstyle=\color{deepred},
  rulecolor=\color{gray},
}

\begin{document}

\preprint{UCI-TR-2020-04}
\preprint{IPMU20-0030}

\title{Strong constraints on thermal relic dark matter from Fermi-LAT observations of the Galactic Center}

\author{Kevork N.~Abazajian}
\affiliation{Center for Cosmology, Department of Physics and Astronomy, University of California, Irvine, CA 92697, USA}

\author{Shunsaku Horiuchi}
\affiliation{Center for Neutrino Physics, Department of Physics, Virginia Tech, Blacksburg, VA 24061, USA}

\author{Manoj Kaplinghat}
\affiliation{Center for Cosmology, Department of Physics and Astronomy, University of California, Irvine, CA 92697, USA}
 
\author{Ryan E.~Keeley}
\affiliation{Center for Cosmology, Department of Physics and Astronomy, University of California, Irvine, CA 92697, USA}
\affiliation{Korea Astronomy and Space Science Institute, Daejeon 34055, Korea}

\author{Oscar Macias}
\affiliation{Kavli Institute for the Physics and Mathematics of the
Universe (WPI), University of Tokyo, Kashiwa, Chiba 277-8583, Japan}
\affiliation{GRAPPA Institute, University of Amsterdam, 1098 XH
Amsterdam, The Netherlands}

\date{\today}

\newcommand{\mk}[1]{{\bf #1}}
\newcommand{\om}[1]{\textcolor{blue}{#1}}
\newcommand{\sh}[1]{\textcolor{red}{#1}}
\newcommand{\aap}{Astronomy and Astrophysics}
\newcommand{\mnras}{Monthly Notices of the RAS}

\begin{abstract}

The extended excess toward the Galactic Center (GC) in gamma rays inferred from Fermi-LAT observations has been interpreted as being due to dark matter (DM) annihilation. Here, we perform new likelihood analyses of the GC and show that, when including templates for the stellar galactic and nuclear bulges, the GC shows no significant detection of a DM annihilation template, even after generous variations in the Galactic diffuse emission models and a wide range of DM halo profiles. 
We include Galactic diffuse emission models with combinations of three-dimensional inverse Compton maps, variations of interstellar gas maps, and a central source of electrons. For the DM profile, we include both spherical and ellipsoidal DM morphologies and a range of radial profiles from steep cusps to kiloparsec-sized cores, motivated in part by hydrodynamical simulations. Our derived upper limits on the dark matter annihilation flux place strong constraints on DM properties. In the case of the pure $b$-quark annihilation channel, our limits on the annihilation cross section  
are more stringent than those from the Milky Way dwarfs up to DM masses of approximately TeV and rule out the thermal relic cross section up to approximately 300 GeV. Better understanding of the DM profile, as well as the Fermi-LAT data at its highest energies, would further improve the sensitivity to DM properties.

\end{abstract}

\pacs{PACS}
\maketitle


\section{Introduction}\label{sec:Introduction}

The particle nature of dark matter (DM) remains one of the most important unresolved questions in astrophysics, cosmology, and particle physics. Hierarchical structure formation with cold collisionless or self-interacting dark matter predicts that the Milky Way (MW) Galactic Center (GC) would contain a large concentration of DM \cite{Navarro:1996gj,Kaplinghat:2013xca}, providing an avenue for stringent tests of DM annihilation \cite{Hooper:2002ru,Hooper:2011ti,Hooper:2012sr}. After the launch of the {\it Fermi Gamma-Ray Space Telescope}, an extended source of gamma ray emission was quickly identified toward the GC and shown to be consistent with the annihilation of thermal weak-interaction-scale DM producing gamma rays \cite{Hooper:2010mq}. This GC excess (GCE) has since been detected by many follow-up analyses, which also indicated its potential association with unresolved point sources or new diffuse emission processes \cite{Abazajian:2010zy,Abazajian:2012pn,Gordon:2013vta,Daylan:2014rsa,TheFermi-LAT:2015kwa}. 

A major challenge for establishing DM signals in our MW's GC is the abundant astrophysical activity in the GC region. For example, part of the GCE signal could be explained by gamma-ray emission induced by cosmic rays injected by ongoing star formation activity in the GC region \cite{Carlson:2014cwa}, cosmic-ray bremsstrahlung off of molecular gas \cite{Abazajian:2012pn}, or inverse-Compton emission from leptonic cosmic rays \cite{Abazajian:2014hsa,Gaggero:2015nsa}. However, these studies are not able to completely explain the data, and still leave the need for a spherical GCE. 

Recently, Refs.~\cite{Macias:2016nev,Bartels:2017vsx} showed that the GCE overwhelmingly prefers the spatially \textit{asymmetric} morphology of the Galactic stellar bulge---a triaxial barlike structure extending a few kiloparsecs in the GC \cite{2016ARA&A..54..529B}---over the spherically symmetric morphology assumed by a DM origin. The bulge, which includes a concentrated ``nuclear" component and an extended boxy component, has a radially varying asymmetry that was not captured in earlier elliptical shape tests conducted on the GCE \cite{Abazajian:2012pn,Daylan:2014rsa}. A detailed study of the robustness for the detection of the bulge, including systematic uncertainties arising from background emissions and other gamma-ray sources, was presented in Ref.~\cite{Macias:2019omb}. Since the bulge contains a broad mix of star-forming and old stellar populations, this motivates a population of astrophysical gamma-ray emitters such as young pulsars and millisecond pulsars (MSPs) as the source of the excess gamma rays. Most significantly, the inclusion of the asymmetric bulge model completely eliminates the need for a spherically symmetric DM component of the GCE  \cite{Macias:2016nev,Bartels:2017vsx}. This provides an opportunity to substantially improve the sensitivity to test DM properties.

In this article, we present stringent DM limits incorporating recent  developments in modeling the bulge and other astrophysical gamma-ray sources in the GC region. To ensure that our limits are robust, we use results from galaxy formation simulations to inform our DM templates, which provides a significant point of departure from previous work. Furthermore, we explore generous variations in models of the gamma-ray emission from cosmic-ray interactions. Even with the considerably larger freedom for the astrophysical emission and DM profiles, our results show that the {\it Fermi} Large Area Telescope ({\it Fermi}-LAT) observations of the GC provide very stringent constraints on DM annihilation. For two-body final states with hadronic components, we are able to rule out thermal DM up to approximately 300 GeV in mass, surpassing the reach from dwarf satellites of the MW for DM particles with masses less than a TeV.

\section{Dark matter limits}\label{sec:limits}

To calculate the limits on DM annihilation cross section, we must first generate a likelihood profile for the DM annihilation intensity for a given DM halo model. We consider four classes of MW DM profiles, described in the next section. The likelihood for each value of the DM annihilation intensity is computed by varying the fluxes of all the background templates such that the log-likelihood is maximized. We use the \textit{Fermi} \textsc{UpperLimits} tool\footnote{\url{https://fermi.gsfc.nasa.gov/ssc/data/analysis/scitools/upper_limits.html}} to perform this maximization and generate the likelihood profile for the DM annihilation intensity. 

Our background model contains templates for the following: hadronic emission traced by HI and H2 gas maps divided in four cylindrical concentric rings and two total dust maps, three-dimensional (3D) inverse Compton (IC) divided into four or six rings and a two-dimensional (2D) IC map with a central source of electrons, an isotropic background, the 4FGL~\cite{Fermi-LAT:4FGL} point sources, Fermi bubbles, Loop I, the Sun, and the Moon. Details of the templates, methods employed, likelihood profiles, and resulting spectra, as well as our comprehensive checks and analyses of the systematic effects, are all presented in the Appendixes. Additional tests of the robustness of the preference for the bulge template are as discussed in Ref.~\cite{Macias:2019omb}, which also showed that the strong preference for the boxy bulge+nuclear bulge model is not dependent on the GDE models adopted---they showed that the preference is present in a standard 2D IC model as well as various 3D IC models for the  $40^\circ\times 40^\circ$ region of interest (RoI) despite the variation in the total log-likelihood values among the GDE models of about 2000 (see Figs.~3 and 5 of Ref.~\cite{Macias:2019omb}). 

The likelihood profile is generated in 15 independent logarithmic-spaced energy bins between 0.667 and 158 GeV, and no broadband spectral shape is assumed for any of the templates. Following this methodology, we are able to marginalize over the uncertainties in the astrophysical backgrounds in a manner that is independent of  the uncertainties in the particle physics models. An indicator of the success of our method is that we recover physically consistent, continuous spectra for all the background templates (see Fig.~\ref{fig:total_spectra} in Appendix~\ref{appx:syst_uncertainties_DMlimits}). We adopt a $40^\circ\times 40^\circ$ RoI, and provide results of our tests with a $15^\circ\times 15^\circ$ RoI in Appendix~\ref{appx:RoI_selection}.

With the likelihood profiles in hand, we use Bayes's theorem to calculate a posterior in the annihilation cross section and DM mass parameter space. The flux signal from DM annihilation scales as
\begin{equation}
\frac{d \Phi}{dE} = \frac{\langle \sigma v \rangle}{8\pi} \frac{J}{m_\chi^2} \frac{dN}{dE},
\end{equation}
where $d \Phi/ dE$ is the differential number flux, $\langle \sigma v \rangle$ is the velocity averaged DM cross section times relative velocity, $m_\chi$ is the DM mass, $dN/dE$ is the gamma-ray energy spectrum, and the $J$-factor ($J$) is the integral through the line of sight over the region of interest of the DM density squared, $J = \int d \Omega \int ds \ \rho^2(r(s,\Omega))$.
We need to marginalize over this $J$-factor in order to calculate the posterior for the DM mass and annihilation cross section. 

We assume that the dark matter is single component when calculating the $J$-factor. If this is not the case, our constraints on $\langle \sigma v \rangle$ should be recast as constraints on $f_{\rm DM}^2\langle \sigma v \rangle$, where $f_{\rm DM}$ is the fraction of cosmological dark matter density in the model being constrained. This is important for thermal relics because $f_{\rm DM}$ scales inversely with the total annihilation cross section in the early Universe and hence the flux decreases for $s$-wave cross sections larger than the thermal relic cross section.

For comparison purposes, we also calculate the posterior distribution of the DM mass and annihilation cross section for the eight classical MW dwarf spheroidals, with well-determined J-factors. We use the likelihood profiles for the classical dwarfs from Ref.~\cite{Fermi-LAT:2016uux} and the uncertainties in the J-factors of the dwarfs are taken from Ref.~\cite{Geringer-Sameth:2014yza}, 
which are inferred from fits to the stellar kinematic data using generalized Navarro-Frenk-White (NFW) profiles. Unlike the GC region, the J-factors for the classical dwarfs are well constrained by stellar kinematic data because they are dark matter dominated and the RoI of approximately $0.5^{\circ}$ is well-matched to their stellar half-light radii~\cite{Abdo:2010ex}. 

\begin{figure*}[ht!]
    \centering
    \includegraphics[width=0.48\textwidth]{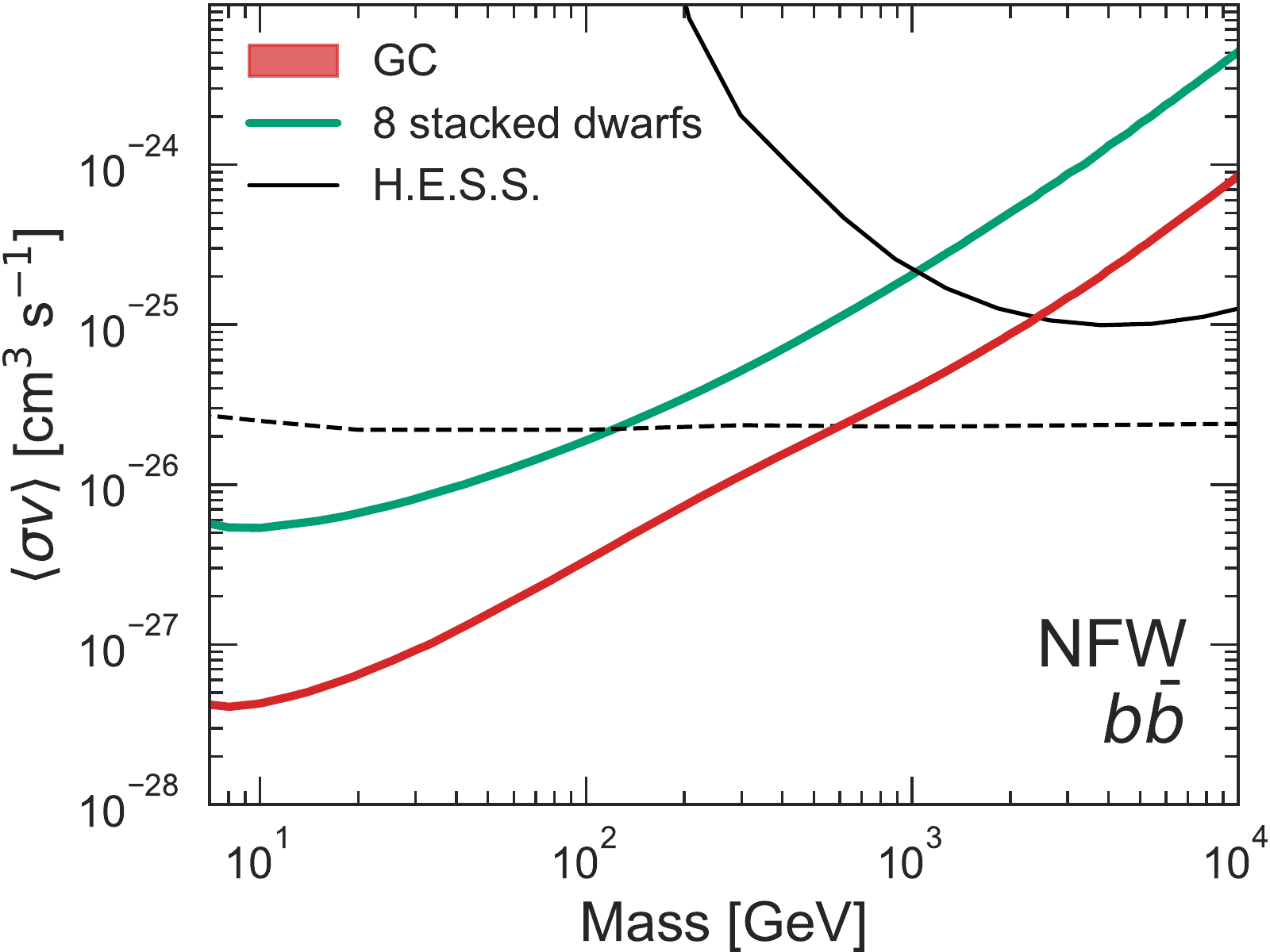}
    \includegraphics[width=0.48\textwidth]{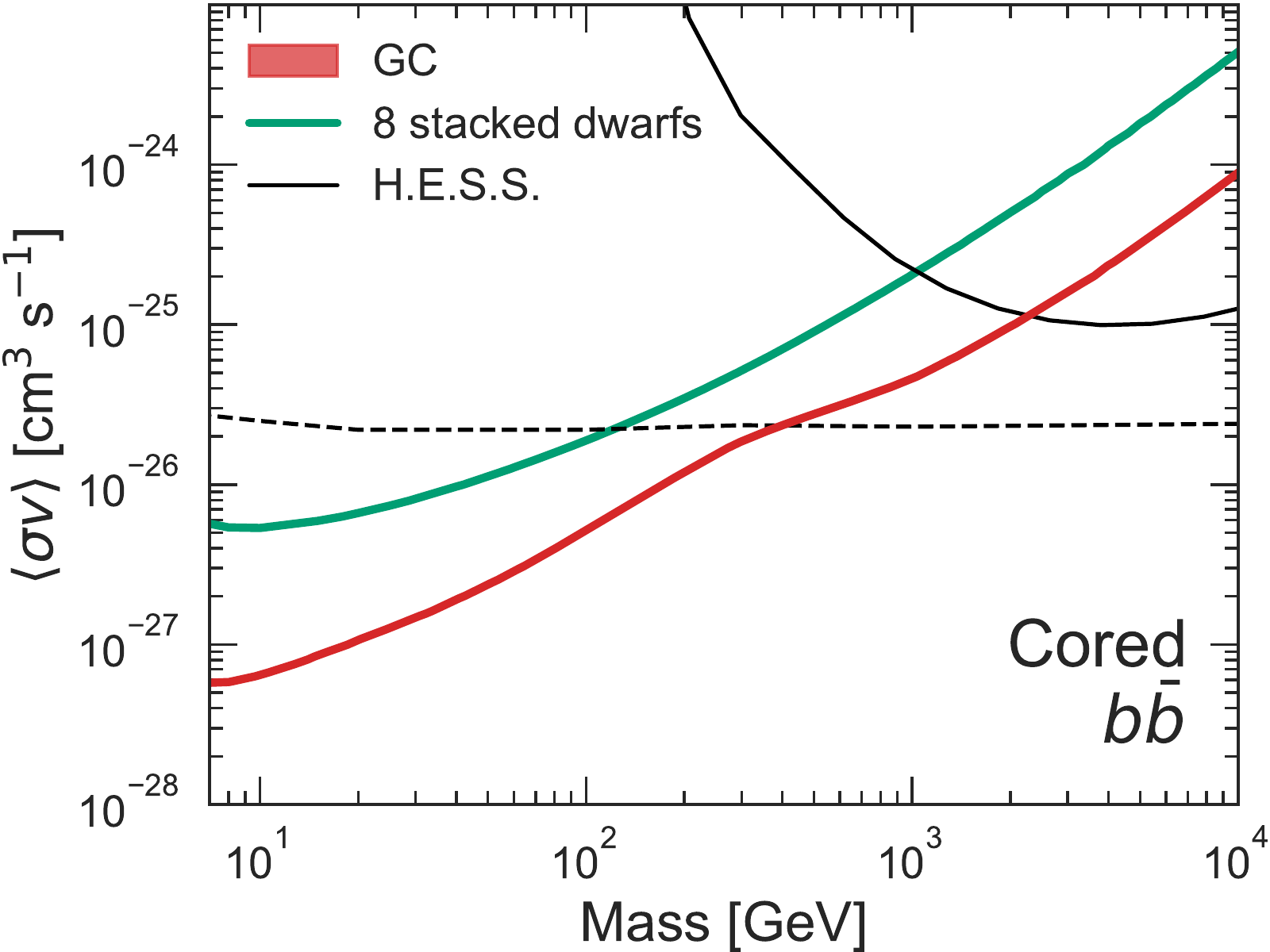}
    \caption{
    The least constraining upper limit (95\% C.L.) on the average DM cross section times relative velocity $\langle \sigma v\rangle$ for annihilation to  $b\bar{b}$, among a large number of GDE models and DM distributions considered. The GDE models allow for changes in the interstellar gas, dust, and IC distributions. For both the gNFW (left) and cored (right) DM profiles, we considered spherical and ellipsoidal shapes. For gNFW, the inner slope was also varied. See text and Fig.~\ref{fig:limitsDiffuse} for details. The dashed black line is the thermal cross section \cite{Steigman2012}. The H.E.S.S.~\cite{Abdallah:2016ygi} and stacked dwarfs limits \cite{Fermi-LAT:2016uux} are shown for comparison and do not reflect the different GDE models and DM profiles. All the constraints shown assume that the DM is entirely made up of one kind of particle. If this assumption is relaxed, then the constraints on $\langle \sigma v\rangle$ should be weakened by the square of the fraction of DM in the component being constrained. The data files and code necessary to reproduce this figure are available at \url{https://github.com/oscar-macias/Fermi_GC_limits}.
    }
    \label{fig:limitsNFW} 
\end{figure*}

\section{Dark matter profiles}\label{sec:profile}
We consider four classes of MW DM profiles: a generalized NFW (gNFW) profile, a cored profile that matches smoothly on to a NFW profile while conserving mass \cite{Read:2015sta}, and ellipsoidal versions of both. 
The gNFW density profile is
\begin{equation}\label{eq:nfw}
\rho(r)= \rho_\odot \left( \frac{R_\odot}{r} \right)^\gamma \left( \frac{r_s+R_\odot}{r_s+r} \right)^{3-\gamma}\, ,
\end{equation}
where $\rho_\odot$ is the local DM density, $R_\odot$ is the solar radius, $r_s$ is the scale radius, and $\gamma$ is the inner slope.  Parametrizing the MW DM profile this way is useful since it allows us to use independent datasets to characterize the uncertainty in each parameter. We allow $\gamma$ to vary between 0.5 and 1.5. Note that if the GCE is to be explained by a gNFW squared (gNFW$^2$) template, then we need $\gamma \approx 1.2$ \cite{Abazajian:2012pn,Gordon:2013vta,Daylan:2014rsa}. We adopt a log-normal prior on $r_s$ with a mean value of $26 \ \rm kpc$ and a width of 0.14 dex, consistent with the $\Lambda$CDM concentration-mass relation~\cite{Sanchez-Conde:2013yxa} for halo mass of $10^{12}\rm \, M_\odot$. We neglect the factor of 2 uncertainty in MW's halo mass since it is subdominant to the adopted spread. For the local density, we take $\rho_\odot = 0.28 \pm 0.08$ GeV/cm$^3$ as a prior from Zhang {\it et al.}~\cite{Zhang:2012rsb}, who constrain the local DM density from the vertical motions of K-dwarfs close to the plane of the MW, independent of the DM density at other radii. We note that the local density constraint agrees very well in the $\rho_s$--$r_s$ plane with the mass constraint at about $20\ \rm kpc$ obtained using Globular cluster proper motions from Gaia \cite{2019ApJ...873..118W,2019A&A...621A..56P}.

Previous hydrodynamical simulations of the MW with cold DM have typically found the profile to be adiabatically contracted \cite{Guedes:2011ux}. Our gNFW profiles with inner slopes of $\gamma > 1$ capture this possibility. However, recent hydrodynamical simulations of MW-like galaxies also show the presence of a core in the DM density profile with a size of roughly a kiloparsec. 
Using the Eris simulation \cite{Guedes:2011ux}, Ref.\ \cite{Kuhlen:2012qw} argued that the core is formed in response to the bar, along the lines of ideas proposed earlier \cite{Weinberg:2001gm,Weinberg:2006ps}, and not due to feedback. They also noted the supporting fact that a roughly same-size core is present in another simulation identical to Eris but with a lower star formation threshold, which reduces feedback effects dramatically. Further evidence supporting the view that the presence of the bulge
can lead to 
kiloparsec-sized cores comes from simulations with a fixed disk and bulge potential that lead to similar cores \cite{Robles:2019mfq}. However, the results from the FIRE cosmological simulations indicate that feedback can also lead to kiloparsec-sized cores in the dark matter halo of the Milky Way \cite{Chan:2015tna}. It is possible that both secular and feedback processes contribute to creating a kiloparsec-sized core. Shallow cusps or cores of this size are consistent with results obtained from equilibrium models fitted to the density profile of Red Clump Giant stars and the stellar kinematics of Bulge stars \cite{2017MNRAS.465.1621P}.


\begin{figure*}[ht!]
    \centering
    \includegraphics[width=0.49\textwidth]{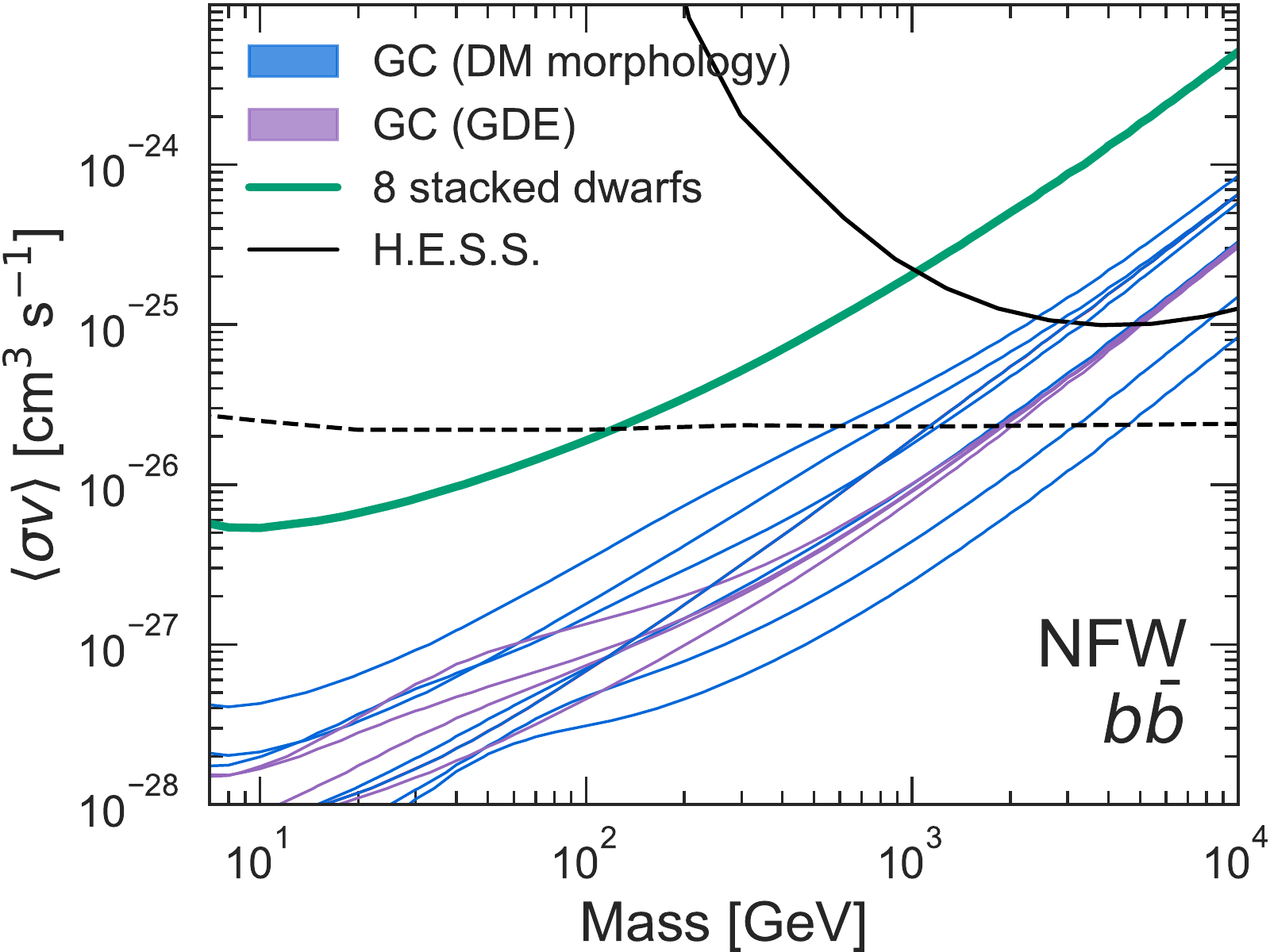}
    \includegraphics[width=0.49\textwidth]{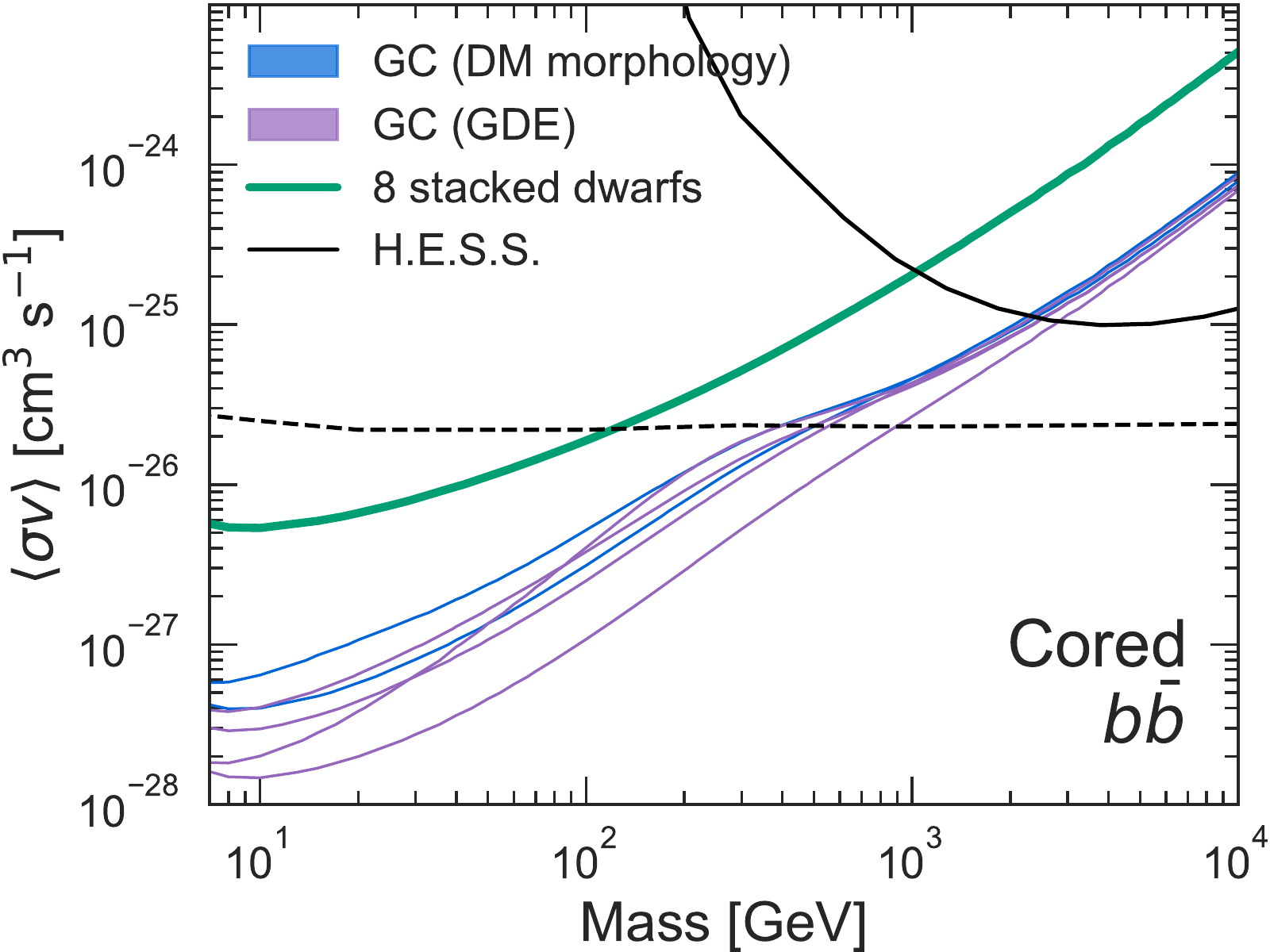}
    \caption{
    Contours of the posterior that contain 95\% of the probability showing the impact of variations in the GDE models and DM spatial morphology. The {\em left panel} shows variations around a gNFW profile; the inner-slope $\gamma$ ranges between [0.5,1.5] and the spatial distribution can be spherical or ellipsoidal (blue lines). The purple lines correspond to the systematic uncertainty arising from different GDE templates. Similarly, the {\em right panel} shows variations around the cored profile (spherical or ellipsoidal; blue) and the GDE models (purple). The data files and code are available at \url{https://github.com/oscar-macias/Fermi_GC_limits}  
    }
    \label{fig:limitsDiffuse} 
\end{figure*}

We use the cored ``Read'' profile~\cite{Read:2015sta} to investigate the effects of a cored dark matter density. It has a core radius $r_c$ that describes the removal of mass from the center to the outer parts due to core formation and the mass asymptotically tends to the NFW profile mass at large radii. The enclosed mass for the cored profile is described by, 
\begin{equation}\label{eq:Read}
    M_c (r)= M_{\rm NFW}(r)\tanh(r/r_c),
\end{equation}
where we take $M_{\rm NFW}(r)$ to be the NFW profile with $\gamma=1$.  We fix the core radius to be 1 kpc in keeping with the discussion of the simulations above, and in order to make a straight-forward one-to-one comparison, we assume the same prior distribution for $r_s$ (a mean of 26 kpc and a scatter of 0.14 dex). Note that this neglects the impact of adiabatic contraction, which would increase the inner core density. A better characterization of the the inner density profile of MW halos is likely to lead to stronger results than those presented here. We then use Monte Carlo sampling to calculate the prior uncertainty on the J-factor from the prior uncertainty on these parameters of the MW's DM profile.  

The presence of the bulge and bar should also have an impact on the axis ratio of the DM template. The expectation is that the DM density profile is an ellipsoid with the short axis perpendicular to the stellar disk \cite{Petersen:2016xtd}. This flattening of the halo should be due, in part, to the formation of the stellar disk. Moreover, there is likely also a perturbative effect of the bar formation on the halo that induces further flattening \cite{Petersen:2016xtd}. The Eris simulation discussed previously finds a minor-to-major axes ratio of about 0.8 at 1 kpc and intermediate-to-major axes ratio of unity ~\cite{Dai:2018ypv}. 

Given the arguments above, a flattened ellipsoid with a mild radial variation in the density is a reasonable description of the inner kiloparsec of the MW halo. This is very different from the spherical gNFW $\gamma=1.2$ profiles that were used by the bulk of the explorations of the GCE and considered to be representative of the expectations for cold DM. To test for the impact of nonspherical DM distribution, we use two different density ellipsoids with axis ratios of 0.7 (somewhat more flattened than the results in Ref.~\cite{Dai:2018ypv}): one in which the radial profile is the same as the gNFW profile with $\gamma=1.2$ and the other in which the density profile is the same as the cored profile with $r_c=1~\rm kpc$. The cored model is favored over the gNFW model by the Bulge modeling in Ref.~\cite{2017MNRAS.465.1621P}.

\section{Results and discussion}\label{sec:discussion}

We find that an emission template that traces stellar mass in the Galactic bulge is preferred in all (independent) energy bins over each of the DM templates considered in this analysis. In none of our maximum likelihood runs---that included a variety of alternative GDE models---was a DM template detected. This allows us to impose strong constraints on  $\langle \sigma v \rangle$ using the flux likelihood profile for each DM template and Galactic diffuse emission (GDE) combination as described in Sec.~\ref{sec:limits}.

Our constraints on $\langle \sigma v \rangle$ are presented in Figs.\ \ref{fig:limitsNFW} and \ref{fig:limitsDiffuse}. The curves correspond to the contours of the posterior in cross section and mass that enclose 95\% of the probability. In Fig.~\ref{fig:limitsNFW}, we show the maximum cross sections from the set of the 95\% limits derived from the gNFW (left panel) and cored(right panel) profiles, while in Fig.~\ref{fig:limitsDiffuse} we show in more detail how the limits are affected by variations of the GDE model and DM morphology. 

First, we explore the impact of the GDE model on the DM limits. The purple lines in both panels of Fig.\ \ref{fig:limitsDiffuse} illustrate the systematic uncertainty that arises from different GDE model assumptions.
We explore alternative dust, interstellar gas, 3D IC map composed of four or six independent rings, and a 2D IC map containing an additional central source population of $e^-$ (model B in Ref.~\cite{Ackermann:2014usa}). 
In particular, interstellar gas maps constructed using hydrodynamic simulations~\cite{Macias:2016nev} or the standard interpolation method~\cite{Casandjian:andFermiLat2016} and dust maps with $E(B-V)$ magnitude cuts of either 2 or 5 mag were considered. Details about these models are provided in the Appendix and in Ref.~\cite{Macias:2019omb}.  
By using 2D IC models with interpolated gas maps fitted in rings, we are able to capture a wide range of models used in the literature to infer the existence of the GCE, including \textsc{p6v11} and other \textit{Fermi} GDE models (e.g.,~\cite{Calore:2015bsx}). Next, we explore the impact of changes to the DM profile. The blue lines in Fig.~\ref{fig:limitsDiffuse} correspond to the systematic uncertainty arising from different MW DM profiles.  For the left panel, the blue lines represent  95\% C.L. upper limits on the cross sections derived from the gNFW DM templates with a generous $\gamma =[0.5,1.5]$ range and an Cored ellipsoidal case, while the same lines in the right panel are determined by the limits from the cored profile with a core size $r_c = 1~\rm kpc$ for spherical and ellipsoidal shapes.

For the gNFW profile, we find that $\gamma = 1.2$ is the value that attributes the most flux to the squared-NFW profile.  This is why we used the $\gamma = 1.2$ profile as our baseline when varying the GDE models. Note that, while $\gamma=1.2$ is the value with the largest flux for that template, smaller values of $\gamma$ have smaller J-factors. These two effects compete, and in the end, the $\gamma=1.0$ value corresponds to the weaker limit for all masses. For all panels shown in Fig.~\ref{fig:limitsNFW} and \ref{fig:limitsDiffuse}, when DM profile variations are studied, we assumed the benchmark GDE model described in the Appendix.

To address potential issues of mismodeling and over- or undersubtraction affecting our limits, we have performed a series of injection tests. We found that in the vast majority of cases, our analysis successfully recovers the correct statistical coverage of constraints. However, we found a systematic bias in the last energy bin ($34.49-158.11$ GeV) of our analysis  for the cored profile cases considered. Therefore, we removed the last energy bin from our upper limits for all the cored profile cases  (see Appendix). This explains why the limits shown in the right panels of Figs.~\ref{fig:limitsNFW} and \ref{fig:limitsDiffuse} are weaker than those on the left, and become comparable to those from dwarfs for DM masses larger than a TeV.

The results for other channels including $W^+W^-$, $ZZ$, $\tau^+\tau^-$, and $HH$ are presented in the appedinx. The qualitative features with respect to variations in the diffuse models and the density profiles are the same as the $b\bar{b}$ channel for these other channels. We do not consider  annihilation to $e^+e^-$ and $\mu^+\mu^-$ since the dominant gamma-ray contribution in these cases will arise from the IC process,  causing the spatial profile to change from the DM density-squared morphology~\cite{Song:2019nrx}. 

A noteworthy aspect of our results is that, despite allowing for extensive systematic uncertainty, they provide strong constraints on thermal relic models with DM particle masses smaller than about 300 GeV and they are comparable to the H.E.S.S.~constraints for masses around a TeV~\cite{Abdallah:2016ygi}. It is interesting to compare our results with those of Refs.~\cite{Hooper:2011ti,Hooper:2012sr}, which showed that the inner Galaxy constraints could be better than those arising from MW dwarfs, even with a kiloparsec-sized cored DM template. More data, better models for the point sources and the diffuse emission, and the inclusion of the bulge templates have all contributed to making our constraints stronger and more robust. 
We note that our limits are in reasonable agreement with expected {\it Fermi}-LAT limits for the inner Galaxy~\cite{Charles:2016pgz}. 

Three advances in the future will make our results even more powerful. The first is a deeper understanding of the central density profile of DM in the Milky Way and its correlation with the stellar bulge and disk, which could remove the uncertainty arising from the radial distribution and shape of the DM template. This could allow the properties of the MW dark matter profile in the inner kpc, such as the core size and ellipsoidal shape, to be constrained based on the bulge and disk. In addition, such a study will allow us to include the effect of adiabatic contraction, which has been neglected in our study for the cored profile and could increase the J-factor by up to a factor of 2 (estimated by varying $r_s$ by a factor of 2 to account adiabatic contraction). We note that if the core radius $r_c$ were larger by a factor of 2, the $J$-factor for our RoI would only decrease by about 30\%. This is because the $J$-factor for our RoI with a cored profile is dominated by contributions from $r>1\ \rm kpc$.

The second important advance would be a clear determination of the point source nature of the GCE. While this will not quantitatively change our constraints, it will provide corroborating evidence for the Bulge-GCE connection that our analysis clearly prefers. 
This may be possible through the non-Poissonian template fitting procedure~\cite{Lee:2015fea,Leane:2019xiy,Chang:2019ars,Leane:2020pfc,Leane:2020nmi,Buschmann:2020adf} and wavelet techniques~\cite{Bartels:2015aea,Balaji:2018rwz,Zhong:2019ycb} to detect clustering of photons or radio detection of point sources responsible for the bulge emission~\cite{Calore:2015bsx} or detection of a significant number of millisecond pulsars (putative sources for the bulge gamma-ray emission) with radio telescopes \cite{2015ApJ...805..172M,2016ApJ...827..143C,2017MNRAS.471..730R,2019ApJ...876...20H}. 

The third is further improvements in 3D models of the gas and Interstellar Radiation Fields (ISRF) maps, which directly feed into the diffuse emission models and determine the residuals from fitting to {\it Fermi}-LAT data. For the cored profile, the upper limits for the six-ring 3D ISRF model are evidently more stringent than those for the other GDE models, which exhibit more subtle differences. Since we have chosen our limits to be the weakest among the GDE models, and not the best fitting, a study that includes a more extensive set of background models may be able to improve upon our $\langle \sigma v\rangle$ limits at high masses by a factor of few.

\section{Conclusion}\label{sec:conclusion}

The detection in the {\it Fermi}-LAT data of a spatially concentrated excess of gamma-ray emission in the MW potentially consistent with DM annihilation \cite{Hooper:2010mq,Abazajian:2012pn,Daylan:2014rsa} has sparked great interest in the sources of high-energy emission in the GC. At the same time, the {\it Fermi}-LAT data have spurred steady progress in our understanding of the gamma-ray emission from our Galaxy over the past decade. With detected sources that are consistent with the Fermi bubbles; 4FGL point sources; detailed IC emission maps; disk gas; and, most importantly, the emission from the stellar Galactic bulge and nuclear bulge, there is no significant excess in the GC that may be attributed to DM annihilation. This result is robust to a wide range of variations in the GDE model and DM profiles. Although we cannot test for all possible GDE models and DM profiles, the important point is that our approach covers the wide range that has been used to infer the existence of the GCE, and go beyond them.

Our results strongly favor the hypothesis that the excess emission in the GC at GeV energies is dominantly of astrophysical origin related to the stellar bulge. While gamma-ray emission from DM annihilation in the GC is still possible, the flux would have to be below that of the GCE, and with parameters consistent with the exclusion regions of Fig.~\ref{fig:limitsNFW}. In arriving at this conclusion, we allowed for a variety of DM templates. These include ellipsoidal profiles with a kiloparsec-sized core that we suggest, based on existing simulations of the MW, are closest to the true prediction for the density profile of cold dark matter. We explored in detail the robustness of our results to variations in the GDE models arising from new sources of relativistic $e^\pm$, new 3D IC templates, and changes to the standard gas maps. Our results provide stringent constraints on models of thermal relic dark matter with masses up to a few hundred GeV and prompt annihilation to Standard Model particles.

\section{Acknowledgments} 
We are grateful to Troy Porter for his very helpful comments about GALPROP version 56 and for making the new 3D ISRF data publicly available.   We thank Shin'ichiro Ando, Francesca Calore, Roland Crocker, Douglas Finkbeiner, Chris Gordon, Rebecca Leane, Mariangela Lisanti, Dmitry Malyshev, Shigeki Matsumoto, Nicholas L. Rodd, Satoshi Shirai, Tracy Slatyer, Masahiro Takada and Christoph Weniger for fruitful discussions and feedback. K.N.A. and M.K. are supported by NSF Theoretical Physics Grants No.~PHY-1620638 \& PHY-1915005. S.H. is supported by the U.S.~Department of Energy Office of
Science under award number de-sc0018327 and NSF Grants No.~AST-1908960 \& PHY-1914409.  O.M. acknowledges support by JSPS KAKENHI Grant Numbers JP17H04836, JP18H04340, JP18H04578, and JP20K14463.
This work was supported by World Premier International Research Center Initiative (WPI Initiative), MEXT, Japan. 

\bibliography{reference}	

\begin{thebibliography}{74}%
\makeatletter
\providecommand \@ifxundefined [1]{%
 \@ifx{#1\undefined}
}%
\providecommand \@ifnum [1]{%
 \ifnum #1\expandafter \@firstoftwo
 \else \expandafter \@secondoftwo
 \fi
}%
\providecommand \@ifx [1]{%
 \ifx #1\expandafter \@firstoftwo
 \else \expandafter \@secondoftwo
 \fi
}%
\providecommand \natexlab [1]{#1}%
\providecommand \enquote  [1]{``#1''}%
\providecommand \bibnamefont  [1]{#1}%
\providecommand \bibfnamefont [1]{#1}%
\providecommand \citenamefont [1]{#1}%
\providecommand \href@noop [0]{\@secondoftwo}%
\providecommand \href [0]{\begingroup \@sanitize@url \@href}%
\providecommand \@href[1]{\@@startlink{#1}\@@href}%
\providecommand \@@href[1]{\endgroup#1\@@endlink}%
\providecommand \@sanitize@url [0]{\catcode `\\12\catcode `\$12\catcode
  `\&12\catcode `\#12\catcode `\^12\catcode `\_12\catcode `\%12\relax}%
\providecommand \@@startlink[1]{}%
\providecommand \@@endlink[0]{}%
\providecommand \url  [0]{\begingroup\@sanitize@url \@url }%
\providecommand \@url [1]{\endgroup\@href {#1}{\urlprefix }}%
\providecommand \urlprefix  [0]{URL }%
\providecommand \Eprint [0]{\href }%
\providecommand \doibase [0]{http://dx.doi.org/}%
\providecommand \selectlanguage [0]{\@gobble}%
\providecommand \bibinfo  [0]{\@secondoftwo}%
\providecommand \bibfield  [0]{\@secondoftwo}%
\providecommand \translation [1]{[#1]}%
\providecommand \BibitemOpen [0]{}%
\providecommand \bibitemStop [0]{}%
\providecommand \bibitemNoStop [0]{.\EOS\space}%
\providecommand \EOS [0]{\spacefactor3000\relax}%
\providecommand \BibitemShut  [1]{\csname bibitem#1\endcsname}%
\let\auto@bib@innerbib\@empty
\bibitem [{\citenamefont {Navarro}\ \emph {et~al.}(1997)\citenamefont
  {Navarro}, \citenamefont {Frenk},\ and\ \citenamefont
  {White}}]{Navarro:1996gj}%
  \BibitemOpen
  \bibfield  {author} {\bibinfo {author} {\bibfnamefont {Julio~F.}\
  \bibnamefont {Navarro}}, \bibinfo {author} {\bibfnamefont {Carlos~S.}\
  \bibnamefont {Frenk}}, \ and\ \bibinfo {author} {\bibfnamefont {Simon D.~M.}\
  \bibnamefont {White}},\ }\bibfield  {title} {\enquote {\bibinfo {title} {{A
  Universal density profile from hierarchical clustering}},}\ }\href {\doibase
  10.1086/304888} {\bibfield  {journal} {\bibinfo  {journal} {Astrophys. J.}\
  }\textbf {\bibinfo {volume} {490}},\ \bibinfo {pages} {493--508} (\bibinfo
  {year} {1997})},\ \Eprint {http://arxiv.org/abs/astro-ph/9611107}
  {arXiv:astro-ph/9611107 [astro-ph]} \BibitemShut {NoStop}%
\bibitem [{\citenamefont {Kaplinghat}\ \emph {et~al.}(2014)\citenamefont
  {Kaplinghat}, \citenamefont {Keeley}, \citenamefont {Linden},\ and\
  \citenamefont {Yu}}]{Kaplinghat:2013xca}%
  \BibitemOpen
  \bibfield  {author} {\bibinfo {author} {\bibfnamefont {Manoj}\ \bibnamefont
  {Kaplinghat}}, \bibinfo {author} {\bibfnamefont {Ryan~E.}\ \bibnamefont
  {Keeley}}, \bibinfo {author} {\bibfnamefont {Tim}\ \bibnamefont {Linden}}, \
  and\ \bibinfo {author} {\bibfnamefont {Hai-Bo}\ \bibnamefont {Yu}},\
  }\bibfield  {title} {\enquote {\bibinfo {title} {{Tying Dark Matter to
  Baryons with Self-interactions}},}\ }\href {\doibase
  10.1103/PhysRevLett.113.021302} {\bibfield  {journal} {\bibinfo  {journal}
  {Phys. Rev. Lett.}\ }\textbf {\bibinfo {volume} {113}},\ \bibinfo {pages}
  {021302} (\bibinfo {year} {2014})},\ \Eprint {http://arxiv.org/abs/1311.6524}
  {arXiv:1311.6524 [astro-ph.CO]} \BibitemShut {NoStop}%
\bibitem [{\citenamefont {Hooper}\ and\ \citenamefont
  {Dingus}(2004)}]{Hooper:2002ru}%
  \BibitemOpen
  \bibfield  {author} {\bibinfo {author} {\bibfnamefont {Dan}\ \bibnamefont
  {Hooper}}\ and\ \bibinfo {author} {\bibfnamefont {Brenda~L.}\ \bibnamefont
  {Dingus}},\ }\bibfield  {title} {\enquote {\bibinfo {title} {{Limits on
  supersymmetric dark matter from EGRET observations of the galactic center
  region}},}\ }\href {\doibase 10.1103/PhysRevD.70.113007} {\bibfield
  {journal} {\bibinfo  {journal} {Phys. Rev.}\ }\textbf {\bibinfo {volume}
  {D70}},\ \bibinfo {pages} {113007} (\bibinfo {year} {2004})},\ \Eprint
  {http://arxiv.org/abs/astro-ph/0210617} {arXiv:astro-ph/0210617 [astro-ph]}
  \BibitemShut {NoStop}%
\bibitem [{\citenamefont {Hooper}\ and\ \citenamefont
  {Linden}(2011)}]{Hooper:2011ti}%
  \BibitemOpen
  \bibfield  {author} {\bibinfo {author} {\bibfnamefont {Dan}\ \bibnamefont
  {Hooper}}\ and\ \bibinfo {author} {\bibfnamefont {Tim}\ \bibnamefont
  {Linden}},\ }\bibfield  {title} {\enquote {\bibinfo {title} {{On The Origin
  Of The Gamma Rays From The Galactic Center}},}\ }\href {\doibase
  10.1103/PhysRevD.84.123005} {\bibfield  {journal} {\bibinfo  {journal} {Phys.
  Rev.}\ }\textbf {\bibinfo {volume} {D84}},\ \bibinfo {pages} {123005}
  (\bibinfo {year} {2011})},\ \Eprint {http://arxiv.org/abs/1110.0006}
  {arXiv:1110.0006 [astro-ph.HE]} \BibitemShut {NoStop}%
\bibitem [{\citenamefont {Hooper}\ \emph {et~al.}(2013)\citenamefont {Hooper},
  \citenamefont {Kelso},\ and\ \citenamefont {Queiroz}}]{Hooper:2012sr}%
  \BibitemOpen
  \bibfield  {author} {\bibinfo {author} {\bibfnamefont {Dan}\ \bibnamefont
  {Hooper}}, \bibinfo {author} {\bibfnamefont {Chris}\ \bibnamefont {Kelso}}, \
  and\ \bibinfo {author} {\bibfnamefont {Farinaldo~S.}\ \bibnamefont
  {Queiroz}},\ }\bibfield  {title} {\enquote {\bibinfo {title} {{Stringent and
  Robust Constraints on the Dark Matter Annihilation Cross Section From the
  Region of the Galactic Center}},}\ }\href {\doibase
  10.1016/j.astropartphys.2013.04.007} {\bibfield  {journal} {\bibinfo
  {journal} {Astropart. Phys.}\ }\textbf {\bibinfo {volume} {46}},\ \bibinfo
  {pages} {55--70} (\bibinfo {year} {2013})},\ \Eprint
  {http://arxiv.org/abs/1209.3015} {arXiv:1209.3015 [astro-ph.HE]} \BibitemShut
  {NoStop}%
\bibitem [{\citenamefont {Hooper}\ and\ \citenamefont
  {Goodenough}(2011)}]{Hooper:2010mq}%
  \BibitemOpen
  \bibfield  {author} {\bibinfo {author} {\bibfnamefont {Dan}\ \bibnamefont
  {Hooper}}\ and\ \bibinfo {author} {\bibfnamefont {Lisa}\ \bibnamefont
  {Goodenough}},\ }\bibfield  {title} {\enquote {\bibinfo {title} {{Dark Matter
  Annihilation in The Galactic Center As Seen by the Fermi Gamma Ray Space
  Telescope}},}\ }\href {\doibase 10.1016/j.physletb.2011.02.029} {\bibfield
  {journal} {\bibinfo  {journal} {Phys. Lett.}\ }\textbf {\bibinfo {volume}
  {B697}},\ \bibinfo {pages} {412--428} (\bibinfo {year} {2011})},\ \Eprint
  {http://arxiv.org/abs/1010.2752} {arXiv:1010.2752 [hep-ph]} \BibitemShut
  {NoStop}%
\bibitem [{\citenamefont {Abazajian}(2011)}]{Abazajian:2010zy}%
  \BibitemOpen
  \bibfield  {author} {\bibinfo {author} {\bibfnamefont {Kevork~N.}\
  \bibnamefont {Abazajian}},\ }\bibfield  {title} {\enquote {\bibinfo {title}
  {{The Consistency of Fermi-LAT Observations of the Galactic Center with a
  Millisecond Pulsar Population in the Central Stellar Cluster}},}\ }\href
  {\doibase 10.1088/1475-7516/2011/03/010} {\bibfield  {journal} {\bibinfo
  {journal} {JCAP}\ }\textbf {\bibinfo {volume} {1103}},\ \bibinfo {pages}
  {010} (\bibinfo {year} {2011})},\ \Eprint {http://arxiv.org/abs/1011.4275}
  {arXiv:1011.4275 [astro-ph.HE]} \BibitemShut {NoStop}%
\bibitem [{\citenamefont {Abazajian}\ and\ \citenamefont
  {Kaplinghat}(2012)}]{Abazajian:2012pn}%
  \BibitemOpen
  \bibfield  {author} {\bibinfo {author} {\bibfnamefont {Kevork~N.}\
  \bibnamefont {Abazajian}}\ and\ \bibinfo {author} {\bibfnamefont {Manoj}\
  \bibnamefont {Kaplinghat}},\ }\bibfield  {title} {\enquote {\bibinfo {title}
  {{Detection of a Gamma-Ray Source in the Galactic Center Consistent with
  Extended Emission from Dark Matter Annihilation and Concentrated
  Astrophysical Emission}},}\ }\href {\doibase 10.1103/PhysRevD.86.083511,
  10.1103/PhysRevD.87.129902} {\bibfield  {journal} {\bibinfo  {journal} {Phys.
  Rev.}\ }\textbf {\bibinfo {volume} {D86}},\ \bibinfo {pages} {083511}
  (\bibinfo {year} {2012})},\ \bibinfo {note} {[Erratum: Phys.
  Rev.D87,129902(2013)]},\ \Eprint {http://arxiv.org/abs/1207.6047}
  {arXiv:1207.6047 [astro-ph.HE]} \BibitemShut {NoStop}%
\bibitem [{\citenamefont {Gordon}\ and\ \citenamefont
  {Macias}(2013)}]{Gordon:2013vta}%
  \BibitemOpen
  \bibfield  {author} {\bibinfo {author} {\bibfnamefont {Chris}\ \bibnamefont
  {Gordon}}\ and\ \bibinfo {author} {\bibfnamefont {Oscar}\ \bibnamefont
  {Macias}},\ }\bibfield  {title} {\enquote {\bibinfo {title} {{Dark Matter and
  Pulsar Model Constraints from Galactic Center Fermi-LAT Gamma Ray
  Observations}},}\ }\href {\doibase 10.1103/PhysRevD.88.083521,
  10.1103/PhysRevD.89.049901} {\bibfield  {journal} {\bibinfo  {journal} {Phys.
  Rev.}\ }\textbf {\bibinfo {volume} {D88}},\ \bibinfo {pages} {083521}
  (\bibinfo {year} {2013})},\ \bibinfo {note} {[Erratum: Phys.
  Rev.D89,no.4,049901(2014)]},\ \Eprint {http://arxiv.org/abs/1306.5725}
  {arXiv:1306.5725 [astro-ph.HE]} \BibitemShut {NoStop}%
\bibitem [{\citenamefont {Daylan}\ \emph {et~al.}(2016)\citenamefont {Daylan},
  \citenamefont {Finkbeiner}, \citenamefont {Hooper}, \citenamefont {Linden},
  \citenamefont {Portillo}, \citenamefont {Rodd},\ and\ \citenamefont
  {Slatyer}}]{Daylan:2014rsa}%
  \BibitemOpen
  \bibfield  {author} {\bibinfo {author} {\bibfnamefont {Tansu}\ \bibnamefont
  {Daylan}}, \bibinfo {author} {\bibfnamefont {Douglas~P.}\ \bibnamefont
  {Finkbeiner}}, \bibinfo {author} {\bibfnamefont {Dan}\ \bibnamefont
  {Hooper}}, \bibinfo {author} {\bibfnamefont {Tim}\ \bibnamefont {Linden}},
  \bibinfo {author} {\bibfnamefont {Stephen K.~N.}\ \bibnamefont {Portillo}},
  \bibinfo {author} {\bibfnamefont {Nicholas~L.}\ \bibnamefont {Rodd}}, \ and\
  \bibinfo {author} {\bibfnamefont {Tracy~R.}\ \bibnamefont {Slatyer}},\
  }\bibfield  {title} {\enquote {\bibinfo {title} {{The characterization of the
  gamma-ray signal from the central Milky Way: A case for annihilating dark
  matter}},}\ }\href {\doibase 10.1016/j.dark.2015.12.005} {\bibfield
  {journal} {\bibinfo  {journal} {Phys. Dark Univ.}\ }\textbf {\bibinfo
  {volume} {12}},\ \bibinfo {pages} {1--23} (\bibinfo {year} {2016})},\ \Eprint
  {http://arxiv.org/abs/1402.6703} {arXiv:1402.6703 [astro-ph.HE]} \BibitemShut
  {NoStop}%
\bibitem [{\citenamefont {Ajello}\ \emph {et~al.}(2016)\citenamefont {Ajello}
  \emph {et~al.}}]{TheFermi-LAT:2015kwa}%
  \BibitemOpen
  \bibfield  {author} {\bibinfo {author} {\bibfnamefont {M.}~\bibnamefont
  {Ajello}} \emph {et~al.} (\bibinfo {collaboration} {Fermi-LAT}),\ }\bibfield
  {title} {\enquote {\bibinfo {title} {{Fermi-LAT Observations of High-Energy
  $\gamma$-Ray Emission Toward the Galactic Center}},}\ }\href {\doibase
  10.3847/0004-637X/819/1/44} {\bibfield  {journal} {\bibinfo  {journal}
  {Astrophys. J.}\ }\textbf {\bibinfo {volume} {819}},\ \bibinfo {pages} {44}
  (\bibinfo {year} {2016})},\ \Eprint {http://arxiv.org/abs/1511.02938}
  {arXiv:1511.02938 [astro-ph.HE]} \BibitemShut {NoStop}%
\bibitem [{\citenamefont {Carlson}\ and\ \citenamefont
  {Profumo}(2014)}]{Carlson:2014cwa}%
  \BibitemOpen
  \bibfield  {author} {\bibinfo {author} {\bibfnamefont {Eric}\ \bibnamefont
  {Carlson}}\ and\ \bibinfo {author} {\bibfnamefont {Stefano}\ \bibnamefont
  {Profumo}},\ }\bibfield  {title} {\enquote {\bibinfo {title} {{Cosmic Ray
  Protons in the Inner Galaxy and the Galactic Center Gamma-Ray Excess}},}\
  }\href {\doibase 10.1103/PhysRevD.90.023015} {\bibfield  {journal} {\bibinfo
  {journal} {Phys. Rev.}\ }\textbf {\bibinfo {volume} {D90}},\ \bibinfo {pages}
  {023015} (\bibinfo {year} {2014})},\ \Eprint {http://arxiv.org/abs/1405.7685}
  {arXiv:1405.7685 [astro-ph.HE]} \BibitemShut {NoStop}%
\bibitem [{\citenamefont {Abazajian}\ \emph {et~al.}(2015)\citenamefont
  {Abazajian}, \citenamefont {Canac}, \citenamefont {Horiuchi}, \citenamefont
  {Kaplinghat},\ and\ \citenamefont {Kwa}}]{Abazajian:2014hsa}%
  \BibitemOpen
  \bibfield  {author} {\bibinfo {author} {\bibfnamefont {Kevork~N.}\
  \bibnamefont {Abazajian}}, \bibinfo {author} {\bibfnamefont {Nicolas}\
  \bibnamefont {Canac}}, \bibinfo {author} {\bibfnamefont {Shunsaku}\
  \bibnamefont {Horiuchi}}, \bibinfo {author} {\bibfnamefont {Manoj}\
  \bibnamefont {Kaplinghat}}, \ and\ \bibinfo {author} {\bibfnamefont {Anna}\
  \bibnamefont {Kwa}},\ }\bibfield  {title} {\enquote {\bibinfo {title}
  {{Discovery of a New Galactic Center Excess Consistent with Upscattered
  Starlight}},}\ }\href {\doibase 10.1088/1475-7516/2015/07/013} {\bibfield
  {journal} {\bibinfo  {journal} {JCAP}\ }\textbf {\bibinfo {volume} {1507}},\
  \bibinfo {pages} {013} (\bibinfo {year} {2015})},\ \Eprint
  {http://arxiv.org/abs/1410.6168} {arXiv:1410.6168 [astro-ph.HE]} \BibitemShut
  {NoStop}%
\bibitem [{\citenamefont {Gaggero}\ \emph {et~al.}(2015)\citenamefont
  {Gaggero}, \citenamefont {Taoso}, \citenamefont {Urbano}, \citenamefont
  {Valli},\ and\ \citenamefont {Ullio}}]{Gaggero:2015nsa}%
  \BibitemOpen
  \bibfield  {author} {\bibinfo {author} {\bibfnamefont {Daniele}\ \bibnamefont
  {Gaggero}}, \bibinfo {author} {\bibfnamefont {Marco}\ \bibnamefont {Taoso}},
  \bibinfo {author} {\bibfnamefont {Alfredo}\ \bibnamefont {Urbano}}, \bibinfo
  {author} {\bibfnamefont {Mauro}\ \bibnamefont {Valli}}, \ and\ \bibinfo
  {author} {\bibfnamefont {Piero}\ \bibnamefont {Ullio}},\ }\bibfield  {title}
  {\enquote {\bibinfo {title} {{Towards a realistic astrophysical
  interpretation of the gamma-ray Galactic center excess}},}\ }\href {\doibase
  10.1088/1475-7516-2015-12-056, 10.1088/1475-7516/2015/12/056} {\bibfield
  {journal} {\bibinfo  {journal} {JCAP}\ }\textbf {\bibinfo {volume} {1512}},\
  \bibinfo {pages} {056} (\bibinfo {year} {2015})},\ \Eprint
  {http://arxiv.org/abs/1507.06129} {arXiv:1507.06129 [astro-ph.HE]}
  \BibitemShut {NoStop}%
\bibitem [{\citenamefont {Macias}\ \emph {et~al.}(2018)\citenamefont {Macias},
  \citenamefont {Gordon}, \citenamefont {Crocker}, \citenamefont {Coleman},
  \citenamefont {Paterson}, \citenamefont {Horiuchi},\ and\ \citenamefont
  {Pohl}}]{Macias:2016nev}%
  \BibitemOpen
  \bibfield  {author} {\bibinfo {author} {\bibfnamefont {Oscar}\ \bibnamefont
  {Macias}}, \bibinfo {author} {\bibfnamefont {Chris}\ \bibnamefont {Gordon}},
  \bibinfo {author} {\bibfnamefont {Roland~M.}\ \bibnamefont {Crocker}},
  \bibinfo {author} {\bibfnamefont {Brendan}\ \bibnamefont {Coleman}}, \bibinfo
  {author} {\bibfnamefont {Dylan}\ \bibnamefont {Paterson}}, \bibinfo {author}
  {\bibfnamefont {Shunsaku}\ \bibnamefont {Horiuchi}}, \ and\ \bibinfo {author}
  {\bibfnamefont {Martin}\ \bibnamefont {Pohl}},\ }\bibfield  {title} {\enquote
  {\bibinfo {title} {{Galactic bulge preferred over dark matter for the
  Galactic centre gamma-ray excess}},}\ }\href {\doibase
  10.1038/s41550-018-0414-3} {\bibfield  {journal} {\bibinfo  {journal} {Nat.
  Astron.}\ }\textbf {\bibinfo {volume} {2}},\ \bibinfo {pages} {387--392}
  (\bibinfo {year} {2018})},\ \Eprint {http://arxiv.org/abs/1611.06644}
  {arXiv:1611.06644 [astro-ph.HE]} \BibitemShut {NoStop}%
\bibitem [{\citenamefont {Bartels}\ \emph {et~al.}(2018)\citenamefont
  {Bartels}, \citenamefont {Storm}, \citenamefont {Weniger},\ and\
  \citenamefont {Calore}}]{Bartels:2017vsx}%
  \BibitemOpen
  \bibfield  {author} {\bibinfo {author} {\bibfnamefont {Richard}\ \bibnamefont
  {Bartels}}, \bibinfo {author} {\bibfnamefont {Emma}\ \bibnamefont {Storm}},
  \bibinfo {author} {\bibfnamefont {Christoph}\ \bibnamefont {Weniger}}, \ and\
  \bibinfo {author} {\bibfnamefont {Francesca}\ \bibnamefont {Calore}},\
  }\bibfield  {title} {\enquote {\bibinfo {title} {{The Fermi-LAT GeV excess as
  a tracer of stellar mass in the Galactic bulge}},}\ }\href {\doibase
  10.1038/s41550-018-0531-z} {\bibfield  {journal} {\bibinfo  {journal} {Nat.
  Astron.}\ }\textbf {\bibinfo {volume} {2}},\ \bibinfo {pages} {819--828}
  (\bibinfo {year} {2018})},\ \Eprint {http://arxiv.org/abs/1711.04778}
  {arXiv:1711.04778 [astro-ph.HE]} \BibitemShut {NoStop}%
\bibitem [{\citenamefont {{Bland-Hawthorn}}\ and\ \citenamefont
  {{Gerhard}}(2016{\natexlab{a}})}]{2016ARA&A..54..529B}%
  \BibitemOpen
  \bibfield  {author} {\bibinfo {author} {\bibfnamefont {Joss}\ \bibnamefont
  {{Bland-Hawthorn}}}\ and\ \bibinfo {author} {\bibfnamefont {Ortwin}\
  \bibnamefont {{Gerhard}}},\ }\bibfield  {title} {\enquote {\bibinfo {title}
  {{The Galaxy in Context: Structural, Kinematic, and Integrated
  Properties}},}\ }\href {\doibase 10.1146/annurev-astro-081915-023441}
  {\bibfield  {journal} {\bibinfo  {journal} {Annual Review of Astronomy and
  Astrophysics}\ }\textbf {\bibinfo {volume} {54}},\ \bibinfo {pages}
  {529--596} (\bibinfo {year} {2016}{\natexlab{a}})},\ \Eprint
  {http://arxiv.org/abs/1602.07702} {arXiv:1602.07702 [astro-ph.GA]}
  \BibitemShut {NoStop}%
\bibitem [{\citenamefont {Macias}\ \emph {et~al.}(2019)\citenamefont {Macias},
  \citenamefont {Horiuchi}, \citenamefont {Kaplinghat}, \citenamefont {Gordon},
  \citenamefont {Crocker},\ and\ \citenamefont {Nataf}}]{Macias:2019omb}%
  \BibitemOpen
  \bibfield  {author} {\bibinfo {author} {\bibfnamefont {Oscar}\ \bibnamefont
  {Macias}}, \bibinfo {author} {\bibfnamefont {Shunsaku}\ \bibnamefont
  {Horiuchi}}, \bibinfo {author} {\bibfnamefont {Manoj}\ \bibnamefont
  {Kaplinghat}}, \bibinfo {author} {\bibfnamefont {Chris}\ \bibnamefont
  {Gordon}}, \bibinfo {author} {\bibfnamefont {Roland~M.}\ \bibnamefont
  {Crocker}}, \ and\ \bibinfo {author} {\bibfnamefont {David~M.}\ \bibnamefont
  {Nataf}},\ }\bibfield  {title} {\enquote {\bibinfo {title} {{Strong Evidence
  that the Galactic Bulge is Shining in Gamma Rays}},}\ }\href@noop {} {\
  (\bibinfo {year} {2019})},\ \Eprint {http://arxiv.org/abs/1901.03822}
  {arXiv:1901.03822 [astro-ph.HE]} \BibitemShut {NoStop}%
\bibitem [{\citenamefont {Abdollahi}\ \emph {et~al.}(2019)\citenamefont
  {Abdollahi} \emph {et~al.}}]{Fermi-LAT:4FGL}%
  \BibitemOpen
  \bibfield  {author} {\bibinfo {author} {\bibfnamefont {S.}~\bibnamefont
  {Abdollahi}} \emph {et~al.} (\bibinfo {collaboration} {Fermi-LAT}),\
  }\bibfield  {title} {\enquote {\bibinfo {title} {{Fermi Large Area Telescope
  Fourth Source Catalog}},}\ }\href@noop {} {\  (\bibinfo {year} {2019})},\
  \Eprint {http://arxiv.org/abs/1902.10045} {arXiv:1902.10045 [astro-ph.HE]}
  \BibitemShut {NoStop}%
\bibitem [{\citenamefont {Albert}\ \emph {et~al.}(2017)\citenamefont {Albert}
  \emph {et~al.}}]{Fermi-LAT:2016uux}%
  \BibitemOpen
  \bibfield  {author} {\bibinfo {author} {\bibfnamefont {A.}~\bibnamefont
  {Albert}} \emph {et~al.} (\bibinfo {collaboration} {DES, Fermi-LAT}),\
  }\bibfield  {title} {\enquote {\bibinfo {title} {{Searching for Dark Matter
  Annihilation in Recently Discovered Milky Way Satellites with Fermi-LAT}},}\
  }\href {\doibase 10.3847/1538-4357/834/2/110} {\bibfield  {journal} {\bibinfo
   {journal} {Astrophys. J.}\ }\textbf {\bibinfo {volume} {834}},\ \bibinfo
  {pages} {110} (\bibinfo {year} {2017})},\ \Eprint
  {http://arxiv.org/abs/1611.03184} {arXiv:1611.03184 [astro-ph.HE]}
  \BibitemShut {NoStop}%
\bibitem [{\citenamefont {Geringer-Sameth}\ \emph {et~al.}(2015)\citenamefont
  {Geringer-Sameth}, \citenamefont {Koushiappas},\ and\ \citenamefont
  {Walker}}]{Geringer-Sameth:2014yza}%
  \BibitemOpen
  \bibfield  {author} {\bibinfo {author} {\bibfnamefont {Alex}\ \bibnamefont
  {Geringer-Sameth}}, \bibinfo {author} {\bibfnamefont {Savvas~M.}\
  \bibnamefont {Koushiappas}}, \ and\ \bibinfo {author} {\bibfnamefont
  {Matthew}\ \bibnamefont {Walker}},\ }\bibfield  {title} {\enquote {\bibinfo
  {title} {{Dwarf galaxy annihilation and decay emission profiles for dark
  matter experiments}},}\ }\href {\doibase 10.1088/0004-637X/801/2/74}
  {\bibfield  {journal} {\bibinfo  {journal} {Astrophys. J.}\ }\textbf
  {\bibinfo {volume} {801}},\ \bibinfo {pages} {74} (\bibinfo {year} {2015})},\
  \Eprint {http://arxiv.org/abs/1408.0002} {arXiv:1408.0002 [astro-ph.CO]}
  \BibitemShut {NoStop}%
\bibitem [{\citenamefont {Abdo}\ \emph {et~al.}(2010)\citenamefont {Abdo} \emph
  {et~al.}}]{Abdo:2010ex}%
  \BibitemOpen
  \bibfield  {author} {\bibinfo {author} {\bibfnamefont {A.~A.}\ \bibnamefont
  {Abdo}} \emph {et~al.} (\bibinfo {collaboration} {Fermi-LAT}),\ }\bibfield
  {title} {\enquote {\bibinfo {title} {{Observations of Milky Way Dwarf
  Spheroidal galaxies with the Fermi-LAT detector and constraints on Dark
  Matter models}},}\ }\href {\doibase 10.1088/0004-637X/712/1/147} {\bibfield
  {journal} {\bibinfo  {journal} {Astrophys. J.}\ }\textbf {\bibinfo {volume}
  {712}},\ \bibinfo {pages} {147--158} (\bibinfo {year} {2010})},\ \Eprint
  {http://arxiv.org/abs/1001.4531} {arXiv:1001.4531 [astro-ph.CO]} \BibitemShut
  {NoStop}%
\bibitem [{\citenamefont {{Steigman}}\ \emph {et~al.}(2012)\citenamefont
  {{Steigman}}, \citenamefont {{Dasgupta}},\ and\ \citenamefont
  {{Beacom}}}]{Steigman2012}%
  \BibitemOpen
  \bibfield  {author} {\bibinfo {author} {\bibfnamefont {Gary}\ \bibnamefont
  {{Steigman}}}, \bibinfo {author} {\bibfnamefont {Basudeb}\ \bibnamefont
  {{Dasgupta}}}, \ and\ \bibinfo {author} {\bibfnamefont {John~F.}\
  \bibnamefont {{Beacom}}},\ }\bibfield  {title} {\enquote {\bibinfo {title}
  {{Precise relic WIMP abundance and its impact on searches for dark matter
  annihilation}},}\ }\href {\doibase 10.1103/PhysRevD.86.023506} {\bibfield
  {journal} {\bibinfo  {journal} {\prd}\ }\textbf {\bibinfo {volume} {86}},\
  \bibinfo {eid} {023506} (\bibinfo {year} {2012})},\ \Eprint
  {http://arxiv.org/abs/1204.3622} {arXiv:1204.3622 [hep-ph]} \BibitemShut
  {NoStop}%
\bibitem [{\citenamefont {Abdallah}\ \emph {et~al.}(2016)\citenamefont
  {Abdallah} \emph {et~al.}}]{Abdallah:2016ygi}%
  \BibitemOpen
  \bibfield  {author} {\bibinfo {author} {\bibfnamefont {H.}~\bibnamefont
  {Abdallah}} \emph {et~al.} (\bibinfo {collaboration} {H.E.S.S.}),\ }\bibfield
   {title} {\enquote {\bibinfo {title} {{Search for dark matter annihilations
  towards the inner Galactic halo from 10 years of observations with
  H.E.S.S}},}\ }\href {\doibase 10.1103/PhysRevLett.117.111301} {\bibfield
  {journal} {\bibinfo  {journal} {Phys. Rev. Lett.}\ }\textbf {\bibinfo
  {volume} {117}},\ \bibinfo {pages} {111301} (\bibinfo {year} {2016})},\
  \Eprint {http://arxiv.org/abs/1607.08142} {arXiv:1607.08142 [astro-ph.HE]}
  \BibitemShut {NoStop}%
\bibitem [{\citenamefont {Read}\ \emph {et~al.}(2016)\citenamefont {Read},
  \citenamefont {Agertz},\ and\ \citenamefont {Collins}}]{Read:2015sta}%
  \BibitemOpen
  \bibfield  {author} {\bibinfo {author} {\bibfnamefont {J.~I.}\ \bibnamefont
  {Read}}, \bibinfo {author} {\bibfnamefont {O.}~\bibnamefont {Agertz}}, \ and\
  \bibinfo {author} {\bibfnamefont {M.~L.~M.}\ \bibnamefont {Collins}},\
  }\bibfield  {title} {\enquote {\bibinfo {title} {{Dark matter cores all the
  way down}},}\ }\href {\doibase 10.1093/mnras/stw713} {\bibfield  {journal}
  {\bibinfo  {journal} {Mon. Not. Roy. Astron. Soc.}\ }\textbf {\bibinfo
  {volume} {459}},\ \bibinfo {pages} {2573--2590} (\bibinfo {year} {2016})},\
  \Eprint {http://arxiv.org/abs/1508.04143} {arXiv:1508.04143 [astro-ph.GA]}
  \BibitemShut {NoStop}%
\bibitem [{\citenamefont {Sánchez-Conde}\ and\ \citenamefont
  {Prada}(2014)}]{Sanchez-Conde:2013yxa}%
  \BibitemOpen
  \bibfield  {author} {\bibinfo {author} {\bibfnamefont {Miguel~A.}\
  \bibnamefont {Sánchez-Conde}}\ and\ \bibinfo {author} {\bibfnamefont
  {Francisco}\ \bibnamefont {Prada}},\ }\bibfield  {title} {\enquote {\bibinfo
  {title} {{The flattening of the concentration–mass relation towards low
  halo masses and its implications for the annihilation signal boost}},}\
  }\href {\doibase 10.1093/mnras/stu1014} {\bibfield  {journal} {\bibinfo
  {journal} {Mon. Not. Roy. Astron. Soc.}\ }\textbf {\bibinfo {volume} {442}},\
  \bibinfo {pages} {2271--2277} (\bibinfo {year} {2014})},\ \Eprint
  {http://arxiv.org/abs/1312.1729} {arXiv:1312.1729 [astro-ph.CO]} \BibitemShut
  {NoStop}%
\bibitem [{\citenamefont {Zhang}\ \emph {et~al.}(2013)\citenamefont {Zhang},
  \citenamefont {Rix}, \citenamefont {van~de Ven}, \citenamefont {Bovy},
  \citenamefont {Liu},\ and\ \citenamefont {Zhao}}]{Zhang:2012rsb}%
  \BibitemOpen
  \bibfield  {author} {\bibinfo {author} {\bibfnamefont {Lan}\ \bibnamefont
  {Zhang}}, \bibinfo {author} {\bibfnamefont {Hans-Walter}\ \bibnamefont
  {Rix}}, \bibinfo {author} {\bibfnamefont {Glenn}\ \bibnamefont {van~de Ven}},
  \bibinfo {author} {\bibfnamefont {Jo}~\bibnamefont {Bovy}}, \bibinfo {author}
  {\bibfnamefont {Chao}\ \bibnamefont {Liu}}, \ and\ \bibinfo {author}
  {\bibfnamefont {Gang}\ \bibnamefont {Zhao}},\ }\bibfield  {title} {\enquote
  {\bibinfo {title} {{The Gravitational Potential Near the Sun From SEGUE
  K-dwarf Kinematics}},}\ }\href {\doibase 10.1088/0004-637X/772/2/108}
  {\bibfield  {journal} {\bibinfo  {journal} {Astrophys. J.}\ }\textbf
  {\bibinfo {volume} {772}},\ \bibinfo {pages} {108} (\bibinfo {year}
  {2013})},\ \Eprint {http://arxiv.org/abs/1209.0256} {arXiv:1209.0256
  [astro-ph.GA]} \BibitemShut {NoStop}%
\bibitem [{\citenamefont {{Watkins}}\ \emph {et~al.}(2019)\citenamefont
  {{Watkins}}, \citenamefont {{van der Marel}}, \citenamefont {{Sohn}},\ and\
  \citenamefont {{Evans}}}]{2019ApJ...873..118W}%
  \BibitemOpen
  \bibfield  {author} {\bibinfo {author} {\bibfnamefont {Laura~L.}\
  \bibnamefont {{Watkins}}}, \bibinfo {author} {\bibfnamefont {Roeland~P.}\
  \bibnamefont {{van der Marel}}}, \bibinfo {author} {\bibfnamefont
  {Sangmo~Tony}\ \bibnamefont {{Sohn}}}, \ and\ \bibinfo {author}
  {\bibfnamefont {N.~Wyn}\ \bibnamefont {{Evans}}},\ }\bibfield  {title}
  {\enquote {\bibinfo {title} {{Evidence for an Intermediate-mass Milky Way
  from Gaia DR2 Halo Globular Cluster Motions}},}\ }\href {\doibase
  10.3847/1538-4357/ab089f} {\bibfield  {journal} {\bibinfo  {journal} {\apj}\
  }\textbf {\bibinfo {volume} {873}},\ \bibinfo {eid} {118} (\bibinfo {year}
  {2019})},\ \Eprint {http://arxiv.org/abs/1804.11348} {arXiv:1804.11348
  [astro-ph.GA]} \BibitemShut {NoStop}%
\bibitem [{\citenamefont {{Posti}}\ and\ \citenamefont
  {{Helmi}}(2019)}]{2019A&A...621A..56P}%
  \BibitemOpen
  \bibfield  {author} {\bibinfo {author} {\bibfnamefont {Lorenzo}\ \bibnamefont
  {{Posti}}}\ and\ \bibinfo {author} {\bibfnamefont {Amina}\ \bibnamefont
  {{Helmi}}},\ }\bibfield  {title} {\enquote {\bibinfo {title} {{Mass and shape
  of the Milky Way's dark matter halo with globular clusters from Gaia and
  Hubble}},}\ }\href {\doibase 10.1051/0004-6361/201833355} {\bibfield
  {journal} {\bibinfo  {journal} {\aap}\ }\textbf {\bibinfo {volume} {621}},\
  \bibinfo {eid} {A56} (\bibinfo {year} {2019})},\ \Eprint
  {http://arxiv.org/abs/1805.01408} {arXiv:1805.01408 [astro-ph.GA]}
  \BibitemShut {NoStop}%
\bibitem [{\citenamefont {Guedes}\ \emph {et~al.}(2011)\citenamefont {Guedes},
  \citenamefont {Callegari}, \citenamefont {Madau},\ and\ \citenamefont
  {Mayer}}]{Guedes:2011ux}%
  \BibitemOpen
  \bibfield  {author} {\bibinfo {author} {\bibfnamefont {Javiera}\ \bibnamefont
  {Guedes}}, \bibinfo {author} {\bibfnamefont {Simone}\ \bibnamefont
  {Callegari}}, \bibinfo {author} {\bibfnamefont {Piero}\ \bibnamefont
  {Madau}}, \ and\ \bibinfo {author} {\bibfnamefont {Lucio}\ \bibnamefont
  {Mayer}},\ }\bibfield  {title} {\enquote {\bibinfo {title} {{Forming
  Realistic Late-Type Spirals in a LCDM Universe: The Eris Simulation}},}\
  }\href {\doibase 10.1088/0004-637X/742/2/76} {\bibfield  {journal} {\bibinfo
  {journal} {Astrophys. J.}\ }\textbf {\bibinfo {volume} {742}},\ \bibinfo
  {pages} {76} (\bibinfo {year} {2011})},\ \Eprint
  {http://arxiv.org/abs/1103.6030} {arXiv:1103.6030 [astro-ph.CO]} \BibitemShut
  {NoStop}%
\bibitem [{\citenamefont {Kuhlen}\ \emph {et~al.}(2013)\citenamefont {Kuhlen},
  \citenamefont {Guedes}, \citenamefont {Pillepich}, \citenamefont {Madau},\
  and\ \citenamefont {Mayer}}]{Kuhlen:2012qw}%
  \BibitemOpen
  \bibfield  {author} {\bibinfo {author} {\bibfnamefont {Michael}\ \bibnamefont
  {Kuhlen}}, \bibinfo {author} {\bibfnamefont {Javiera}\ \bibnamefont
  {Guedes}}, \bibinfo {author} {\bibfnamefont {Annalisa}\ \bibnamefont
  {Pillepich}}, \bibinfo {author} {\bibfnamefont {Piero}\ \bibnamefont
  {Madau}}, \ and\ \bibinfo {author} {\bibfnamefont {Lucio}\ \bibnamefont
  {Mayer}},\ }\bibfield  {title} {\enquote {\bibinfo {title} {{An Off-center
  Density Peak in the Milky Way's Dark Matter Halo?}}}\ }\href {\doibase
  10.1088/0004-637X/765/1/10} {\bibfield  {journal} {\bibinfo  {journal}
  {Astrophys. J.}\ }\textbf {\bibinfo {volume} {765}},\ \bibinfo {pages} {10}
  (\bibinfo {year} {2013})},\ \Eprint {http://arxiv.org/abs/1208.4844}
  {arXiv:1208.4844 [astro-ph.GA]} \BibitemShut {NoStop}%
\bibitem [{\citenamefont {Weinberg}\ and\ \citenamefont
  {Katz}(2002)}]{Weinberg:2001gm}%
  \BibitemOpen
  \bibfield  {author} {\bibinfo {author} {\bibfnamefont {Martin~D.}\
  \bibnamefont {Weinberg}}\ and\ \bibinfo {author} {\bibfnamefont {Neal}\
  \bibnamefont {Katz}},\ }\bibfield  {title} {\enquote {\bibinfo {title}
  {{Bar-driven dark halo evolution: a resolution of the cusp-core
  controversy}},}\ }\href {\doibase 10.1086/343847} {\bibfield  {journal}
  {\bibinfo  {journal} {Astrophys. J.}\ }\textbf {\bibinfo {volume} {580}},\
  \bibinfo {pages} {627--633} (\bibinfo {year} {2002})},\ \Eprint
  {http://arxiv.org/abs/astro-ph/0110632} {arXiv:astro-ph/0110632 [astro-ph]}
  \BibitemShut {NoStop}%
\bibitem [{\citenamefont {Weinberg}\ and\ \citenamefont
  {Katz}(2007)}]{Weinberg:2006ps}%
  \BibitemOpen
  \bibfield  {author} {\bibinfo {author} {\bibfnamefont {Martin~D.}\
  \bibnamefont {Weinberg}}\ and\ \bibinfo {author} {\bibfnamefont {Neal}\
  \bibnamefont {Katz}},\ }\bibfield  {title} {\enquote {\bibinfo {title} {{The
  bar-halo interaction. 2. secular evolution and the religion of n-body
  simulations}},}\ }\href {\doibase 10.1111/j.1365-2966.2006.11307.x}
  {\bibfield  {journal} {\bibinfo  {journal} {Mon. Not. Roy. Astron. Soc.}\
  }\textbf {\bibinfo {volume} {375}},\ \bibinfo {pages} {460--476} (\bibinfo
  {year} {2007})},\ \Eprint {http://arxiv.org/abs/astro-ph/0601138}
  {arXiv:astro-ph/0601138 [astro-ph]} \BibitemShut {NoStop}%
\bibitem [{\citenamefont {Robles}\ \emph {et~al.}(2019)\citenamefont {Robles},
  \citenamefont {Kelley}, \citenamefont {Bullock},\ and\ \citenamefont
  {Kaplinghat}}]{Robles:2019mfq}%
  \BibitemOpen
  \bibfield  {author} {\bibinfo {author} {\bibfnamefont {Victor~H.}\
  \bibnamefont {Robles}}, \bibinfo {author} {\bibfnamefont {Tyler}\
  \bibnamefont {Kelley}}, \bibinfo {author} {\bibfnamefont {James~S.}\
  \bibnamefont {Bullock}}, \ and\ \bibinfo {author} {\bibfnamefont {Manoj}\
  \bibnamefont {Kaplinghat}},\ }\bibfield  {title} {\enquote {\bibinfo {title}
  {{The Milky Way's Halo and Subhalos in Self-Interacting Dark Matter}},}\
  }\href@noop {} {\  (\bibinfo {year} {2019})},\ \Eprint
  {http://arxiv.org/abs/1903.01469} {arXiv:1903.01469 [astro-ph.GA]}
  \BibitemShut {NoStop}%
\bibitem [{\citenamefont {Chan}\ \emph {et~al.}(2015)\citenamefont {Chan},
  \citenamefont {Kere\v~s}, \citenamefont {Oñorbe}, \citenamefont {Hopkins},
  \citenamefont {Muratov}, \citenamefont {Faucher-Giguère},\ and\
  \citenamefont {Quataert}}]{Chan:2015tna}%
  \BibitemOpen
  \bibfield  {author} {\bibinfo {author} {\bibfnamefont {T.K.}\ \bibnamefont
  {Chan}}, \bibinfo {author} {\bibfnamefont {D.}~\bibnamefont {Kere\v~s}},
  \bibinfo {author} {\bibfnamefont {J.}~\bibnamefont {Oñorbe}}, \bibinfo
  {author} {\bibfnamefont {P.F.}\ \bibnamefont {Hopkins}}, \bibinfo {author}
  {\bibfnamefont {A.L.}\ \bibnamefont {Muratov}}, \bibinfo {author}
  {\bibfnamefont {C.~A.}\ \bibnamefont {Faucher-Giguère}}, \ and\ \bibinfo
  {author} {\bibfnamefont {E.}~\bibnamefont {Quataert}},\ }\bibfield  {title}
  {\enquote {\bibinfo {title} {{The impact of baryonic physics on the structure
  of dark matter haloes: the view from the FIRE cosmological simulations}},}\
  }\href {\doibase 10.1093/mnras/stv2165} {\bibfield  {journal} {\bibinfo
  {journal} {Mon. Not. Roy. Astron. Soc.}\ }\textbf {\bibinfo {volume} {454}},\
  \bibinfo {pages} {2981--3001} (\bibinfo {year} {2015})},\ \Eprint
  {http://arxiv.org/abs/1507.02282} {arXiv:1507.02282 [astro-ph.GA]}
  \BibitemShut {NoStop}%
\bibitem [{\citenamefont {{Portail}}\ \emph {et~al.}(2017)\citenamefont
  {{Portail}}, \citenamefont {{Gerhard}}, \citenamefont {{Wegg}},\ and\
  \citenamefont {{Ness}}}]{2017MNRAS.465.1621P}%
  \BibitemOpen
  \bibfield  {author} {\bibinfo {author} {\bibfnamefont {Matthieu}\
  \bibnamefont {{Portail}}}, \bibinfo {author} {\bibfnamefont {Ortwin}\
  \bibnamefont {{Gerhard}}}, \bibinfo {author} {\bibfnamefont {Christopher}\
  \bibnamefont {{Wegg}}}, \ and\ \bibinfo {author} {\bibfnamefont {Melissa}\
  \bibnamefont {{Ness}}},\ }\bibfield  {title} {\enquote {\bibinfo {title}
  {{Dynamical modelling of the galactic bulge and bar: the Milky Way's pattern
  speed, stellar and dark matter mass distribution}},}\ }\href {\doibase
  10.1093/mnras/stw2819} {\bibfield  {journal} {\bibinfo  {journal} {\mnras}\
  }\textbf {\bibinfo {volume} {465}},\ \bibinfo {pages} {1621--1644} (\bibinfo
  {year} {2017})},\ \Eprint {http://arxiv.org/abs/1608.07954} {arXiv:1608.07954
  [astro-ph.GA]} \BibitemShut {NoStop}%
\bibitem [{\citenamefont {Petersen}\ \emph {et~al.}(2016)\citenamefont
  {Petersen}, \citenamefont {Weinberg},\ and\ \citenamefont
  {Katz}}]{Petersen:2016xtd}%
  \BibitemOpen
  \bibfield  {author} {\bibinfo {author} {\bibfnamefont {Michael~S.}\
  \bibnamefont {Petersen}}, \bibinfo {author} {\bibfnamefont {Martin~D.}\
  \bibnamefont {Weinberg}}, \ and\ \bibinfo {author} {\bibfnamefont {Neal}\
  \bibnamefont {Katz}},\ }\bibfield  {title} {\enquote {\bibinfo {title} {{Dark
  Matter Trapping by Stellar Bars: The Shadow Bar}},}\ }\href {\doibase
  10.1093/mnras/stw2141} {\bibfield  {journal} {\bibinfo  {journal} {Mon. Not.
  Roy. Astron. Soc.}\ }\textbf {\bibinfo {volume} {463}},\ \bibinfo {pages}
  {1952--1967} (\bibinfo {year} {2016})},\ \Eprint
  {http://arxiv.org/abs/1602.04826} {arXiv:1602.04826 [astro-ph.GA]}
  \BibitemShut {NoStop}%
\bibitem [{\citenamefont {Dai}\ \emph {et~al.}(2018)\citenamefont {Dai},
  \citenamefont {Robertson},\ and\ \citenamefont {Madau}}]{Dai:2018ypv}%
  \BibitemOpen
  \bibfield  {author} {\bibinfo {author} {\bibfnamefont {Biwei}\ \bibnamefont
  {Dai}}, \bibinfo {author} {\bibfnamefont {Brant~E.}\ \bibnamefont
  {Robertson}}, \ and\ \bibinfo {author} {\bibfnamefont {Piero}\ \bibnamefont
  {Madau}},\ }\bibfield  {title} {\enquote {\bibinfo {title} {{Around The Way:
  Testing $\Lambda$CDM with Milky Way Stellar Stream Constraints}},}\ }\href
  {\doibase 10.3847/1538-4357/aabb06} {\bibfield  {journal} {\bibinfo
  {journal} {Astrophys. J.}\ }\textbf {\bibinfo {volume} {858}},\ \bibinfo
  {pages} {73} (\bibinfo {year} {2018})},\ \Eprint
  {http://arxiv.org/abs/1804.00669} {arXiv:1804.00669 [astro-ph.GA]}
  \BibitemShut {NoStop}%
\bibitem [{\citenamefont {Ackermann}\ \emph {et~al.}(2015)\citenamefont
  {Ackermann} \emph {et~al.}}]{Ackermann:2014usa}%
  \BibitemOpen
  \bibfield  {author} {\bibinfo {author} {\bibfnamefont {M.}~\bibnamefont
  {Ackermann}} \emph {et~al.} (\bibinfo {collaboration} {Fermi-LAT}),\
  }\bibfield  {title} {\enquote {\bibinfo {title} {{The spectrum of isotropic
  diffuse gamma-ray emission between 100 MeV and 820 GeV}},}\ }\href {\doibase
  10.1088/0004-637X/799/1/86} {\bibfield  {journal} {\bibinfo  {journal}
  {Astrophys. J.}\ }\textbf {\bibinfo {volume} {799}},\ \bibinfo {pages} {86}
  (\bibinfo {year} {2015})},\ \Eprint {http://arxiv.org/abs/1410.3696}
  {arXiv:1410.3696 [astro-ph.HE]} \BibitemShut {NoStop}%
\bibitem [{\citenamefont {{Acero}}\ \emph {et~al.}(2016)\citenamefont
  {{Acero}}, \citenamefont {{Ackermann}}, \citenamefont {{Ajello}},
  \citenamefont {{Albert}}, \citenamefont {{Baldini}}, \citenamefont
  {{Ballet}}, \citenamefont {{Barbiellini}}, \citenamefont {{Bastieri}},
  \citenamefont {{Bellazzini}}, \citenamefont {{Bissaldi}}, \citenamefont
  {{Bloom}}, \citenamefont {{Bonino}}, \citenamefont {{Bottacini}},
  \citenamefont {{Brandt}}, \citenamefont {{Bregeon}}, \citenamefont
  {{Bruel}},\ and\ \citenamefont {et~al.}}]{Casandjian:andFermiLat2016}%
  \BibitemOpen
  \bibfield  {author} {\bibinfo {author} {\bibfnamefont {F.}~\bibnamefont
  {{Acero}}}, \bibinfo {author} {\bibfnamefont {M.}~\bibnamefont
  {{Ackermann}}}, \bibinfo {author} {\bibfnamefont {M.}~\bibnamefont
  {{Ajello}}}, \bibinfo {author} {\bibfnamefont {A.}~\bibnamefont {{Albert}}},
  \bibinfo {author} {\bibfnamefont {L.}~\bibnamefont {{Baldini}}}, \bibinfo
  {author} {\bibfnamefont {J.}~\bibnamefont {{Ballet}}}, \bibinfo {author}
  {\bibfnamefont {G.}~\bibnamefont {{Barbiellini}}}, \bibinfo {author}
  {\bibfnamefont {D.}~\bibnamefont {{Bastieri}}}, \bibinfo {author}
  {\bibfnamefont {R.}~\bibnamefont {{Bellazzini}}}, \bibinfo {author}
  {\bibfnamefont {E.}~\bibnamefont {{Bissaldi}}}, \bibinfo {author}
  {\bibfnamefont {E.~D.}\ \bibnamefont {{Bloom}}}, \bibinfo {author}
  {\bibfnamefont {R.}~\bibnamefont {{Bonino}}}, \bibinfo {author}
  {\bibfnamefont {E.}~\bibnamefont {{Bottacini}}}, \bibinfo {author}
  {\bibfnamefont {T.~J.}\ \bibnamefont {{Brandt}}}, \bibinfo {author}
  {\bibfnamefont {J.}~\bibnamefont {{Bregeon}}}, \bibinfo {author}
  {\bibfnamefont {P.}~\bibnamefont {{Bruel}}}, \ and\ \bibinfo {author}
  {\bibnamefont {et~al.}},\ }\bibfield  {title} {\enquote {\bibinfo {title}
  {{Development of the Model of Galactic Interstellar Emission for Standard
  Point-source Analysis of Fermi Large Area Telescope Data}},}\ }\href
  {\doibase 10.3847/0067-0049/223/2/26} {\bibfield  {journal} {\bibinfo
  {journal} {Astrophys. J. Supp.}\ }\textbf {\bibinfo {volume} {223}},\
  \bibinfo {eid} {26} (\bibinfo {year} {2016})},\ \Eprint
  {http://arxiv.org/abs/1602.07246} {arXiv:1602.07246 [astro-ph.HE]}
  \BibitemShut {NoStop}%
\bibitem [{\citenamefont {Calore}\ \emph {et~al.}(2016)\citenamefont {Calore},
  \citenamefont {Di~Mauro}, \citenamefont {Donato}, \citenamefont {Hessels},\
  and\ \citenamefont {Weniger}}]{Calore:2015bsx}%
  \BibitemOpen
  \bibfield  {author} {\bibinfo {author} {\bibfnamefont {Francesca}\
  \bibnamefont {Calore}}, \bibinfo {author} {\bibfnamefont {Mattia}\
  \bibnamefont {Di~Mauro}}, \bibinfo {author} {\bibfnamefont {Fiorenza}\
  \bibnamefont {Donato}}, \bibinfo {author} {\bibfnamefont {Jason W.~T.}\
  \bibnamefont {Hessels}}, \ and\ \bibinfo {author} {\bibfnamefont {Christoph}\
  \bibnamefont {Weniger}},\ }\bibfield  {title} {\enquote {\bibinfo {title}
  {{Radio detection prospects for a bulge population of millisecond pulsars as
  suggested by Fermi LAT observations of the inner Galaxy}},}\ }\href {\doibase
  10.3847/0004-637X/827/2/143} {\bibfield  {journal} {\bibinfo  {journal}
  {Astrophys. J.}\ }\textbf {\bibinfo {volume} {827}},\ \bibinfo {pages} {143}
  (\bibinfo {year} {2016})},\ \Eprint {http://arxiv.org/abs/1512.06825}
  {arXiv:1512.06825 [astro-ph.HE]} \BibitemShut {NoStop}%
\bibitem [{\citenamefont {Song}\ \emph {et~al.}(2019)\citenamefont {Song},
  \citenamefont {Macias},\ and\ \citenamefont {Horiuchi}}]{Song:2019nrx}%
  \BibitemOpen
  \bibfield  {author} {\bibinfo {author} {\bibfnamefont {Deheng}\ \bibnamefont
  {Song}}, \bibinfo {author} {\bibfnamefont {Oscar}\ \bibnamefont {Macias}}, \
  and\ \bibinfo {author} {\bibfnamefont {Shunsaku}\ \bibnamefont {Horiuchi}},\
  }\bibfield  {title} {\enquote {\bibinfo {title} {{Inverse Compton emission
  from millisecond pulsars in the Galactic bulge}},}\ }\href {\doibase
  10.1103/PhysRevD.99.123020} {\bibfield  {journal} {\bibinfo  {journal} {Phys.
  Rev.}\ }\textbf {\bibinfo {volume} {D99}},\ \bibinfo {pages} {123020}
  (\bibinfo {year} {2019})},\ \Eprint {http://arxiv.org/abs/1901.07025}
  {arXiv:1901.07025 [astro-ph.HE]} \BibitemShut {NoStop}%
\bibitem [{\citenamefont {Charles}\ \emph {et~al.}(2016)\citenamefont {Charles}
  \emph {et~al.}}]{Charles:2016pgz}%
  \BibitemOpen
  \bibfield  {author} {\bibinfo {author} {\bibfnamefont {E.}~\bibnamefont
  {Charles}} \emph {et~al.} (\bibinfo {collaboration} {Fermi-LAT}),\ }\bibfield
   {title} {\enquote {\bibinfo {title} {{Sensitivity Projections for Dark
  Matter Searches with the Fermi Large Area Telescope}},}\ }\href {\doibase
  10.1016/j.physrep.2016.05.001} {\bibfield  {journal} {\bibinfo  {journal}
  {Phys. Rept.}\ }\textbf {\bibinfo {volume} {636}},\ \bibinfo {pages} {1--46}
  (\bibinfo {year} {2016})},\ \Eprint {http://arxiv.org/abs/1605.02016}
  {arXiv:1605.02016 [astro-ph.HE]} \BibitemShut {NoStop}%
\bibitem [{\citenamefont {Lee}\ \emph {et~al.}(2016)\citenamefont {Lee},
  \citenamefont {Lisanti}, \citenamefont {Safdi}, \citenamefont {Slatyer},\
  and\ \citenamefont {Xue}}]{Lee:2015fea}%
  \BibitemOpen
  \bibfield  {author} {\bibinfo {author} {\bibfnamefont {Samuel~K.}\
  \bibnamefont {Lee}}, \bibinfo {author} {\bibfnamefont {Mariangela}\
  \bibnamefont {Lisanti}}, \bibinfo {author} {\bibfnamefont {Benjamin~R.}\
  \bibnamefont {Safdi}}, \bibinfo {author} {\bibfnamefont {Tracy~R.}\
  \bibnamefont {Slatyer}}, \ and\ \bibinfo {author} {\bibfnamefont {Wei}\
  \bibnamefont {Xue}},\ }\bibfield  {title} {\enquote {\bibinfo {title}
  {{Evidence for Unresolved $\gamma$-Ray Point Sources in the Inner Galaxy}},}\
  }\href {\doibase 10.1103/PhysRevLett.116.051103} {\bibfield  {journal}
  {\bibinfo  {journal} {Phys. Rev. Lett.}\ }\textbf {\bibinfo {volume} {116}},\
  \bibinfo {pages} {051103} (\bibinfo {year} {2016})},\ \Eprint
  {http://arxiv.org/abs/1506.05124} {arXiv:1506.05124 [astro-ph.HE]}
  \BibitemShut {NoStop}%
\bibitem [{\citenamefont {Leane}\ and\ \citenamefont
  {Slatyer}(2019)}]{Leane:2019xiy}%
  \BibitemOpen
  \bibfield  {author} {\bibinfo {author} {\bibfnamefont {Rebecca~K.}\
  \bibnamefont {Leane}}\ and\ \bibinfo {author} {\bibfnamefont {Tracy~R.}\
  \bibnamefont {Slatyer}},\ }\bibfield  {title} {\enquote {\bibinfo {title}
  {{Dark Matter Strikes Back at the Galactic Center}},}\ }\href@noop {} {\
  (\bibinfo {year} {2019})},\ \Eprint {http://arxiv.org/abs/1904.08430}
  {arXiv:1904.08430 [astro-ph.HE]} \BibitemShut {NoStop}%
\bibitem [{\citenamefont {Chang}\ \emph {et~al.}(2020)\citenamefont {Chang},
  \citenamefont {Mishra-Sharma}, \citenamefont {Lisanti}, \citenamefont
  {Buschmann}, \citenamefont {Rodd},\ and\ \citenamefont
  {Safdi}}]{Chang:2019ars}%
  \BibitemOpen
  \bibfield  {author} {\bibinfo {author} {\bibfnamefont {Laura~J.}\
  \bibnamefont {Chang}}, \bibinfo {author} {\bibfnamefont {Siddharth}\
  \bibnamefont {Mishra-Sharma}}, \bibinfo {author} {\bibfnamefont {Mariangela}\
  \bibnamefont {Lisanti}}, \bibinfo {author} {\bibfnamefont {Malte}\
  \bibnamefont {Buschmann}}, \bibinfo {author} {\bibfnamefont {Nicholas~L.}\
  \bibnamefont {Rodd}}, \ and\ \bibinfo {author} {\bibfnamefont {Benjamin~R.}\
  \bibnamefont {Safdi}},\ }\bibfield  {title} {\enquote {\bibinfo {title}
  {{Characterizing the Nature of the Unresolved Point Sources in the Galactic
  Center}},}\ }\href {\doibase 10.1103/PhysRevD.101.023014} {\bibfield
  {journal} {\bibinfo  {journal} {Phys. Rev.}\ }\textbf {\bibinfo {volume}
  {D101}},\ \bibinfo {pages} {023014} (\bibinfo {year} {2020})},\ \Eprint
  {http://arxiv.org/abs/1908.10874} {arXiv:1908.10874 [astro-ph.CO]}
  \BibitemShut {NoStop}%
\bibitem [{\citenamefont {Leane}\ and\ \citenamefont
  {Slatyer}(2020{\natexlab{a}})}]{Leane:2020pfc}%
  \BibitemOpen
  \bibfield  {author} {\bibinfo {author} {\bibfnamefont {Rebecca~K.}\
  \bibnamefont {Leane}}\ and\ \bibinfo {author} {\bibfnamefont {Tracy~R.}\
  \bibnamefont {Slatyer}},\ }\bibfield  {title} {\enquote {\bibinfo {title}
  {{The Enigmatic Galactic Center Excess: Spurious Point Sources and Signal
  Mismodeling}},}\ }\href@noop {} {\  (\bibinfo {year} {2020}{\natexlab{a}})},\
  \Eprint {http://arxiv.org/abs/2002.12371} {arXiv:2002.12371 [astro-ph.HE]}
  \BibitemShut {NoStop}%
\bibitem [{\citenamefont {Leane}\ and\ \citenamefont
  {Slatyer}(2020{\natexlab{b}})}]{Leane:2020nmi}%
  \BibitemOpen
  \bibfield  {author} {\bibinfo {author} {\bibfnamefont {Rebecca~K.}\
  \bibnamefont {Leane}}\ and\ \bibinfo {author} {\bibfnamefont {Tracy~R.}\
  \bibnamefont {Slatyer}},\ }\bibfield  {title} {\enquote {\bibinfo {title}
  {{Spurious Point Source Signals in the Galactic Center Excess}},}\
  }\href@noop {} {\  (\bibinfo {year} {2020}{\natexlab{b}})},\ \Eprint
  {http://arxiv.org/abs/2002.12370} {arXiv:2002.12370 [astro-ph.HE]}
  \BibitemShut {NoStop}%
\bibitem [{\citenamefont {Buschmann}\ \emph {et~al.}(2020)\citenamefont
  {Buschmann}, \citenamefont {Rodd}, \citenamefont {Safdi}, \citenamefont
  {Chang}, \citenamefont {Mishra-Sharma}, \citenamefont {Lisanti},\ and\
  \citenamefont {Macias}}]{Buschmann:2020adf}%
  \BibitemOpen
  \bibfield  {author} {\bibinfo {author} {\bibfnamefont {Malte}\ \bibnamefont
  {Buschmann}}, \bibinfo {author} {\bibfnamefont {Nicholas~L.}\ \bibnamefont
  {Rodd}}, \bibinfo {author} {\bibfnamefont {Benjamin~R.}\ \bibnamefont
  {Safdi}}, \bibinfo {author} {\bibfnamefont {Laura~J.}\ \bibnamefont {Chang}},
  \bibinfo {author} {\bibfnamefont {Siddharth}\ \bibnamefont {Mishra-Sharma}},
  \bibinfo {author} {\bibfnamefont {Mariangela}\ \bibnamefont {Lisanti}}, \
  and\ \bibinfo {author} {\bibfnamefont {Oscar}\ \bibnamefont {Macias}},\
  }\bibfield  {title} {\enquote {\bibinfo {title} {{Foreground Mismodeling and
  the Point Source Explanation of the Fermi Galactic Center Excess}},}\
  }\href@noop {} {\  (\bibinfo {year} {2020})},\ \Eprint
  {http://arxiv.org/abs/2002.12373} {arXiv:2002.12373 [astro-ph.HE]}
  \BibitemShut {NoStop}%
\bibitem [{\citenamefont {Bartels}\ \emph {et~al.}(2016)\citenamefont
  {Bartels}, \citenamefont {Krishnamurthy},\ and\ \citenamefont
  {Weniger}}]{Bartels:2015aea}%
  \BibitemOpen
  \bibfield  {author} {\bibinfo {author} {\bibfnamefont {Richard}\ \bibnamefont
  {Bartels}}, \bibinfo {author} {\bibfnamefont {Suraj}\ \bibnamefont
  {Krishnamurthy}}, \ and\ \bibinfo {author} {\bibfnamefont {Christoph}\
  \bibnamefont {Weniger}},\ }\bibfield  {title} {\enquote {\bibinfo {title}
  {{Strong support for the millisecond pulsar origin of the Galactic center GeV
  excess}},}\ }\href {\doibase 10.1103/PhysRevLett.116.051102} {\bibfield
  {journal} {\bibinfo  {journal} {Phys. Rev. Lett.}\ }\textbf {\bibinfo
  {volume} {116}},\ \bibinfo {pages} {051102} (\bibinfo {year} {2016})},\
  \Eprint {http://arxiv.org/abs/1506.05104} {arXiv:1506.05104 [astro-ph.HE]}
  \BibitemShut {NoStop}%
\bibitem [{\citenamefont {Balaji}\ \emph {et~al.}(2018)\citenamefont {Balaji},
  \citenamefont {Cholis}, \citenamefont {Fox},\ and\ \citenamefont
  {McDermott}}]{Balaji:2018rwz}%
  \BibitemOpen
  \bibfield  {author} {\bibinfo {author} {\bibfnamefont {Bhaskaran}\
  \bibnamefont {Balaji}}, \bibinfo {author} {\bibfnamefont {Ilias}\
  \bibnamefont {Cholis}}, \bibinfo {author} {\bibfnamefont {Patrick~J.}\
  \bibnamefont {Fox}}, \ and\ \bibinfo {author} {\bibfnamefont {Samuel~D.}\
  \bibnamefont {McDermott}},\ }\bibfield  {title} {\enquote {\bibinfo {title}
  {{Analyzing the Gamma-Ray Sky with Wavelets}},}\ }\href {\doibase
  10.1103/PhysRevD.98.043009} {\bibfield  {journal} {\bibinfo  {journal} {Phys.
  Rev.}\ }\textbf {\bibinfo {volume} {D98}},\ \bibinfo {pages} {043009}
  (\bibinfo {year} {2018})},\ \Eprint {http://arxiv.org/abs/1803.01952}
  {arXiv:1803.01952 [astro-ph.HE]} \BibitemShut {NoStop}%
\bibitem [{\citenamefont {Zhong}\ \emph {et~al.}(2019)\citenamefont {Zhong},
  \citenamefont {McDermott}, \citenamefont {Cholis},\ and\ \citenamefont
  {Fox}}]{Zhong:2019ycb}%
  \BibitemOpen
  \bibfield  {author} {\bibinfo {author} {\bibfnamefont {Yi-Ming}\ \bibnamefont
  {Zhong}}, \bibinfo {author} {\bibfnamefont {Samuel~D.}\ \bibnamefont
  {McDermott}}, \bibinfo {author} {\bibfnamefont {Ilias}\ \bibnamefont
  {Cholis}}, \ and\ \bibinfo {author} {\bibfnamefont {Patrick~J.}\ \bibnamefont
  {Fox}},\ }\bibfield  {title} {\enquote {\bibinfo {title} {{A New Mask for An
  Old Suspect: Testing the Sensitivity of the Galactic Center Excess to the
  Point Source Mask}},}\ }\href@noop {} {\  (\bibinfo {year} {2019})},\ \Eprint
  {http://arxiv.org/abs/1911.12369} {arXiv:1911.12369 [astro-ph.HE]}
  \BibitemShut {NoStop}%
\bibitem [{\citenamefont {{Macquart}}\ and\ \citenamefont
  {{Kanekar}}(2015)}]{2015ApJ...805..172M}%
  \BibitemOpen
  \bibfield  {author} {\bibinfo {author} {\bibfnamefont {Jean-Pierre}\
  \bibnamefont {{Macquart}}}\ and\ \bibinfo {author} {\bibfnamefont {Nissim}\
  \bibnamefont {{Kanekar}}},\ }\bibfield  {title} {\enquote {\bibinfo {title}
  {{On Detecting Millisecond Pulsars at the Galactic Center}},}\ }\href
  {\doibase 10.1088/0004-637X/805/2/172} {\bibfield  {journal} {\bibinfo
  {journal} {\apj}\ }\textbf {\bibinfo {volume} {805}},\ \bibinfo {eid} {172}
  (\bibinfo {year} {2015})},\ \Eprint {http://arxiv.org/abs/1504.02492}
  {arXiv:1504.02492 [astro-ph.HE]} \BibitemShut {NoStop}%
\bibitem [{\citenamefont {{Calore}}\ \emph {et~al.}(2016)\citenamefont
  {{Calore}}, \citenamefont {{Di Mauro}}, \citenamefont {{Donato}},
  \citenamefont {{Hessels}},\ and\ \citenamefont
  {{Weniger}}}]{2016ApJ...827..143C}%
  \BibitemOpen
  \bibfield  {author} {\bibinfo {author} {\bibfnamefont {F.}~\bibnamefont
  {{Calore}}}, \bibinfo {author} {\bibfnamefont {M.}~\bibnamefont {{Di
  Mauro}}}, \bibinfo {author} {\bibfnamefont {F.}~\bibnamefont {{Donato}}},
  \bibinfo {author} {\bibfnamefont {J.~W.~T.}\ \bibnamefont {{Hessels}}}, \
  and\ \bibinfo {author} {\bibfnamefont {C.}~\bibnamefont {{Weniger}}},\
  }\bibfield  {title} {\enquote {\bibinfo {title} {{Radio Detection Prospects
  for a Bulge Population of Millisecond Pulsars as Suggested by Fermi-LAT
  Observations of the Inner Galaxy}},}\ }\href {\doibase
  10.3847/0004-637X/827/2/143} {\bibfield  {journal} {\bibinfo  {journal}
  {\apj}\ }\textbf {\bibinfo {volume} {827}},\ \bibinfo {eid} {143} (\bibinfo
  {year} {2016})},\ \Eprint {http://arxiv.org/abs/1512.06825} {arXiv:1512.06825
  [astro-ph.HE]} \BibitemShut {NoStop}%
\bibitem [{\citenamefont {{Rajwade}}\ \emph {et~al.}(2017)\citenamefont
  {{Rajwade}}, \citenamefont {{Lorimer}},\ and\ \citenamefont
  {{Anderson}}}]{2017MNRAS.471..730R}%
  \BibitemOpen
  \bibfield  {author} {\bibinfo {author} {\bibfnamefont {K.~M.}\ \bibnamefont
  {{Rajwade}}}, \bibinfo {author} {\bibfnamefont {D.~R.}\ \bibnamefont
  {{Lorimer}}}, \ and\ \bibinfo {author} {\bibfnamefont {L.~D.}\ \bibnamefont
  {{Anderson}}},\ }\bibfield  {title} {\enquote {\bibinfo {title} {{Detecting
  pulsars in the Galactic Centre}},}\ }\href {\doibase 10.1093/mnras/stx1661}
  {\bibfield  {journal} {\bibinfo  {journal} {\mnras}\ }\textbf {\bibinfo
  {volume} {471}},\ \bibinfo {pages} {730--739} (\bibinfo {year} {2017})},\
  \Eprint {http://arxiv.org/abs/1611.06977} {arXiv:1611.06977 [astro-ph.HE]}
  \BibitemShut {NoStop}%
\bibitem [{\citenamefont {{Hyman}}\ \emph {et~al.}(2019)\citenamefont
  {{Hyman}}, \citenamefont {{Frail}}, \citenamefont {{Deneva}}, \citenamefont
  {{Kassim}}, \citenamefont {{McLaughlin}}, \citenamefont {{Kooi}},
  \citenamefont {{Ray}},\ and\ \citenamefont
  {{Polisensky}}}]{2019ApJ...876...20H}%
  \BibitemOpen
  \bibfield  {author} {\bibinfo {author} {\bibfnamefont {S.~D.}\ \bibnamefont
  {{Hyman}}}, \bibinfo {author} {\bibfnamefont {D.~A.}\ \bibnamefont
  {{Frail}}}, \bibinfo {author} {\bibfnamefont {J.~S.}\ \bibnamefont
  {{Deneva}}}, \bibinfo {author} {\bibfnamefont {N.~E.}\ \bibnamefont
  {{Kassim}}}, \bibinfo {author} {\bibfnamefont {M.~A.}\ \bibnamefont
  {{McLaughlin}}}, \bibinfo {author} {\bibfnamefont {J.~E.}\ \bibnamefont
  {{Kooi}}}, \bibinfo {author} {\bibfnamefont {P.~S.}\ \bibnamefont {{Ray}}}, \
  and\ \bibinfo {author} {\bibfnamefont {E.~J.}\ \bibnamefont {{Polisensky}}},\
  }\bibfield  {title} {\enquote {\bibinfo {title} {{A Search for Pulsars in
  Steep-spectrum Radio Sources toward the Galactic Center}},}\ }\href {\doibase
  10.3847/1538-4357/ab11c8} {\bibfield  {journal} {\bibinfo  {journal} {\apj}\
  }\textbf {\bibinfo {volume} {876}},\ \bibinfo {eid} {20} (\bibinfo {year}
  {2019})}\BibitemShut {NoStop}%
\bibitem [{\citenamefont {Porter}\ \emph {et~al.}(2017)\citenamefont {Porter},
  \citenamefont {Johannesson},\ and\ \citenamefont
  {Moskalenko}}]{Porter:2017vaa}%
  \BibitemOpen
  \bibfield  {author} {\bibinfo {author} {\bibfnamefont {Troy~A.}\ \bibnamefont
  {Porter}}, \bibinfo {author} {\bibfnamefont {Gudlaugur}\ \bibnamefont
  {Johannesson}}, \ and\ \bibinfo {author} {\bibfnamefont {Igor~V.}\
  \bibnamefont {Moskalenko}},\ }\bibfield  {title} {\enquote {\bibinfo {title}
  {{High-Energy Gamma Rays from the Milky Way: Three-Dimensional Spatial Models
  for the Cosmic-Ray and Radiation Field Densities in the Interstellar
  Medium}},}\ }\href {\doibase 10.3847/1538-4357/aa844d} {\bibfield  {journal}
  {\bibinfo  {journal} {Astrophys. J.}\ }\textbf {\bibinfo {volume} {846}},\
  \bibinfo {pages} {67} (\bibinfo {year} {2017})},\ \Eprint
  {http://arxiv.org/abs/1708.00816} {arXiv:1708.00816 [astro-ph.HE]}
  \BibitemShut {NoStop}%
\bibitem [{\citenamefont {Ackermann}\ \emph {et~al.}(2017)\citenamefont
  {Ackermann} \emph {et~al.}}]{TheFermi-LAT:2017vmf}%
  \BibitemOpen
  \bibfield  {author} {\bibinfo {author} {\bibfnamefont {M.}~\bibnamefont
  {Ackermann}} \emph {et~al.} (\bibinfo {collaboration} {Fermi-LAT}),\
  }\bibfield  {title} {\enquote {\bibinfo {title} {{The Fermi Galactic Center
  GeV Excess and Implications for Dark Matter}},}\ }\href {\doibase
  10.3847/1538-4357/aa6cab} {\bibfield  {journal} {\bibinfo  {journal}
  {Astrophys. J.}\ }\textbf {\bibinfo {volume} {840}},\ \bibinfo {pages} {43}
  (\bibinfo {year} {2017})},\ \Eprint {http://arxiv.org/abs/1704.03910}
  {arXiv:1704.03910 [astro-ph.HE]} \BibitemShut {NoStop}%
\bibitem [{Gal()}]{Galprop}%
  \BibitemOpen
  \href@noop {} {\enquote {\bibinfo {title} {Galprop},}\ }\bibinfo
  {howpublished} {\url{http://galprop.stanford.edu}},\ \bibinfo {note}
  {accessed: 2018-10-15}\BibitemShut {NoStop}%
\bibitem [{\citenamefont {{Wolleben}}(2007)}]{Wolleben:2007}%
  \BibitemOpen
  \bibfield  {author} {\bibinfo {author} {\bibfnamefont {M.}~\bibnamefont
  {{Wolleben}}},\ }\bibfield  {title} {\enquote {\bibinfo {title} {{A New Model
  for the Loop I (North Polar Spur) Region}},}\ }\href {\doibase
  10.1086/518711} {\bibfield  {journal} {\bibinfo  {journal} {Astrophys.~J.}\
  }\textbf {\bibinfo {volume} {664}},\ \bibinfo {pages} {349--356} (\bibinfo
  {year} {2007})},\ \Eprint {http://arxiv.org/abs/0704.0276} {arXiv:0704.0276}
  \BibitemShut {NoStop}%
\bibitem [{\citenamefont {Nishiyama}\ \emph {et~al.}(2013)\citenamefont
  {Nishiyama}, \citenamefont {Yasui}, \citenamefont {Nagata}, \citenamefont
  {Yoshikawa}, \citenamefont {Uchiyama}, \citenamefont {Schdel}, \citenamefont
  {Hatano}, \citenamefont {Sato}, \citenamefont {Sugitani}, \citenamefont
  {Suenaga}, \citenamefont {Kwon},\ and\ \citenamefont
  {Tamura}}]{Nishiyama2015}%
  \BibitemOpen
  \bibfield  {author} {\bibinfo {author} {\bibfnamefont {Shogo}\ \bibnamefont
  {Nishiyama}}, \bibinfo {author} {\bibfnamefont {Kazuki}\ \bibnamefont
  {Yasui}}, \bibinfo {author} {\bibfnamefont {Tetsuya}\ \bibnamefont {Nagata}},
  \bibinfo {author} {\bibfnamefont {Tatsuhito}\ \bibnamefont {Yoshikawa}},
  \bibinfo {author} {\bibfnamefont {Hideki}\ \bibnamefont {Uchiyama}}, \bibinfo
  {author} {\bibfnamefont {Rainer}\ \bibnamefont {Schdel}}, \bibinfo {author}
  {\bibfnamefont {Hirofumi}\ \bibnamefont {Hatano}}, \bibinfo {author}
  {\bibfnamefont {Shuji}\ \bibnamefont {Sato}}, \bibinfo {author}
  {\bibfnamefont {Koji}\ \bibnamefont {Sugitani}}, \bibinfo {author}
  {\bibfnamefont {Takuya}\ \bibnamefont {Suenaga}}, \bibinfo {author}
  {\bibfnamefont {Jungmi}\ \bibnamefont {Kwon}}, \ and\ \bibinfo {author}
  {\bibfnamefont {Motohide}\ \bibnamefont {Tamura}},\ }\bibfield  {title}
  {\enquote {\bibinfo {title} {Magnetically confined interstellar hot plasma in
  the nuclear bulge of our galaxy},}\ }\href
  {http://stacks.iop.org/2041-8205/769/i=2/a=L28} {\bibfield  {journal}
  {\bibinfo  {journal} {ApJ. Lett.}\ }\textbf {\bibinfo {volume} {769}},\
  \bibinfo {pages} {L28} (\bibinfo {year} {2013})}\BibitemShut {NoStop}%
\bibitem [{\citenamefont {Freudenreich}(1998)}]{Freudenreich:1997bx}%
  \BibitemOpen
  \bibfield  {author} {\bibinfo {author} {\bibfnamefont {H.~T.}\ \bibnamefont
  {Freudenreich}},\ }\bibfield  {title} {\enquote {\bibinfo {title} {{Cobe's
  galactic bar and disk}},}\ }\href {\doibase 10.1086/305065} {\bibfield
  {journal} {\bibinfo  {journal} {Astrophys. J.}\ }\textbf {\bibinfo {volume}
  {492}},\ \bibinfo {pages} {495--510} (\bibinfo {year} {1998})},\ \Eprint
  {http://arxiv.org/abs/astro-ph/9707340} {arXiv:astro-ph/9707340 [astro-ph]}
  \BibitemShut {NoStop}%
\bibitem [{\citenamefont {{Freudenreich}}(1998)}]{Freudenreich:1998}%
  \BibitemOpen
  \bibfield  {author} {\bibinfo {author} {\bibfnamefont {H.~T.}\ \bibnamefont
  {{Freudenreich}}},\ }\bibfield  {title} {\enquote {\bibinfo {title} {{A COBE
  Model of the Galactic Bar and Disk}},}\ }\href {\doibase 10.1086/305065}
  {\bibfield  {journal} {\bibinfo  {journal} {\apj}\ }\textbf {\bibinfo
  {volume} {492}},\ \bibinfo {pages} {495--510} (\bibinfo {year} {1998})},\
  \Eprint {http://arxiv.org/abs/astro-ph/9707340} {astro-ph/9707340}
  \BibitemShut {NoStop}%
\bibitem [{\citenamefont {{Acero}}\ \emph {et~al.}(2015)\citenamefont
  {{Acero}}, \citenamefont {{Ackermann}}, \citenamefont {{Ajello}},
  \citenamefont {{Albert}} \emph {et~al.}}]{3FGL}%
  \BibitemOpen
  \bibfield  {author} {\bibinfo {author} {\bibfnamefont {F.}~\bibnamefont
  {{Acero}}}, \bibinfo {author} {\bibfnamefont {M.}~\bibnamefont
  {{Ackermann}}}, \bibinfo {author} {\bibfnamefont {M.}~\bibnamefont
  {{Ajello}}}, \bibinfo {author} {\bibfnamefont {A.}~\bibnamefont {{Albert}}},
  \emph {et~al.} (\bibinfo {collaboration} {Fermi-LAT}),\ }\bibfield  {title}
  {\enquote {\bibinfo {title} {{Fermi Large Area Telescope Third Source
  Catalog}},}\ }\href {\doibase 10.1088/0067-0049/218/2/23} {\bibfield
  {journal} {\bibinfo  {journal} {Astrophys. J. Supp.}\ }\textbf {\bibinfo
  {volume} {218}},\ \bibinfo {eid} {23} (\bibinfo {year} {2015})},\ \Eprint
  {http://arxiv.org/abs/1501.02003} {arXiv:1501.02003 [astro-ph.HE]}
  \BibitemShut {NoStop}%
\bibitem [{\citenamefont {{Pohl}}\ \emph {et~al.}(2008)\citenamefont {{Pohl}},
  \citenamefont {{Englmaier}},\ and\ \citenamefont {{Bissantz}}}]{Pohl2008}%
  \BibitemOpen
  \bibfield  {author} {\bibinfo {author} {\bibfnamefont {M.}~\bibnamefont
  {{Pohl}}}, \bibinfo {author} {\bibfnamefont {P.}~\bibnamefont {{Englmaier}}},
  \ and\ \bibinfo {author} {\bibfnamefont {N.}~\bibnamefont {{Bissantz}}},\
  }\bibfield  {title} {\enquote {\bibinfo {title} {{Three-Dimensional
  Distribution of Molecular Gas in the Barred Milky Way}},}\ }\href {\doibase
  10.1086/529004} {\bibfield  {journal} {\bibinfo  {journal} {Astrophys.~J.}\
  }\textbf {\bibinfo {volume} {677}},\ \bibinfo {eid} {283-291} (\bibinfo
  {year} {2008})},\ \Eprint {http://arxiv.org/abs/0712.4264} {arXiv:0712.4264}
  \BibitemShut {NoStop}%
\bibitem [{\citenamefont {Ackermann}\ \emph {et~al.}(2012)\citenamefont
  {Ackermann} \emph {et~al.}}]{ackermannajelloatwood2012}%
  \BibitemOpen
  \bibfield  {author} {\bibinfo {author} {\bibfnamefont {M.}~\bibnamefont
  {Ackermann}} \emph {et~al.},\ }\bibfield  {title} {\enquote {\bibinfo {title}
  {Fermi-lat observations of the diffuse gamma-ray emission: Implications for
  cosmic rays and the interstellar medium},}\ }\href
  {http://stacks.iop.org/0004-637X/750/i=1/a=3} {\bibfield  {journal} {\bibinfo
   {journal} {Astrophys. J.}\ }\textbf {\bibinfo {volume} {750}},\ \bibinfo
  {pages} {3} (\bibinfo {year} {2012})}\BibitemShut {NoStop}%
\bibitem [{\citenamefont {{Bland-Hawthorn}}\ and\ \citenamefont
  {{Gerhard}}(2016{\natexlab{b}})}]{Bland-Hawthorn2016}%
  \BibitemOpen
  \bibfield  {author} {\bibinfo {author} {\bibfnamefont {J.}~\bibnamefont
  {{Bland-Hawthorn}}}\ and\ \bibinfo {author} {\bibfnamefont {O.}~\bibnamefont
  {{Gerhard}}},\ }\bibfield  {title} {\enquote {\bibinfo {title} {{The Galaxy
  in Context: Structural, Kinematic, and Integrated Properties}},}\ }\href
  {\doibase 10.1146/annurev-astro-081915-023441} {\bibfield  {journal}
  {\bibinfo  {journal} {Ann. Rev. of A \& A}\ }\textbf {\bibinfo {volume}
  {54}},\ \bibinfo {pages} {529--596} (\bibinfo {year} {2016}{\natexlab{b}})},\
  \Eprint {http://arxiv.org/abs/1602.07702} {arXiv:1602.07702} \BibitemShut
  {NoStop}%
\bibitem [{\citenamefont {Frenk}\ \emph {et~al.}(1988)\citenamefont {Frenk},
  \citenamefont {White}, \citenamefont {Davis},\ and\ \citenamefont
  {Efstathiou}}]{Frenk:1988zz}%
  \BibitemOpen
  \bibfield  {author} {\bibinfo {author} {\bibfnamefont {Carlos~S.}\
  \bibnamefont {Frenk}}, \bibinfo {author} {\bibfnamefont {Simon D.~M.}\
  \bibnamefont {White}}, \bibinfo {author} {\bibfnamefont {Marc}\ \bibnamefont
  {Davis}}, \ and\ \bibinfo {author} {\bibfnamefont {George}\ \bibnamefont
  {Efstathiou}},\ }\bibfield  {title} {\enquote {\bibinfo {title} {{The
  formation of dark halos in a universe dominated by cold dark matter}},}\
  }\href {\doibase 10.1086/166213} {\bibfield  {journal} {\bibinfo  {journal}
  {Astrophys. J.}\ }\textbf {\bibinfo {volume} {327}},\ \bibinfo {pages}
  {507--525} (\bibinfo {year} {1988})}\BibitemShut {NoStop}%
\bibitem [{\citenamefont {Vera-Ciro}\ \emph {et~al.}(2011)\citenamefont
  {Vera-Ciro}, \citenamefont {Sales}, \citenamefont {Helmi}, \citenamefont
  {Frenk}, \citenamefont {Navarro}, \citenamefont {Springel}, \citenamefont
  {Vogelsberger},\ and\ \citenamefont {White}}]{VeraCiro:2011nb}%
  \BibitemOpen
  \bibfield  {author} {\bibinfo {author} {\bibfnamefont {Carlos~A.}\
  \bibnamefont {Vera-Ciro}}, \bibinfo {author} {\bibfnamefont {Laura~V.}\
  \bibnamefont {Sales}}, \bibinfo {author} {\bibfnamefont {Amina}\ \bibnamefont
  {Helmi}}, \bibinfo {author} {\bibfnamefont {Carlos~S.}\ \bibnamefont
  {Frenk}}, \bibinfo {author} {\bibfnamefont {Julio~F.}\ \bibnamefont
  {Navarro}}, \bibinfo {author} {\bibfnamefont {Volker}\ \bibnamefont
  {Springel}}, \bibinfo {author} {\bibfnamefont {Mark}\ \bibnamefont
  {Vogelsberger}}, \ and\ \bibinfo {author} {\bibfnamefont {Simon D.~M.}\
  \bibnamefont {White}},\ }\bibfield  {title} {\enquote {\bibinfo {title} {{The
  Shape of Dark Matter Haloes in the Aquarius Simulations: Evolution and
  Memory}},}\ }\href {\doibase 10.1111/j.1365-2966.2011.19134.x} {\bibfield
  {journal} {\bibinfo  {journal} {Mon. Not. Roy. Astron. Soc.}\ }\textbf
  {\bibinfo {volume} {416}},\ \bibinfo {pages} {1377--1391} (\bibinfo {year}
  {2011})},\ \Eprint {http://arxiv.org/abs/1104.1566} {arXiv:1104.1566
  [astro-ph.CO]} \BibitemShut {NoStop}%
\bibitem [{\citenamefont {Tissera}\ \emph {et~al.}(2010)\citenamefont
  {Tissera}, \citenamefont {White}, \citenamefont {Pedrosa},\ and\
  \citenamefont {Scannapieco}}]{Tissera:2009cm}%
  \BibitemOpen
  \bibfield  {author} {\bibinfo {author} {\bibfnamefont {Patricia~B.}\
  \bibnamefont {Tissera}}, \bibinfo {author} {\bibfnamefont {Simon D.~M.}\
  \bibnamefont {White}}, \bibinfo {author} {\bibfnamefont {Susana}\
  \bibnamefont {Pedrosa}}, \ and\ \bibinfo {author} {\bibfnamefont {Cecilia}\
  \bibnamefont {Scannapieco}},\ }\bibfield  {title} {\enquote {\bibinfo {title}
  {{Dark matter response to galaxy formation}},}\ }\href {\doibase
  10.1111/j.1365-2966.2010.16777.x} {\bibfield  {journal} {\bibinfo  {journal}
  {Mon. Not. Roy. Astron. Soc.}\ }\textbf {\bibinfo {volume} {406}},\ \bibinfo
  {pages} {922} (\bibinfo {year} {2010})},\ \Eprint
  {http://arxiv.org/abs/0911.2316} {arXiv:0911.2316 [astro-ph.CO]} \BibitemShut
  {NoStop}%
\bibitem [{\citenamefont {{Kalberla}}\ \emph {et~al.}(2005)\citenamefont
  {{Kalberla}}, \citenamefont {{Burton}}, \citenamefont {{Hartmann}},
  \citenamefont {{Arnal}}, \citenamefont {{Bajaja}}, \citenamefont {{Morras}},\
  and\ \citenamefont {{P{\"o}ppel}}}]{kalberla2005}%
  \BibitemOpen
  \bibfield  {author} {\bibinfo {author} {\bibfnamefont {P.~M.~W.}\
  \bibnamefont {{Kalberla}}}, \bibinfo {author} {\bibfnamefont {W.~B.}\
  \bibnamefont {{Burton}}}, \bibinfo {author} {\bibfnamefont {D.}~\bibnamefont
  {{Hartmann}}}, \bibinfo {author} {\bibfnamefont {E.~M.}\ \bibnamefont
  {{Arnal}}}, \bibinfo {author} {\bibfnamefont {E.}~\bibnamefont {{Bajaja}}},
  \bibinfo {author} {\bibfnamefont {R.}~\bibnamefont {{Morras}}}, \ and\
  \bibinfo {author} {\bibfnamefont {W.~G.~L.}\ \bibnamefont {{P{\"o}ppel}}},\
  }\bibfield  {title} {\enquote {\bibinfo {title} {{The Leiden/Argentine/Bonn
  (LAB) Survey of Galactic HI. Final data release of the combined LDS and IAR
  surveys with improved stray-radiation corrections}},}\ }\href {\doibase
  10.1051/0004-6361:20041864} {\bibfield  {journal} {\bibinfo  {journal}
  {Astron. \& Astrophys.}\ }\textbf {\bibinfo {volume} {440}},\ \bibinfo
  {pages} {775--782} (\bibinfo {year} {2005})},\ \Eprint
  {http://arxiv.org/abs/astro-ph/0504140} {astro-ph/0504140} \BibitemShut
  {NoStop}%
\bibitem [{\citenamefont {Malyshev}(2012)}]{Malyshev:2012mb}%
  \BibitemOpen
  \bibfield  {author} {\bibinfo {author} {\bibfnamefont {Dmitry}\ \bibnamefont
  {Malyshev}},\ }\bibfield  {title} {\enquote {\bibinfo {title} {{Spectral
  components analysis of diffuse emission processes}},}\ }\href@noop {} {\
  (\bibinfo {year} {2012})},\ \Eprint {http://arxiv.org/abs/1202.1034}
  {arXiv:1202.1034 [astro-ph.IM]} \BibitemShut {NoStop}%
\bibitem [{\citenamefont {Chang}\ \emph {et~al.}(2018)\citenamefont {Chang},
  \citenamefont {Lisanti},\ and\ \citenamefont
  {Mishra-Sharma}}]{Chang:2018bpt}%
  \BibitemOpen
  \bibfield  {author} {\bibinfo {author} {\bibfnamefont {Laura~J.}\
  \bibnamefont {Chang}}, \bibinfo {author} {\bibfnamefont {Mariangela}\
  \bibnamefont {Lisanti}}, \ and\ \bibinfo {author} {\bibfnamefont {Siddharth}\
  \bibnamefont {Mishra-Sharma}},\ }\bibfield  {title} {\enquote {\bibinfo
  {title} {{Search for dark matter annihilation in the Milky Way halo}},}\
  }\href {\doibase 10.1103/PhysRevD.98.123004} {\bibfield  {journal} {\bibinfo
  {journal} {Phys. Rev.}\ }\textbf {\bibinfo {volume} {D98}},\ \bibinfo {pages}
  {123004} (\bibinfo {year} {2018})},\ \Eprint
  {http://arxiv.org/abs/1804.04132} {arXiv:1804.04132 [astro-ph.CO]}
  \BibitemShut {NoStop}%
\bibitem [{\citenamefont {Ando}\ \emph {et~al.}(2020)\citenamefont {Ando},
  \citenamefont {Geringer-Sameth}, \citenamefont {Hiroshima}, \citenamefont
  {Hoof}, \citenamefont {Trotta},\ and\ \citenamefont {Walker}}]{Ando:2020yyk}%
  \BibitemOpen
  \bibfield  {author} {\bibinfo {author} {\bibfnamefont {Shin'ichiro}\
  \bibnamefont {Ando}}, \bibinfo {author} {\bibfnamefont {Alex}\ \bibnamefont
  {Geringer-Sameth}}, \bibinfo {author} {\bibfnamefont {Nagisa}\ \bibnamefont
  {Hiroshima}}, \bibinfo {author} {\bibfnamefont {Sebastian}\ \bibnamefont
  {Hoof}}, \bibinfo {author} {\bibfnamefont {Roberto}\ \bibnamefont {Trotta}},
  \ and\ \bibinfo {author} {\bibfnamefont {Matthew~G.}\ \bibnamefont
  {Walker}},\ }\bibfield  {title} {\enquote {\bibinfo {title} {{Structure
  Formation Models Weaken Limits on WIMP Dark Matter from Dwarf Spheroidal
  Galaxies}},}\ }\href@noop {} {\  (\bibinfo {year} {2020})},\ \Eprint
  {http://arxiv.org/abs/2002.11956} {arXiv:2002.11956 [astro-ph.CO]}
  \BibitemShut {NoStop}%
\end{thebibliography}%

\appendix

\section{Methods}\label{appx:methods}

We used eight years (August 4, 2008$-$August 2, 2016) of \textsc{P8R3 ULTRAC.L.EANVETO} data recorded by \textit{Fermi}-LAT. We chose these particular time cuts because these are exactly the same used in the construction of the 4FGL catalog~\cite{Fermi-LAT:4FGL}, thus making our point source modeling completely self-consistent. Note that if a bigger amount of data had been chosen then a dedicated point source search would have been necessary for this work. Events with measured energies between 667 MeV and 158 GeV were considered in the analysis. We binned the data into 14 logarithmic energy bins between 667 MeV and 37.5 GeV plus one additional macroenergy bin for energies between 37.5 and 158 GeV.  In order to minimize contamination from the Earth atmosphere, we only considered photons detected at zenith angles larger than  90$^{\circ}$. Moreover, we employed the recommended data quality filters \textsc{(DATA$_{-}$QUAL$>$0)\&\&(LAT$_{-}$CONFIG==1)} and restricted the analysis to a square region of $40^{\circ}\times40^{\circ}$ around the GC. Our study was carried out using the standard  \textsc{Fermitools v1.0.1}\footnote{\url{https://github.com/fermi-lat/Fermitools-conda/wiki}} analysis framework, and instrument response functions \textsc{P8R3$_{-}$ULTRAC.L.EANVETO$_{-}$V2}. 
The gamma-ray background and foreground model used in this work is similar to that developed in Ref.~\cite{Macias:2016nev}. However, this has been further improved by including new 3D IC maps~\cite{Porter:2017vaa} and a more robust low-latitude Fermi Bubbles (FBs)  template~\cite{TheFermi-LAT:2017vmf, Macias:2019omb}. In particular, the 3D IC maps were modeled using the 3D ISRF data available with the recent \textsc{GALPROP v56}~\cite{Galprop,Porter:2017vaa}, though conventional 2D IC maps were also tested in our analysis of the systematic uncertainties in the GDE model. Furthermore, the 3D IC maps have been divided in several rings (see Table~\ref{Tab:definitions}) and their corresponding normalization floated during the fits to account for the impact of cosmic-ray (CR) density uncertainties. As for the Fermi bubbles component, we have included the map recently developed in  Ref.~\cite{Macias:2019omb}. In that study, the structured FB template of Ref.~\cite{TheFermi-LAT:2017vmf} was further modified by an inpainting algorithm to help restore image processing artifacts due to point source masks used in its derivation.

\begin{table*}[!htbp]\caption{{\bf List of spatial templates considered in our maximum likelihood runs. \label{Tab:definitions}}}
\begin{adjustbox}{width=1.0\textwidth, center}
\centering\begin{threeparttable}
 \scriptsize
\begin{tabular}{llr}
\hline\hline
Component  & Description & Reference\\\hline 
Gas-correlated gamma rays  &  Considered two different versions: (i) hydrodynamical  
  & \\
  &and (ii) interpolated templates. & \\
 &  They consist of H{\tiny I} and H\boldmath$_{2}$ gas  column density maps divided& \\
& in four rings each, and two dust correction maps.\\
& Two different values of $E(B-V)$ magnitude cuts in the dust &\\
& maps were studied as a check of the systematic uncertainties. & ~\cite{Macias:2016nev, Macias:2019omb}\\
 &&\\
Inverse Compton emission & Three different versions were considered: (i) a standard 2D IC map  & \\
&  including a central source of electrons, (ii) a 3D IC map$^\dagger$ divided  &\\
&  in four rings and (iii) a 3D IC map divided in six rings.  &  \cite{Porter:2017vaa}\\
&&\\
\textit{Fermi} bubbles&  Inpainted \textit{Fermi} bubbles template shown to improve the fit &~\cite{Macias:2019omb}\\
&&\\
Loop I& Analytical model & \cite{Wolleben:2007}\\
Point sources & \textit{Fermi} LAT Fourth Source Catalog (4FGL)& \cite{Fermi-LAT:4FGL}\\
Sun and Moon templates& Templates available in the  4FGL catalog&\cite{Fermi-LAT:4FGL} \\
Isotropic emission& \texttt{iso$_{-}$P8R3$_{-}$ULTRAC.L.EANVETO$_{-}$V2$_{-}$v1.txt}&\\\hline
&&\\
Nuclear bulge & Map constructed from stellar counts (near-infrared observations)& ~\cite{Nishiyama2015}\\
Boxy bulge& Model derived from a fit to diffuse infrared data from COBE& ~\cite{Freudenreich:1997bx}\\
Dark matter templates & Considered gNFW profiles with different slopes  &\\
&($\gamma=[0.5,1.5]$) as well as a cored profile. Ellipsoidal versions of    &\\
&these two classes were also included (see Fig.~\ref{fig:SpatialMaps}).&\\
\hline\hline
\end{tabular}
\begin{tablenotes}
\item  $^\dagger$Here, we adopt the Galaxy-wide dust and stellar distribution model based on the Freudenreich~\cite{Freudenreich:1997bx} (F98) stellar bulge model (see Ref.~\cite{Porter:2017vaa}). Two different IC ring subdivisions were considered: four rings ($0-3.5$, $3.5-8.0$, $8.0-10.0$, and $10.0-50.0$ kpc) and six rings ($0-1.5$, $1.5-2.5$ , $2.5-3.5$, $3.5-8.0$, $8.0-10.0$, and $10.0-50.0$ kpc). In the four-ring case, the annular sizes of the 3D IC maps match those used for the interstellar gas maps.
\end{tablenotes}
\end{threeparttable}
\end{adjustbox}
\end{table*}

\begin{figure*}[ht!]
\begin{tabular}{ccc}
\includegraphics[width=0.262\textwidth]{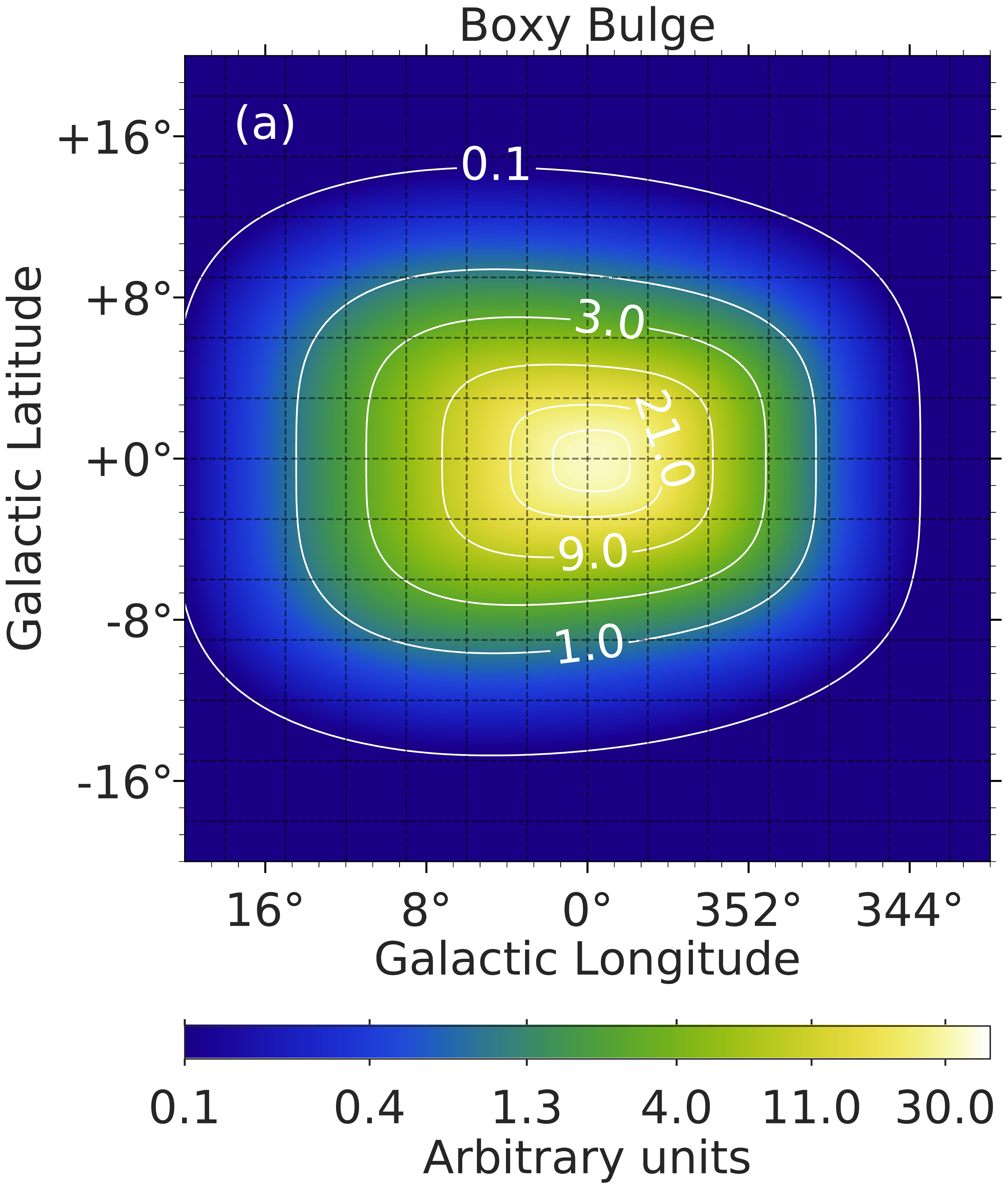} & \includegraphics[width=0.27\textwidth]{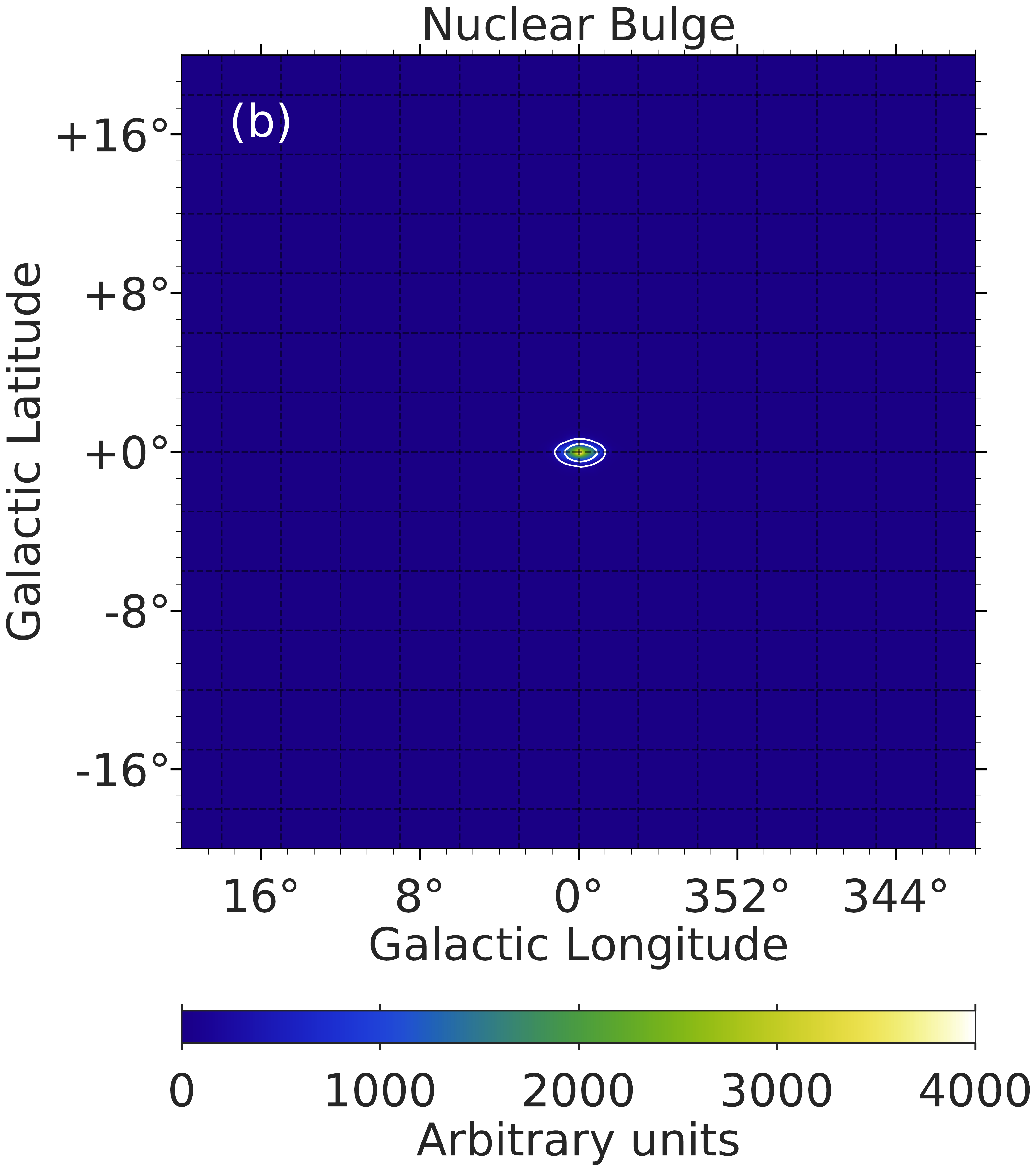}  & \includegraphics[width=0.262\textwidth]{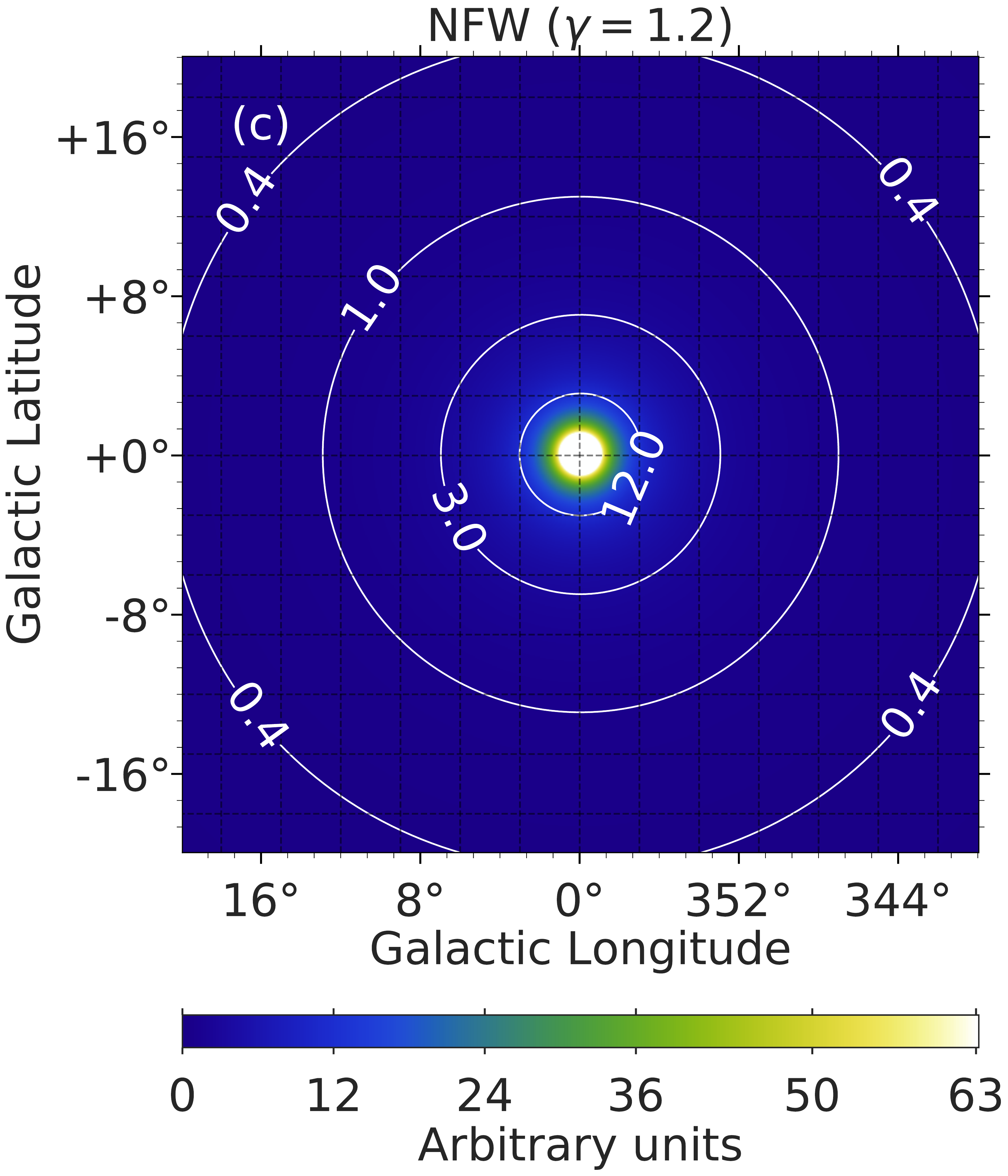}\\
\includegraphics[width=0.268\textwidth]{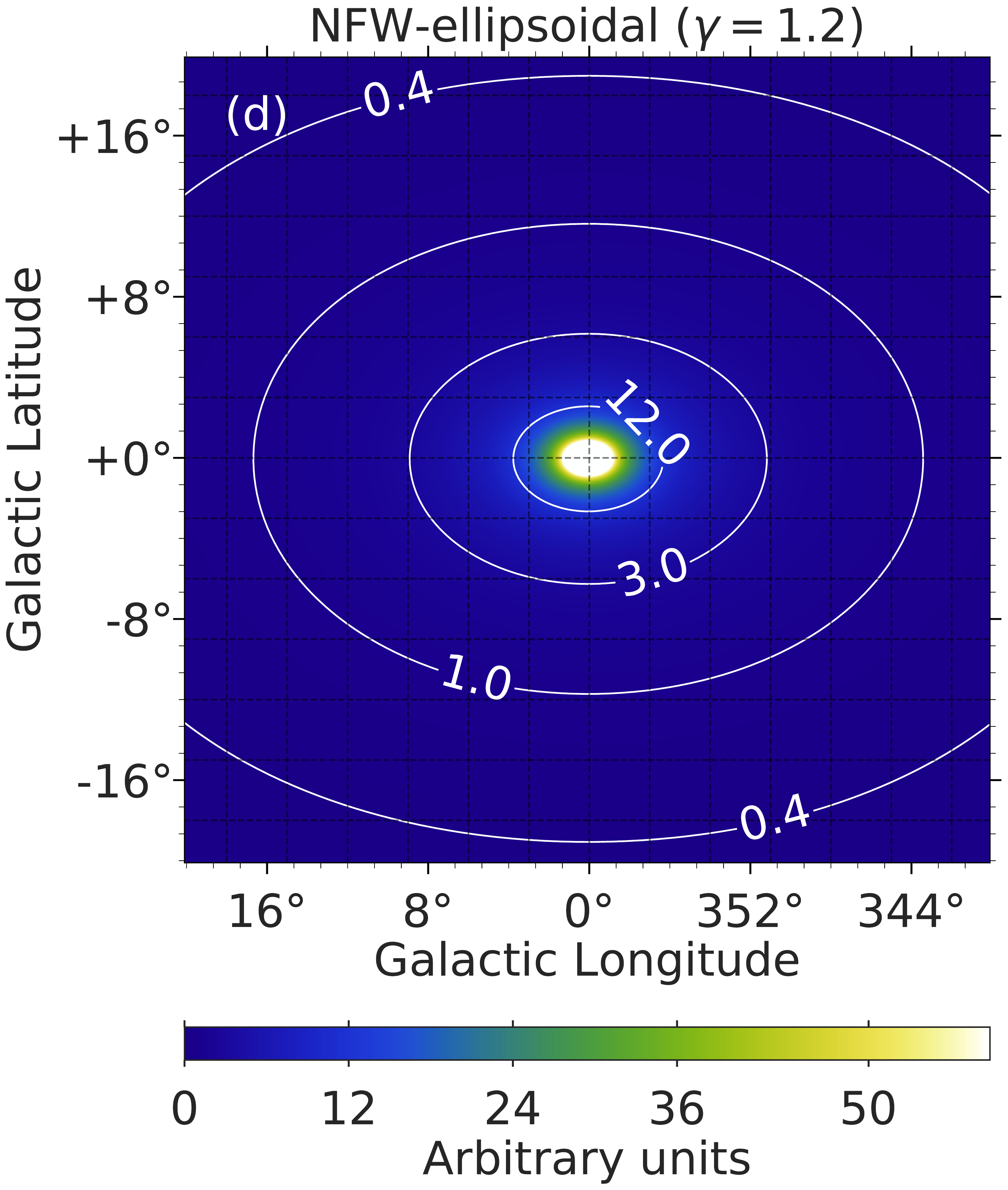} & \includegraphics[width=0.27\textwidth]{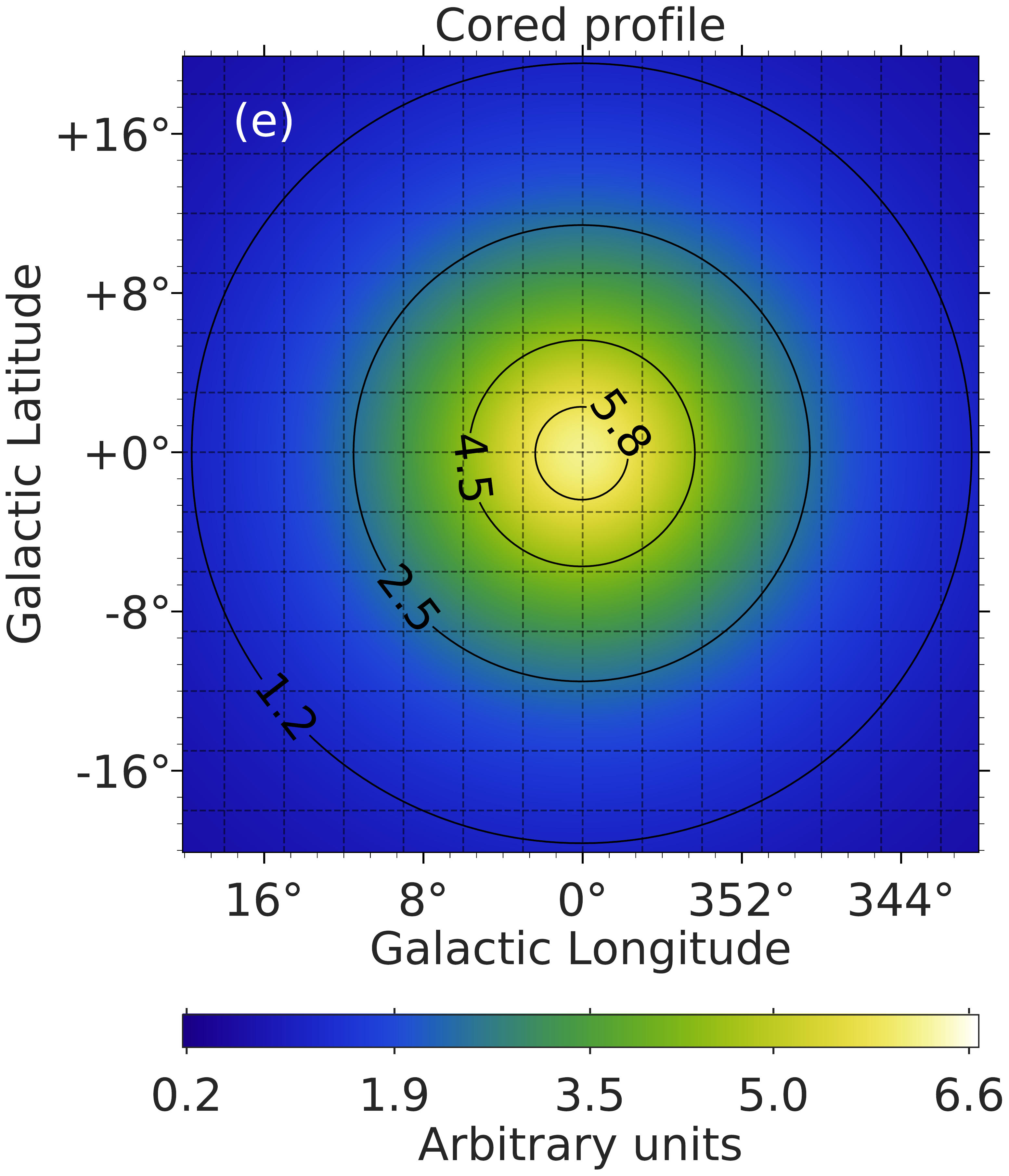}  & \includegraphics[width=0.265\textwidth]{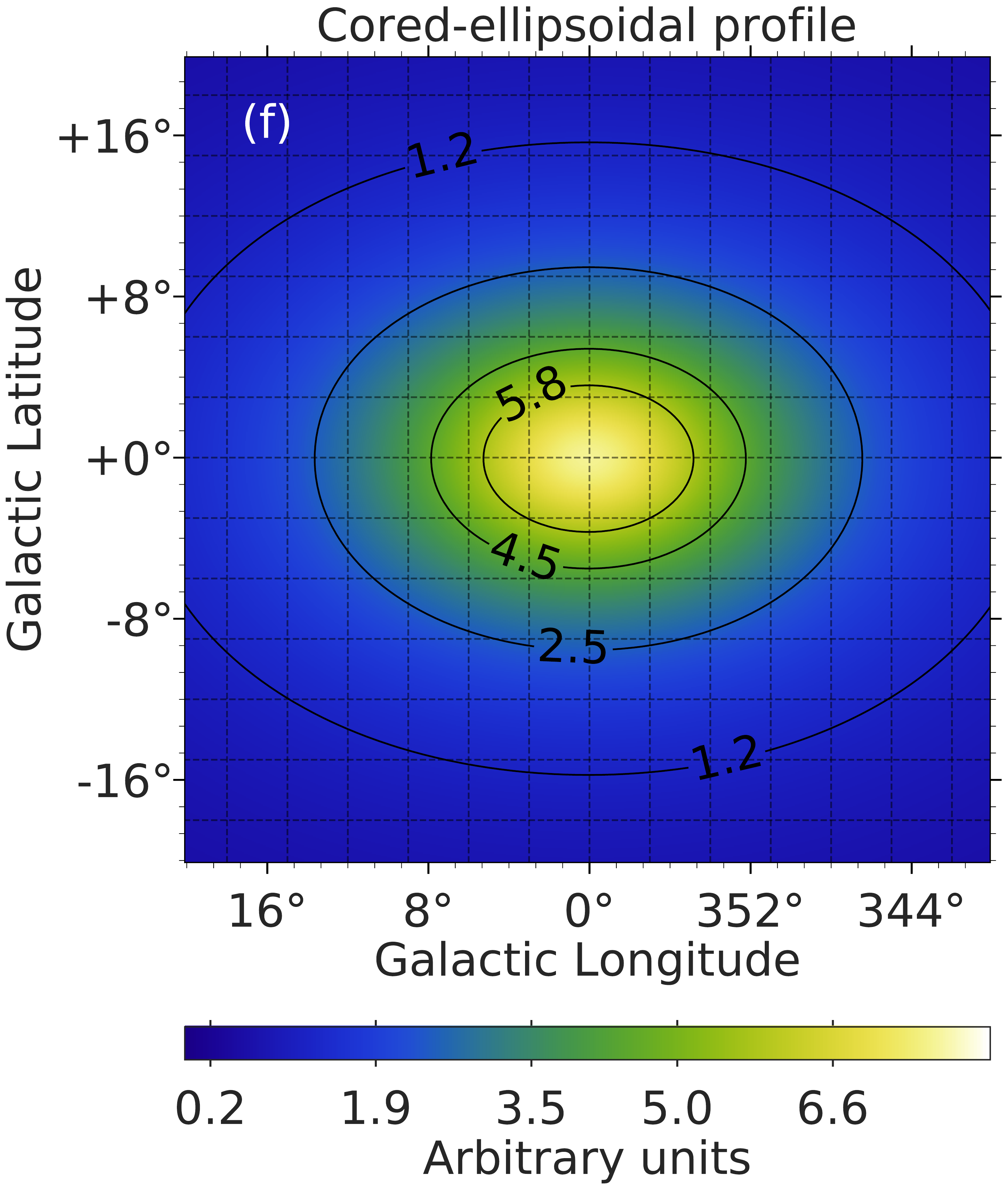}
\end{tabular}\caption{Spatial templates considered for the GCE: (a) The \textit{boxy bulge} model corresponds to the F98~\cite{Freudenreich:1998} stellar density map (see Ref.~\cite{Macias:2016nev,Macias:2019omb} for details). (b) The \textit{nuclear bulge} is a stellar density map of the inner 400 pc of the GC constructed with the use of the NIR Camera SIRIUS in Ref.~\cite{Nishiyama2015}. (c) Generalized spherically symmetric NFW-squared profile with a mild slope ($\gamma=1.2$). (d) gNFW-squared profile ($\gamma=1.2$) with a minor-to-major axis ratio of 0.7 in the $z$ axis. (e) Spherically symmetric Read-squared DM density profile. (f) Read-squared DM density profile with a minor-to-major axis ratio of 0.7 in the $z$ axis. All maps are normalized to unit flux. } 
\label{fig:SpatialMaps}
\end{figure*}

 We also included templates for the \textit{Sun} and the \textit{Moon} that match our photon event class and cuts (available in the \textit{Fermi} fourth catalog of point sources 4FGL~\cite{Fermi-LAT:4FGL}), an isotropic component (\textsc{iso$_{-}$P8R3$_{-}$ULTRAC.L.EANVETO$_{-}$V2$_{-}$v1.txt}) and an emission model map for Loop I~\cite{Wolleben:2007,Macias:2016nev}. The diffuse gamma-ray emission resulting from the interaction of energetic CR particles with the interstellar medium was modeled as a linear combination of atomic and molecular hydrogen gas templates divided into four concentric rings ($0-3.5$, $3.5-8.0$, $8.0-10.0$, and $10.0-50.0$ kpc) plus dust residual maps accounting for dark neutral material in the Galaxy. 

The gamma-ray point sources present in our RoI were modeled using the 4FGL catalog~\cite{Fermi-LAT:4FGL}.  There is a total of 487 pointlike and extended sources inside our RoI. Due to limitations in the maximum number of parameters that can be reliably fitted in a given run within \texttt{Fermitools}, we have employed the hybrid fitting procedure implemented in Ref.~\cite{Macias:2019omb}. Specifically, we varied the normalization of each of the 120 brightest point sources in our RoI, while for the remaining 367 sources, we constructed a point source population template whose normalization was allowed to vary at each energy bin. This is a good approximation given that the amount of data utilized in the present study is the same used in the construction of the 4FGL catalog. The point source population template was constructed by using the best-fit spectra in the 4FGL and convolving it with the \textit{Fermi} point spread function at each energy bin. The convolution was done with the \textsc{gtmodel} tool within \texttt{Fermitools} and the resulting map appropriately normalized for inclusion in the maximum likelihood procedure. Other extended sources (FHES J1723.5-0501, W 28, HESS J1804-216, W 30, HESS J1808-204, HESS J1809-193, HESS J1813-178, HESS J1825-137) inside of our RoI were taken from the 4FGL catalog~\cite{3FGL} and varied independently in the fits.    

 The systematic uncertainties in the gas-correlated emission were studied using alternative model templates. In particular, the interstellar gas maps included in our benchmark model were obtained from a suite of hydrodynamic simulations of interstellar gas flow~\cite{Pohl2008}. However, we also considered interpolated gas templates that reproduce those used in the construction of the official \textit{Fermi} diffuse emission model~\cite{Casandjian:andFermiLat2016}. Reference~\cite{Macias:2016nev} showed in detail that there are important morphological differences between the interpolated and hydrodynamic gas maps and that the latter provides a significantly better fit to the gamma-ray data in the GC region. Note that this result has been independently confirmed with the non-Poissonian template fitting pipeline~\cite{Buschmann:2020adf}.
 
 Some previous GCE analyses estimated the systematic uncertainties associated to the IC component by using the results of a \textsc{GALPROP} propagation parameter scan in Ref.~\cite{ackermannajelloatwood2012}. However, that study was restricted to a selected set of CR injection and propagation scenarios that assumed 2D Galactocentric cylindrically symmetric geometry for the Galaxy. Although this assumption is physically sensible and has allowed to gain deep insights into the gamma-ray sky, it is expected to introduce a bias to GCE studies since the 2D IC models fail to incorporate the nonaxisymmetric characteristics of the stellar distribution in the MW, such as the spiral arms and bar~\cite{Bland-Hawthorn2016,Porter:2017vaa}.  Indeed, the most recent release of the \texttt{GALPROP} code~\cite{Porter:2017vaa} has introduced more realistic 3D spatial models for the CR source and ISRF densities. These include sophisticated templates for the spiral arms, the bulge/bar complex, and warped stellar/dust disk~\cite{Bland-Hawthorn2016}. In the present study we have reproduced the results in Ref.~\cite{Porter:2017vaa} and included in our analysis one of their main 3D IC models named F98-SA50 (see Table 3 of Ref.~\cite{Porter:2017vaa}). The choice of this particular model had no impact in our results since we have divided the 3D IC  map in four or six rings and allowed their normalization to float in the fits in order to account for uncertainties in the CR densities. To allow for a greater range of systematics, we also included a 2D IC map containing an additional central source population of $e^-$ (model B in Ref.~\cite{Ackermann:2014usa}).  A summary of the foreground/background models considered in this study is shown in Table~\ref{Tab:definitions}.

The analysis procedure used here is similar to that of Refs.~\cite{Macias:2016nev, Macias:2019omb}. We employed a bin-by-bin fitting method in which a separate maximum likelihood was run at each energy bin. To obtain the band fluxes for each component, we assumed a power law with a fixed slope of $-2$ and simultaneously varied the normalization of all the sources in each different energy bin. In particular, we varied the normalization of all the GDE templates, the 120 brightest 4FGL point sources, and the point source population template containing the remaining 367 point sources. We used the \textit{pyLikelihood} tool to vary a total of 146 parameters in the fits and ensure they converged.

\section{Spatial maps for the GCE}\label{appx:spatial_maps}

Detailed specifications of the templates for the GCE are given in Sec.~II of Ref.~\cite{Macias:2019omb}. Here we provide a brief description of the templates considered with an emphasis on those that are new in the present work. 

We used two types of spatial models in our analysis of the morphology of the GCE: stellar density and DM density (squared) maps. For the bulge stars, we included the ``boxy bulge'' [Fig.~\ref{fig:SpatialMaps} (a)] model -- obtained in Ref.~\cite{Freudenreich:1998} from a fit to diffuse infrared COBE/DIRBE data -- as well as the ``nuclear bulge''~\cite{Nishiyama2015} [Fig.~\ref{fig:SpatialMaps} (b)], which is a stellar density map of the inner 400 pc. For DM we used the density distributions given by gNFW [Fig.~\ref{fig:SpatialMaps} (c)] or cored [Fig.~\ref{fig:SpatialMaps} (e)] profiles, already described in the main text. However, in our analysis, we also included a DM halo shape that departs from the commonly assumed spherical symmetry; we considered an oblate halo shape with its longer axis aligned with the Galactic disk. Indeed, collisionless N-body simulations predict ellipsoidal halos with density profile minor-to-major axis ratios approximately $ 0.4-0.6$ (e.g., Ref.~\cite{Frenk:1988zz,VeraCiro:2011nb}). Moreover, hydrodynamical simulations have shown that baryonic dissipation can mitigate this halo shape contraction by making the DM halos more spherical (see e.g.,~\cite{Tissera:2009cm}, and references therein).  In practice, the halo shape contraction is implemented in our analysis by making a transformation in the Galactic distance [introduced in Eqs.~\ref{eq:nfw} and~\ref{eq:Read}] of the form $r\rightarrow r^\prime$, where $r^\prime$ is given by
\begin{equation}
    r^{\prime 2}=x^2+\frac{y^2}{(b/a)^2}+\frac{z^2}{(c/a)^2};
\end{equation}
$x$, $y$, and $z$ are Galactocentric Cartesian coordinates; and $a$, $b$, and $c$ are the major, intermediate, and minor axis scale lengths. We have opted for assuming a minor-to-major axis ratio $c/a \sim 0.7$ and intermediate-to-minor $b/a \sim 1$, which are the best values found in a recent study~\cite{Dai:2018ypv} (based on the results of the \textit{Eris} simulations). The actual DM templates included in our maximum likelihood runs were constructed by performing a line-of-sight integral of the density squared profiles. Figure~\ref{fig:SpatialMaps} (d) and 3(f) show the NFW and cored profiles after implementation of the above halo shape contraction.

\begin{table*}[!htbp]\caption{{\bf Log-likelihood values for our baseline astrophysical model.\label{Tab:GCElikelihoods}}}
\begin{adjustbox}{width=1.0\textwidth, center}
\centering
\begin{threeparttable}
\scriptsize
\begin{tabular}{llllccc}
\hline\hline
Base  & Source &  $-\log(\mathcal{L}_{\rm Base})$  & $-\log(\mathcal{L}_{{\rm Base}+{\rm  Source}})$  &  $\mbox{TS}_{\rm Source}$& \textit{d.o.f}  & Significance \\ 
           &              &                                  &                         &                                               & &  \\\hline
Baseline$^\dagger$     &Cored ellipsoidal & -3258814.98     &-3259263.66   &897.4 & 15& $-$ \\ 
Baseline     &Cored & -3258814.98     &  -3259267.33 & 904.7 &15 & $-$ \\
Baseline     &BB & -3258814.98    & -3259417.25 & 1204.5 & 15 & $-$ \\ 
Baseline     &NFW ellipsoidal & -3258814.98     & -3259515.47  & 1401.0 &15 & $-$ \\
Baseline     &NFW & -3258814.98     & -3259619.27  & 1608.6 & 15 & $-$ \\
Baseline     &NB & -3258814.98    & -3259695.78 & 1761.6 & 15 & $-$  \\ \hline
Baseline+NB     & Cored ellipsoidal  & -3259695.78    & -3259702.11  & 12.7 & 15 & $1.6\;\sigma$ \\ 
Baseline+NB     & Cored  & -3259695.78    & -3259705.14  & 18.7 & 15 & $2.4\;\sigma$ \\ 
Baseline+NB     & NFW ellipsoidal  & -3259695.78     & -3259714.55  & 37.5 & 15 & $4.3\;\sigma$ \\ 
Baseline+NB     & NFW  & -3259695.78     & -3259745.66  &99.8 & 15 & $8.4\;\sigma$ \\ 
Baseline+NB     & BB  & -3259695.78     & -3259834.20 & 276.8 &  15 & $15.4\;\sigma$ \\ \hline
Baseline+NB+BB     & Cored ellipsoidal    &-3259834.20     &-3259834.45  & 0.5  &  15 &  $0.0\;\sigma$ \\ 
Baseline+NB+BB     & NFW   &-3259834.20     &-3259837.79  & 7.2 &  15 &   $0.7\;\sigma$\\ 
Baseline+NB+BB     & NFW ellipsoidal   &-3259834.20     & -3259839.66 & 10.9 &  15 &  $1.3\;\sigma$  \\ 
Baseline+NB+BB     & Cored   &-3259834.20     &-3259844.40  & 20.4  &  15 &  $2.6\;\sigma$ \\    \hline

\hline\hline
\end{tabular}
\begin{tablenotes}
\item $^\dagger$The baseline model is a combination of the hydrodynamic gas maps (four rings), IC (four rings), 4FGL point sources, FBs, Sun and Moon, isotropic and Loop I template (see Table~\ref{Tab:definitions}). Additional sources considered in the analysis are: Nuclear bulge (NB)~\cite{Nishiyama2015}, boxy bulge (BB)~\cite{Freudenreich:1997bx},  NFW profile with $\gamma=1.2$, cored dark matter~\cite{Read:2015sta} and ellipsoidal versions of these two DM templates (Fig.~\ref{fig:SpatialMaps}). The maximized likelihoods ($\mathcal{L}$) are given for the Base and Base$+$Source models. The statistical significance for each new source is obtained by computing the TS$_{\rm Source}$ as shown in Eq.~\ref{eq:TS}.
    \end{tablenotes}
\end{threeparttable}
\end{adjustbox}
\end{table*}

\section{Morphological Analysis of the Galactic Center Excess Signal}\label{appx:Morpholical_Analysis_GCE}

 In Refs.~\cite{Macias:2016nev,Macias:2019omb}, a subset of us showed that the GCE spatial morphology was better explained by the stellar nuclear bulge~\cite{Nishiyama2015} and galactic bulge~\cite{Freudenreich:1998} templates than by a spherically symmetric excess map given by a gNFW profile, as would be consistent with annihilating DM \cite{Navarro:1996gj} (e.g., Table~I of Ref.~\cite{Macias:2016nev}). 
 The preference for the bulge model was typically approximately $10 \sigma$ or higher. Similar results have been quantitatively obtained also by Ref.~\cite{Bartels:2017vsx, Macias:2019omb}. 
 
 Given that in the present study we consider a greater variety of dark matter morphologies (i.e., cuspy, cored, and ellipsoidal versions of these), and an improved GDE model for the GC region, here we have undertaken the same kind of statistical procedure utilized in Refs.~\cite{Macias:2016nev,Macias:2019omb} to find out which templates fit best the spatial morphology of the GCE signal.     
 
Table~\ref{Tab:GCElikelihoods} presents a summary of the tests carried out to evaluate whether a new template was required by the data. We used the test statistic (TS) defined as   

\begin{equation}\label{eq:TS}
{\rm TS}_{\rm Source} = 2(\log(\mathcal{L}_{{\rm Base}+{\rm  Source}}) - \log(\mathcal{L}_{\rm Base})),
\end{equation}
where $\mathcal{L}$ is the Poissonian likelihood function. The \textit{Base} and  \textit{Base+Source} models are described in the first and second columns of Table~\ref{Tab:GCElikelihoods}. The third and fourth columns display their respective loglike values (obtained through independent maximum likelihood runs), the fifth column shows the TS value for each new source considered, and the sixth and seventh columns show the number of degrees of freedom (same as the number of energy bins adopted in our analysis), and the statistical significance in sigma units, respectively.

As a first step, we computed the loglike value for the \textit{baseline} background/foreground model and then evaluated the TS values for each new template. The results of this step are shown in the first six rows of Table~\ref{Tab:GCElikelihoods}. For the second step, we added the nuclear bulge template (which was the template found the highest TS up to this point) to our \textit{Base} model and repeated the procedure with this augmented \textit{Base} model. As can be seen in the second set of rows of the table, the boxy bulge template is now the one that improves the fit the most and we have therefore proceeded to append it to our \textit{Base} model. As a final step, we iterated through the remaining templates until the highest TS value of a new template was below the $4\sigma$ detection threshold. The last set of rows shown in Table~\ref{Tab:GCElikelihoods} illustrates how once the stellar templates (NB+BB) are included in the model, the data no longer require a dark matter model to be appended to the \textit{Base} model. We note that in Table~\ref{Tab:GCElikelihoods} we display the results for NFW($\gamma=1.2$), which has been shown to approximately describe the GCE in previous works (e.g. Ref.~\cite{Abazajian:2012pn,Gordon:2013vta, Carlson:2014cwa}). We found that a NFW template with a slope $\gamma$ in the range $[0.5,1.5]$ was not significantly detected ($<4\sigma$) in any of our maximum likelihood runs. Furthermore, in our upper limits procedure the NFW profile slope is a nuisance parameter that is marginalized over.

For each new template there are  $15$ new parameters. The probability distribution is the same as Eq.~(2.5) in Ref.~\cite{Macias:2019omb}. It follows that for one new template being considered (i.e.\ 15 new parameters),  a  $4\sigma$  significance detection amounts to $TS=34.8$.

Given the high significance of these results, we add the nuclear bulge and galactic bulge templates to our astrophysical model for the GC region. 
Note that these templates are detected \textit{in addition to} the 3D IC templates that already contain the galactic bulge as a source of photons and CRs. 
This can be interpreted as gamma-ray sources distributed according to the galactic bulge. For example, in the MSP scenario, the prompt gamma-ray emission would still be required to be accounted for even while their secondary emission is modeled by the 3D IC maps. It is also worth noting that a nuclear bulge component has not yet been included in the \texttt{GALPROP} Galaxy model.

\begin{figure}[t!]
\includegraphics[scale=0.5]{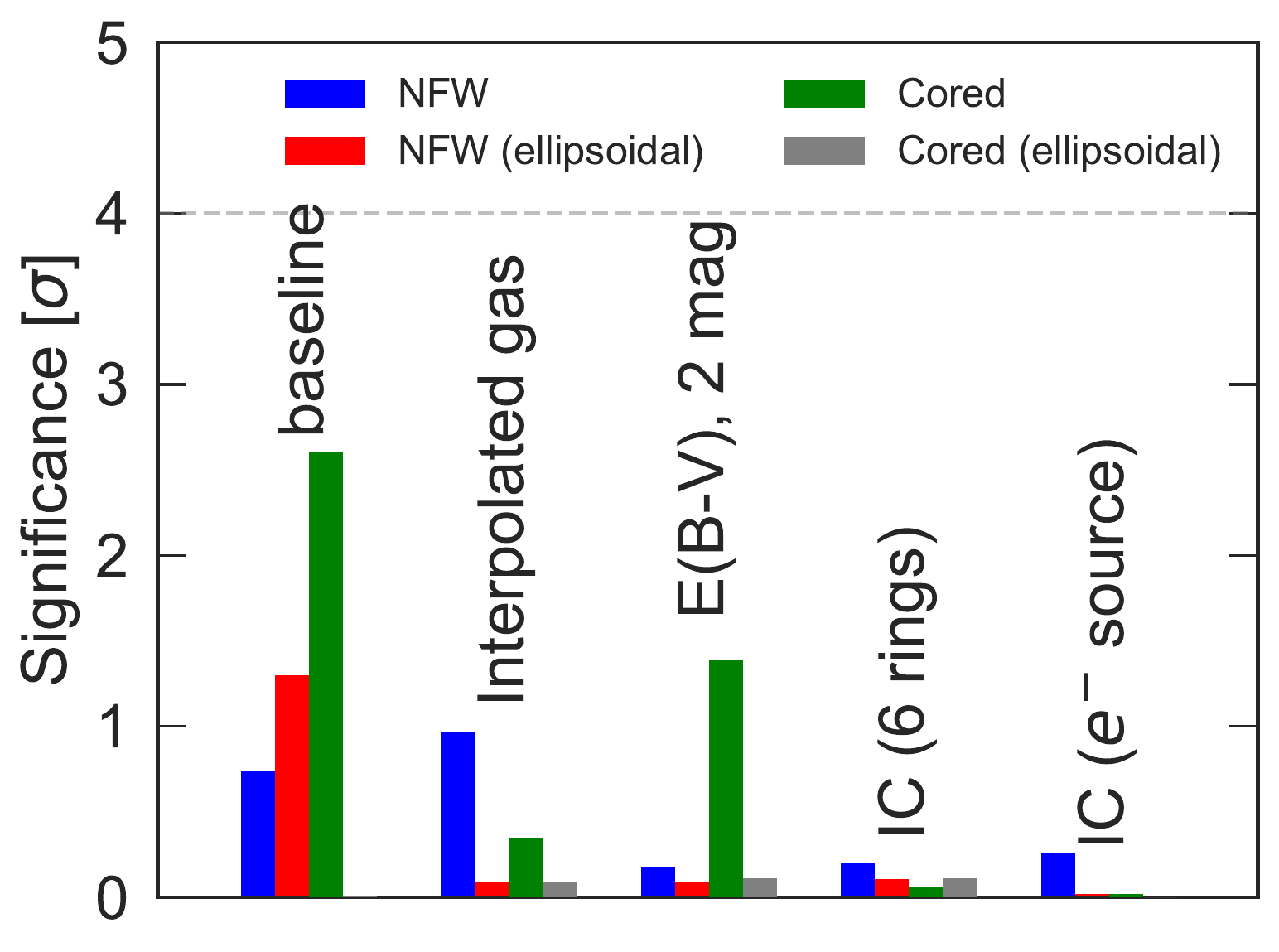}
\caption{Summary of the detection significance (in sigma units) for each of the DM templates considered in this work (see Fig.~\ref{fig:SpatialMaps} and Table~\ref{Tab:definitions}). The baseline background model (Table~\ref{Tab:GCElikelihoods}) includes the hydrodynamic gas maps and 3D IC maps (four rings). For the alternative backgrounds models we switched to the dust residual maps with different interstellar extinction E(B-V) magnitude cut (2 magnitude),  3D IC maps (divided in six rings), and 2D IC map containing a central source of $e^-$~\cite{Ackermann:2014usa}, except in the case of the traditional interpolated gas maps which were paired with the IC maps broken in six rings. When evaluating the significance of the DM templates, the nuclear bulge and boxy bulge maps were included in the fits. The horizontal gray line shows the usual $4\sigma$ detection threshold.}   
\label{Fig:TS_histogram}
\end{figure}

Figure~\ref{Fig:TS_histogram} presents the statistical significance of the main DM maps shown in Fig.~\ref{fig:SpatialMaps}. As can be seen, once the stellar templates are included in the fits, the data no longer requires a DM template for the GCE. This result is robust to a wide range of possible GDE models included in fits. Furthermore, past studies have analyzed the impact of potential degeneracies between the stellar mass and other extended templates included in the fits. Based on a study of the correlation coefficients between these templates, Ref.~\cite{Macias:2016nev} concluded that the impact of degeneracies in the fitted fluxes should be small. 

\section{Evaluation of the Systematic Uncertainties in the DM limits}\label{appx:syst_uncertainties_DMlimits}

The systematic uncertainties were evaluated by repeating our DM limits procedure with variants of the background/foreground emission model. In particular, the log-likelihood scans for the DM source were performed with a bin-by-bin fitting method in which the different templates were fitted independently in small energy bins. This helps to mitigate the impact on the results of the assumed spectrum of the several templates. At each energy bin the differential DM flux was assumed to be described by a simple power law of the form $N_0({E}/{2002.3\;\rm{MeV}})^{-2}$. In our procedure, we first performed a scan of DM flux values in regular steps of $\Delta \log(\mathcal{L})$ using the \texttt{UpperLimits} tool within \texttt{Fermitools}. In particular, with this tool we first obtained the minimum log-likelihood DM flux and then scanned the log-likelihood (with respect to the minimum) in steps of 0.5 until reaching $\Delta \log(\mathcal{L})\sim 6$. The list of $\Delta \log(\mathcal{L})$ were subsequently rescaled by computing the log-likelihood for the null hypothesis (zero DM flux). This last step is necessary for use in our Bayesian procedure. We started the scans with the benchmark model described in the main text, but also applied the procedure to variants of the foreground/background model. Specifically, we ran independent log-likelihood scans in which we replaced the benchmark 3D IC map (divided in four rings) by an alternative 3D IC template (divided in six rings). In addition, we considered a 2D IC model that contains an extra electrons-only~\cite{Ackermann:2014usa} source population in the GC. The spatial distribution of the additional source of electrons used in the construction of this 2D IC model can be seen in Fig. 13 of Ref.\ \cite{Ackermann:2014usa}. The uncertainties introduced by some of the assumptions in the creation of the hydrodynamic gas and dust templates were investigated in the same manner. Since the amount of dust traced by the $E(B-V)$ extinction map is not accurate in regions of high extinction, we utilized dust map templates constructed with two different magnitude cuts; 5 mag (benchmark model) and 2 mag. To encompass a greater range of systematic uncertainties, we also included the interpolated gas maps that reproduce the ones in Ref.~\cite{Casandjian:andFermiLat2016}.

\begin{figure*}[ht!]
\centering
\includegraphics[width=1.0\textwidth]{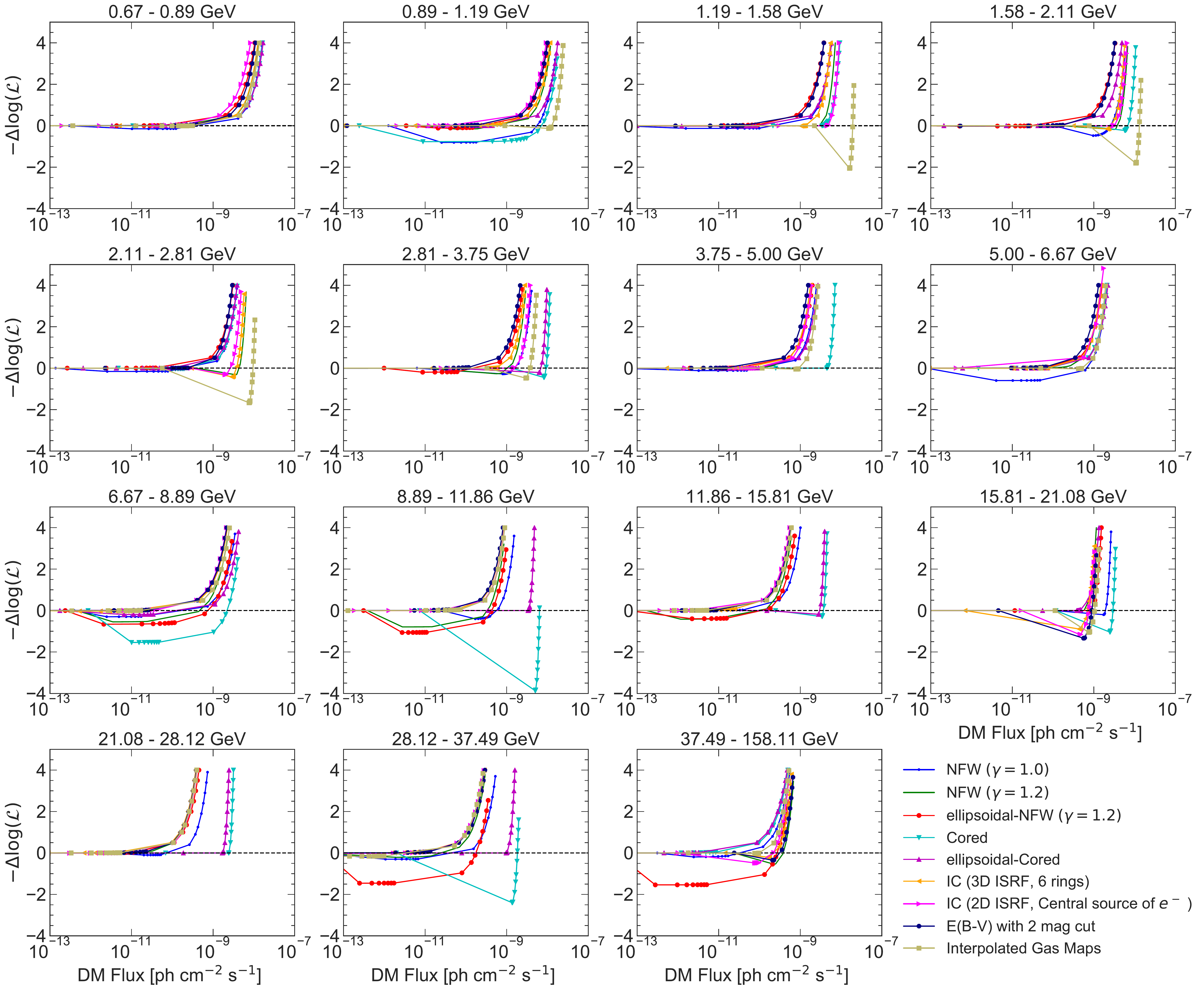}
\caption{Profiles of the bin-by-bin log-likelihood function used to test for a putative DM source in the $40^\circ \times 40^\circ$ region of the GC.  Each profile shows the log-likelihood ratio between ``background/foreground''$+$``DM source'' model and the background/foreground-only model. The bin-by-bin log-likelihood was calculated by scanning the flux normalization of the DM source within each energy bin in the range $10^{-17}$ and $10^{-7}$ ph cm$^{-2}$ s$^{-1}$. When running the scan, the flux normalization of the background sources were varied in the fit, while their spectral slope was fixed to $-2$. The best-fit spectrum of the background/foreground model components included in the fit can be seen in Fig.~\ref{fig:total_spectra}. Within each energy bin, the line colors denote an alternative spatial template for the DM source (cored, Cored ellipsoidal, NFW and Cored ellipsoidal, see Table~\ref{Tab:definitions} for descriptions) or an alternative gamma-ray background/foreground model. We changed the benchmark 3D IC map (divided in four rings) by an alternative 3D IC map divided in six rings, and 2D IC map containing a central source of $e^-$~\cite{Ackermann:2014usa}. We also varied the magnitude cut used in the construction of the gas maps and explored the results obtained with the interpolated gas maps. Unless otherwise stated, we conservatively assume a NFW-squared ($\gamma=1.2$) density profile for the DM map since this model has the largest log-likelihood.    } 
\label{fig:scans}
\end{figure*}

\begin{figure*}
    \centering
    \includegraphics[width=\textwidth]{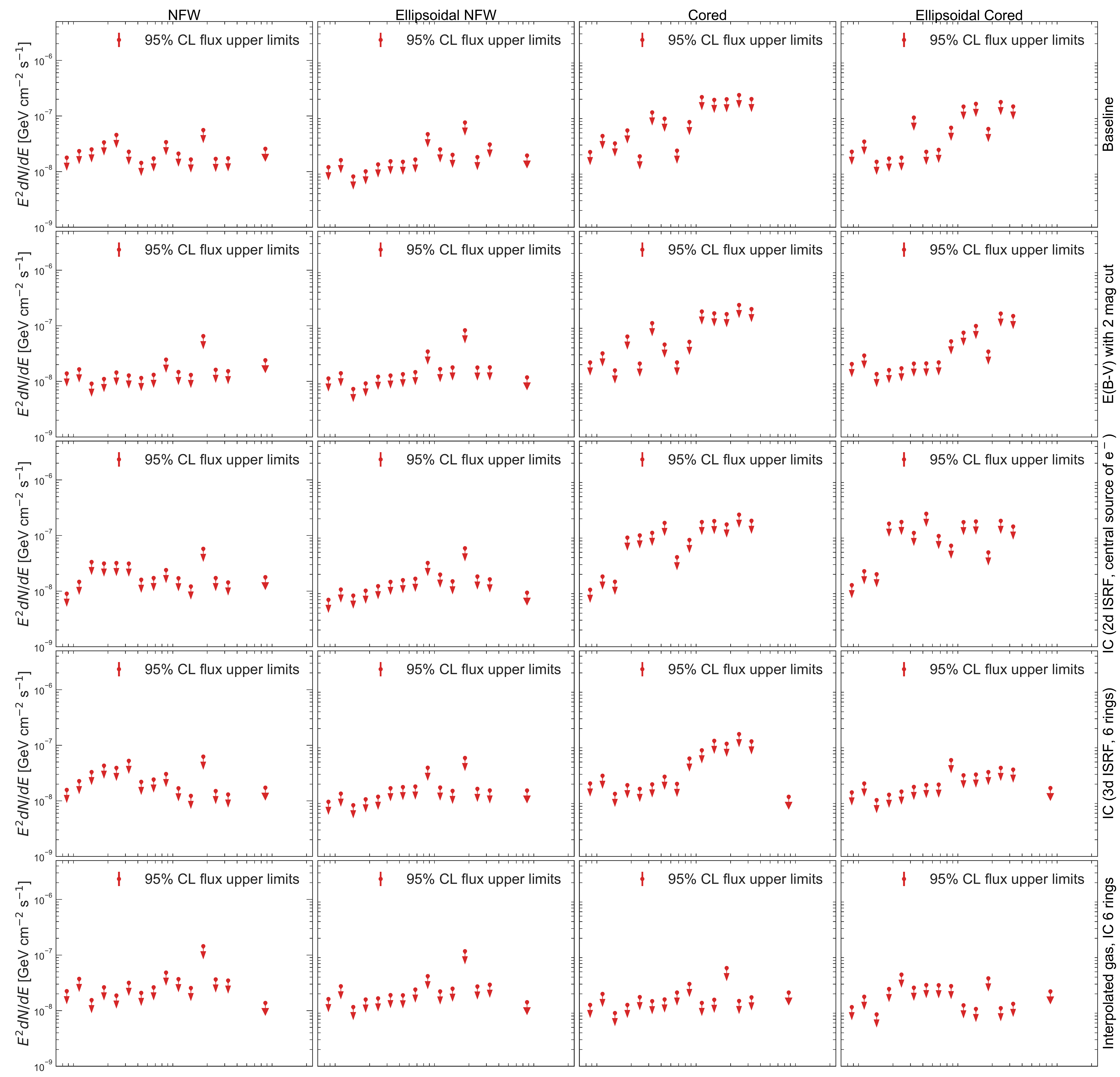}
    \includegraphics[width=\textwidth]{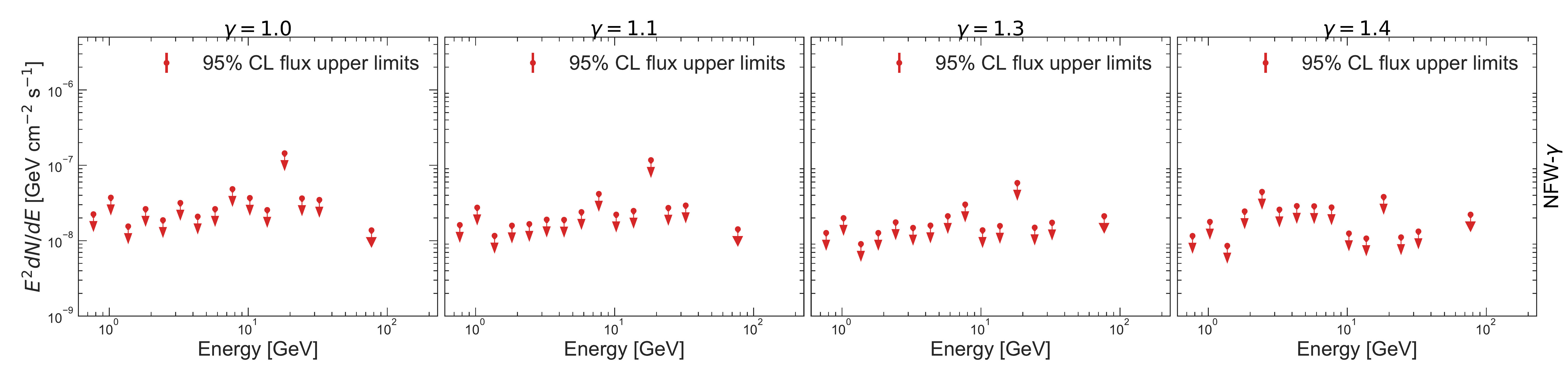}
    \caption{95\% C.L. flux
upper limits for each considered GDE and DM morphology variation. The first five rows correspond to (in order) the baseline, E(B-V) with 2 mag cut, interpolated gas maps, IC (2D ISRF, central source of e$^-$), and IC (3D ISRF, six rings). Unless otherwise stated, all the runs assumed the default IC (3D ISRF, four rings). One additional exception corresponds to the interpolated gas maps which were paired with the IC (3D ISRF, six rings) following Ref.~\cite{TheFermi-LAT:2015kwa}. The columns for these rows correspond to (in order) the NFW ($\gamma=1.2$), Cored ellipsoidal, cored, and Cored ellipsoidal DM morphologies. The last row corresponds to varying the NFW slope, and the columns correspond to $\gamma=1.0$, 1.1, 1.3, and 1.4. We removed from our analysis the last energy bin ($37.49-158.11$ GeV) for the cored cases, since we observed a systematic bias in our DM injection tests (see text).}
    \label{fig:UL_spectra2}
\end{figure*}

The results of our scans for the benchmark and alternative background/foreground models are shown in Fig.~\ref{fig:scans}. Regardless of the background/foreground model assumed, we find that a putative DM source starts to significantly worsen the fits for DM fluxes in the range approximately $5\times 10^{-11}$ ph cm$^{-2}$ s$^{-1}$ and $2\times 10^{-8}$ ph cm$^{-2}$ s$^{-1}$, depending on the energy bin. To have a better understanding of the constraining power of each of our energy bins, we have also displayed the 95\% C.L flux upper limits in Fig.~\ref{fig:UL_spectra2}. These were computed by requiring a change in each profile log-likelihood of 2.71/2 from their maximum.   We remind the reader that our background/foreground model includes templates for the spatial distribution of the bulge stars. As thoroughly discussed in previous studies~\cite{Macias:2016nev, Bartels:2017vsx, Macias:2019omb}, once the stellar bulge models are included to the fits, DM-like spatial models are strongly disfavored. Our current analysis leveraged on this fact to impose some of the strongest constraints on self-annihilating DM models.

\begin{figure*}[ht!]
\centering
\includegraphics[width=0.6\textwidth]{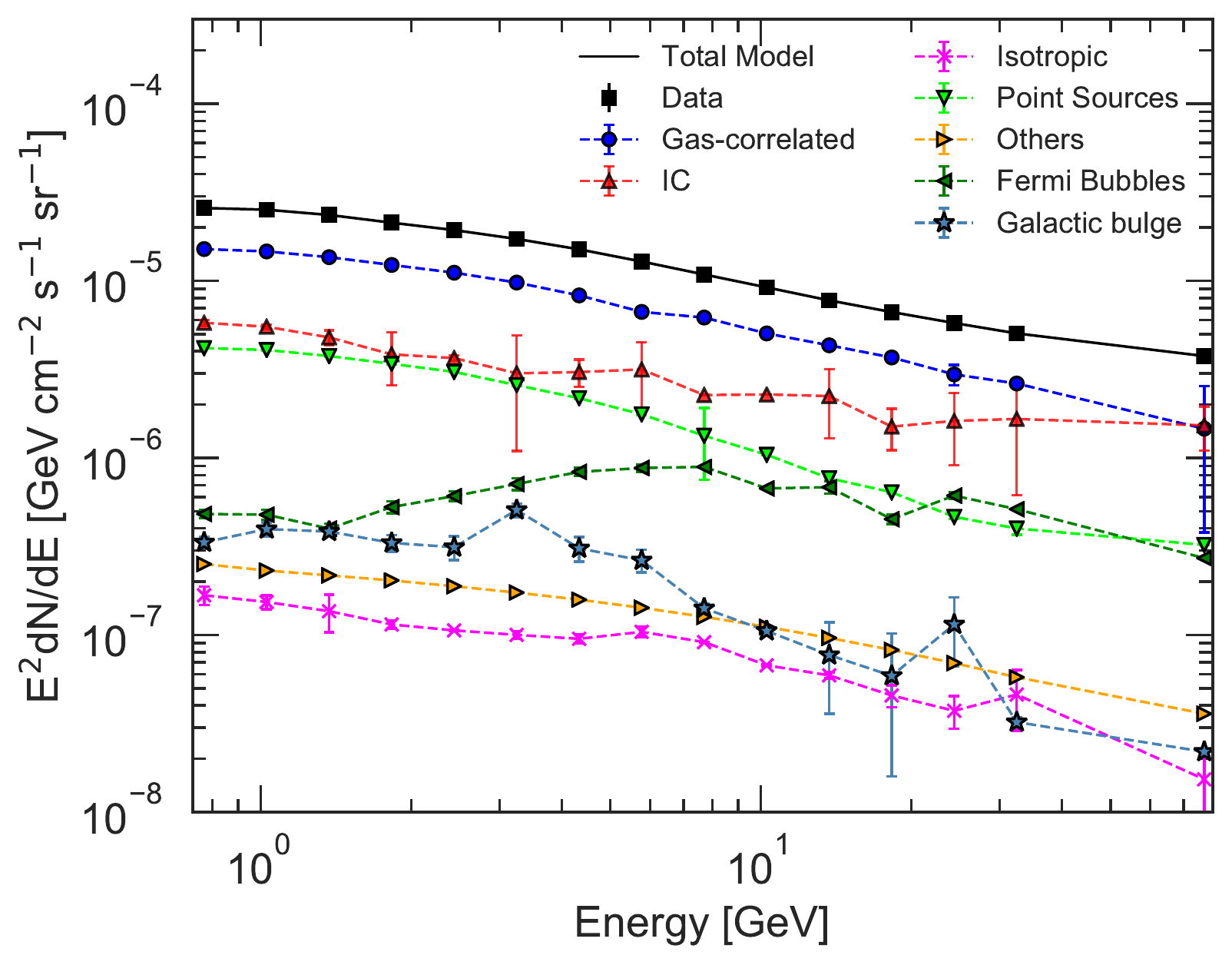}
\caption{Spectral energy distribution of the background/foreground model components included in this study. Shown are the bin-by-bin best fits to \textit{Fermi}-LAT data from the inner $40^{\circ}\times40^{\circ}$ region. The best-fit fluxes of several sources are summed together in assembles for presentation. Gas-correlated gamma-ray emission corresponds to the intensity assigned to the hydrodynamic gas and dust emission templates. The ``Point Sources'' component shows the total spectrum of the 4FGL~\cite{Fermi-LAT:4FGL} point sources in our RoI. The IC template assumed here corresponds to the combined emission of the 3D IC~\cite{Porter:2017vaa} template divided in four rings. The Fermi bubbles map assumed corresponds to an inpainted version of the one constructed in Ref.~\cite{TheFermi-LAT:2017vmf} (see also Fig.1 of Ref.~\cite{Macias:2019omb} for details).``Others'' includes Loop I, Sun, Moon, and extended sources in the 4FGL catalog. The ``galactic bulge'' component is the combined spectrum for the stellar nuclear bulge and galactic bulge templates.} 
\label{fig:total_spectra}
\end{figure*}

 \begin{figure*}[ht!]
    \centering
    \includegraphics[width=0.48\textwidth]{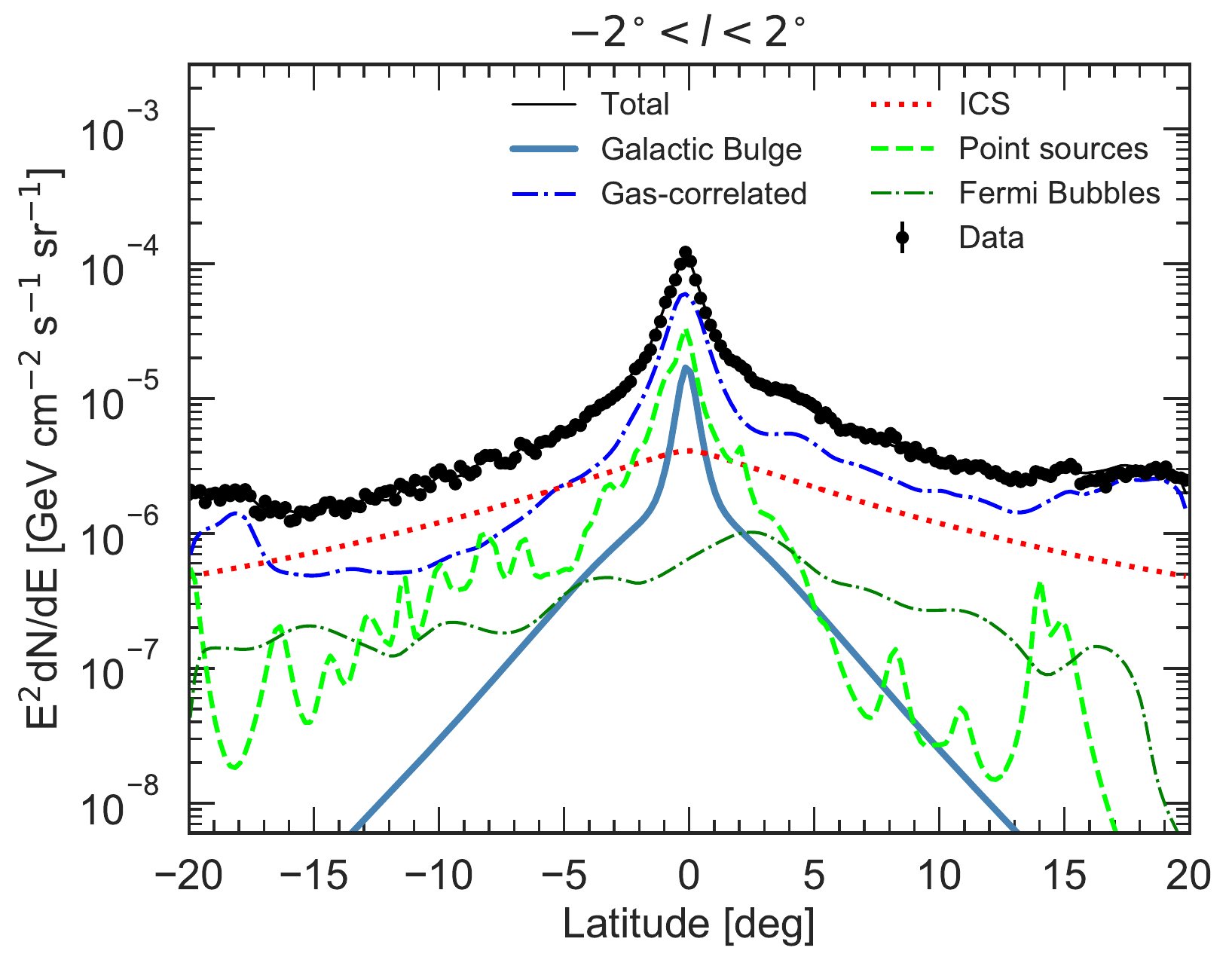}
    \includegraphics[width=0.48\textwidth]{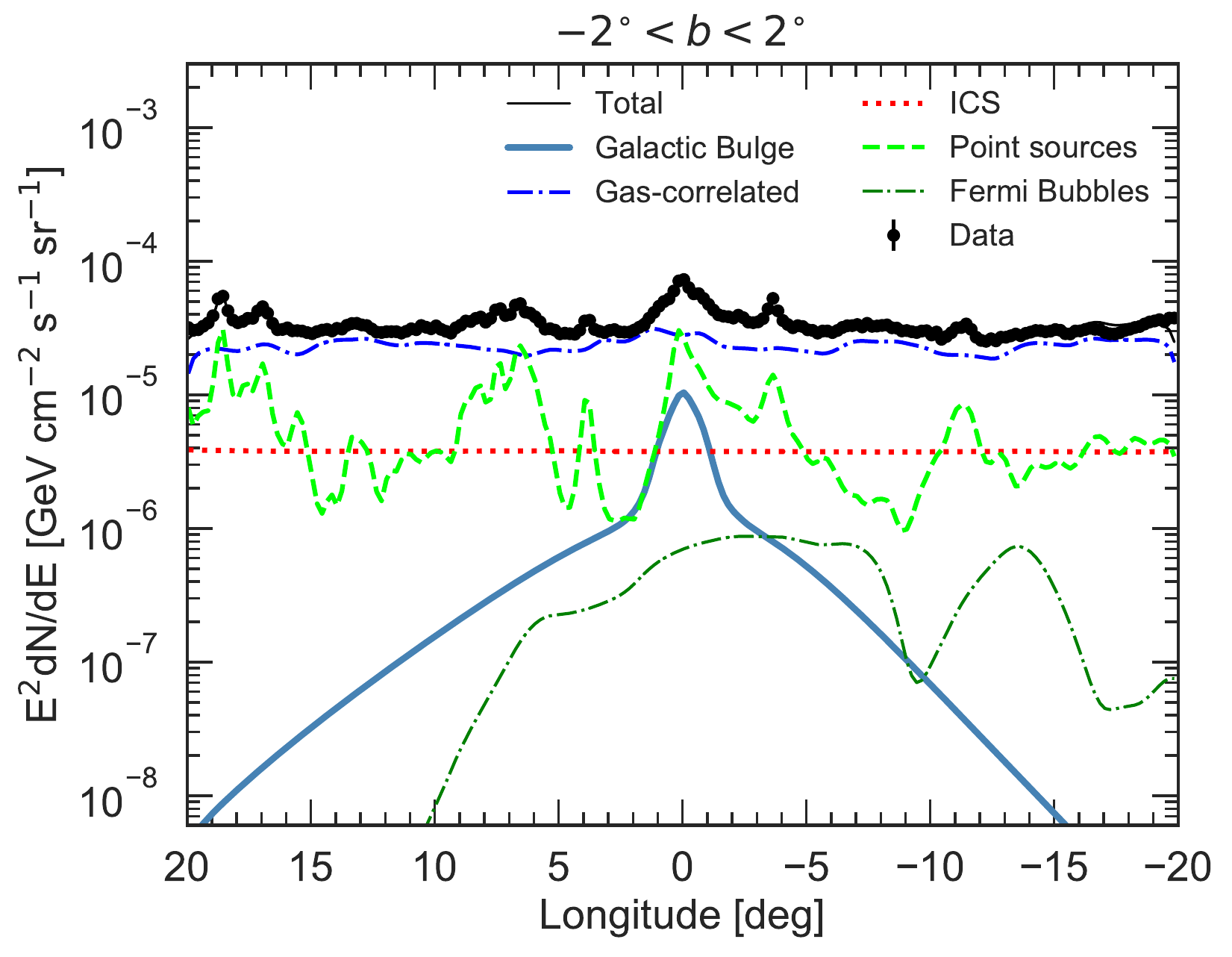}
    \caption{Flux profiles (assuming $|l|<2^\circ$ or $|b|<2^\circ$) of the best-fit model components in the energy range $[1.1,2.8]$ GeV. The black points are the observed $\gamma$-ray flux in the inner $40^\circ\times 40^\circ$ of the GC, while the solid black line represents the superposition of all the best-fit model components. For display purposes, we combine several model components in groups: The galactic bulge is the sum of the boxy bulge and the nuclear bulge. The gas-correlated emission is the sum of all our interstellar gas rings. The best-fit emission of all the 4FGL point sources included in the fit are labeled  point sources. Others [not seen here as they are approximately $\mathcal{O}(1)$ less bright in the profile region] refers to the isotropic, Sun, Moon, and Loop I components. }
    \label{fig:fluxprofiles}
\end{figure*}

\begin{figure*}[ht!]
\begin{tabular}{ccc}
\includegraphics[width=0.24\textwidth]{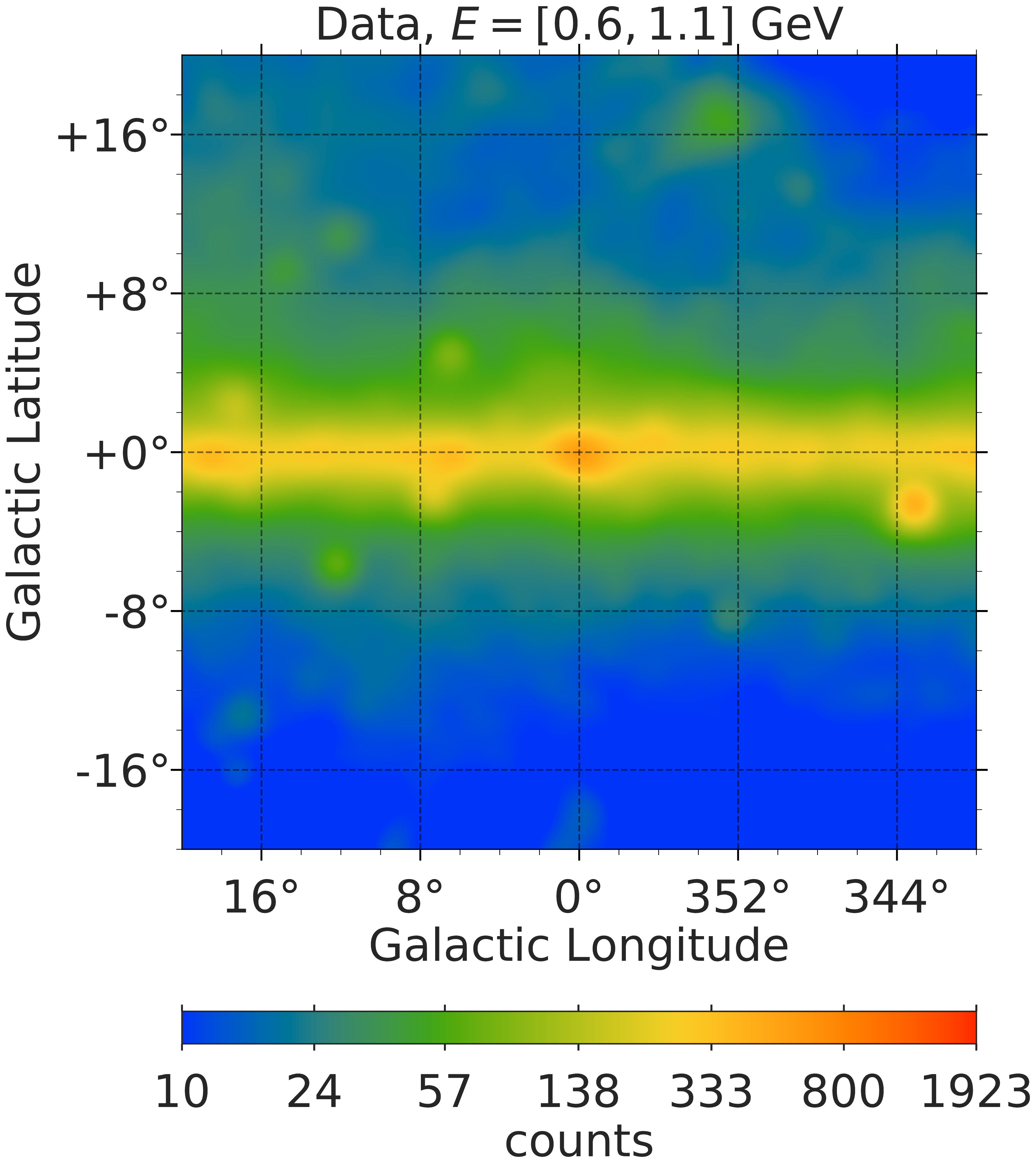} & \includegraphics[width=0.24\textwidth]{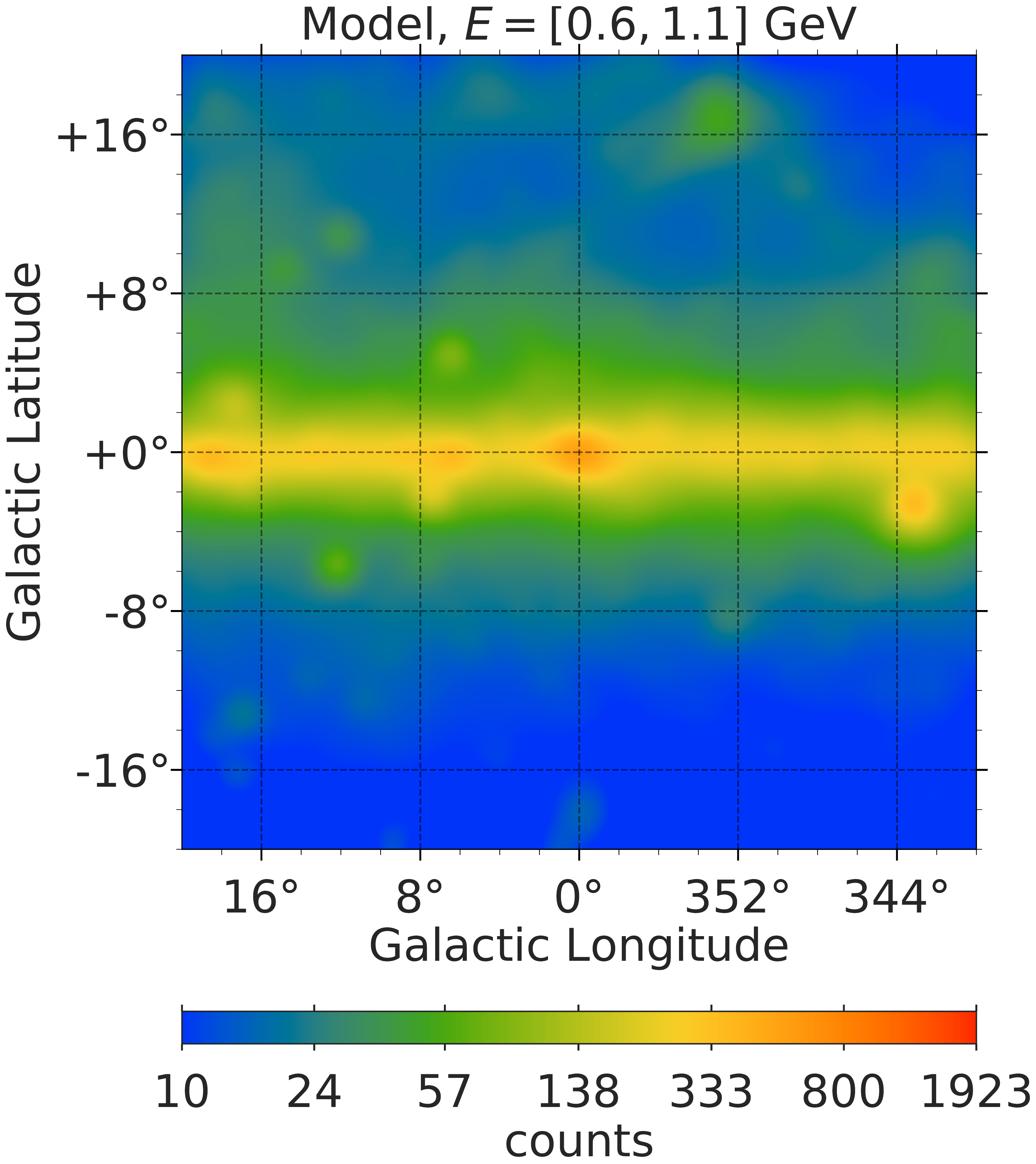}  & \includegraphics[width=0.238\textwidth]{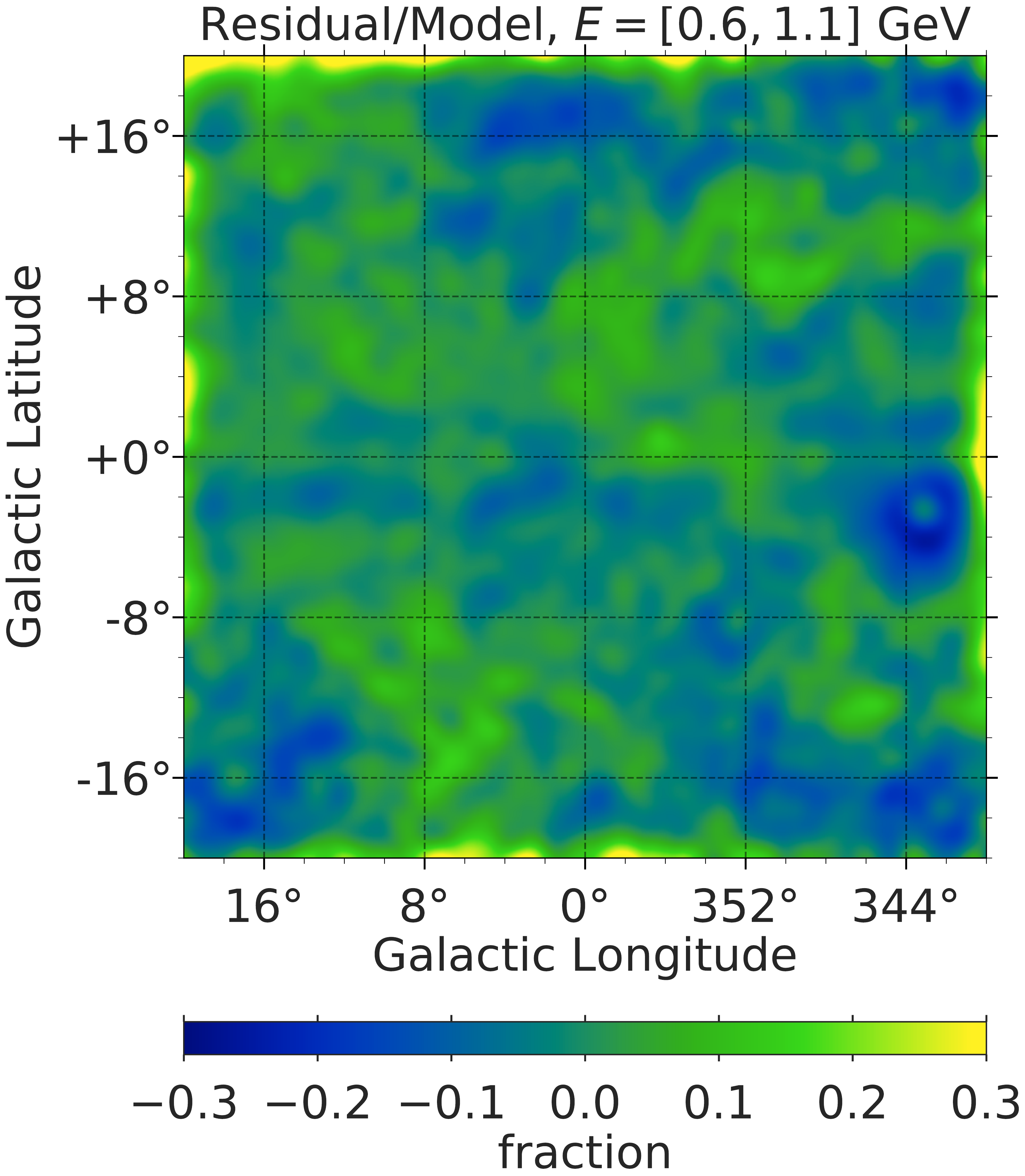}\\
\includegraphics[width=0.24\textwidth]{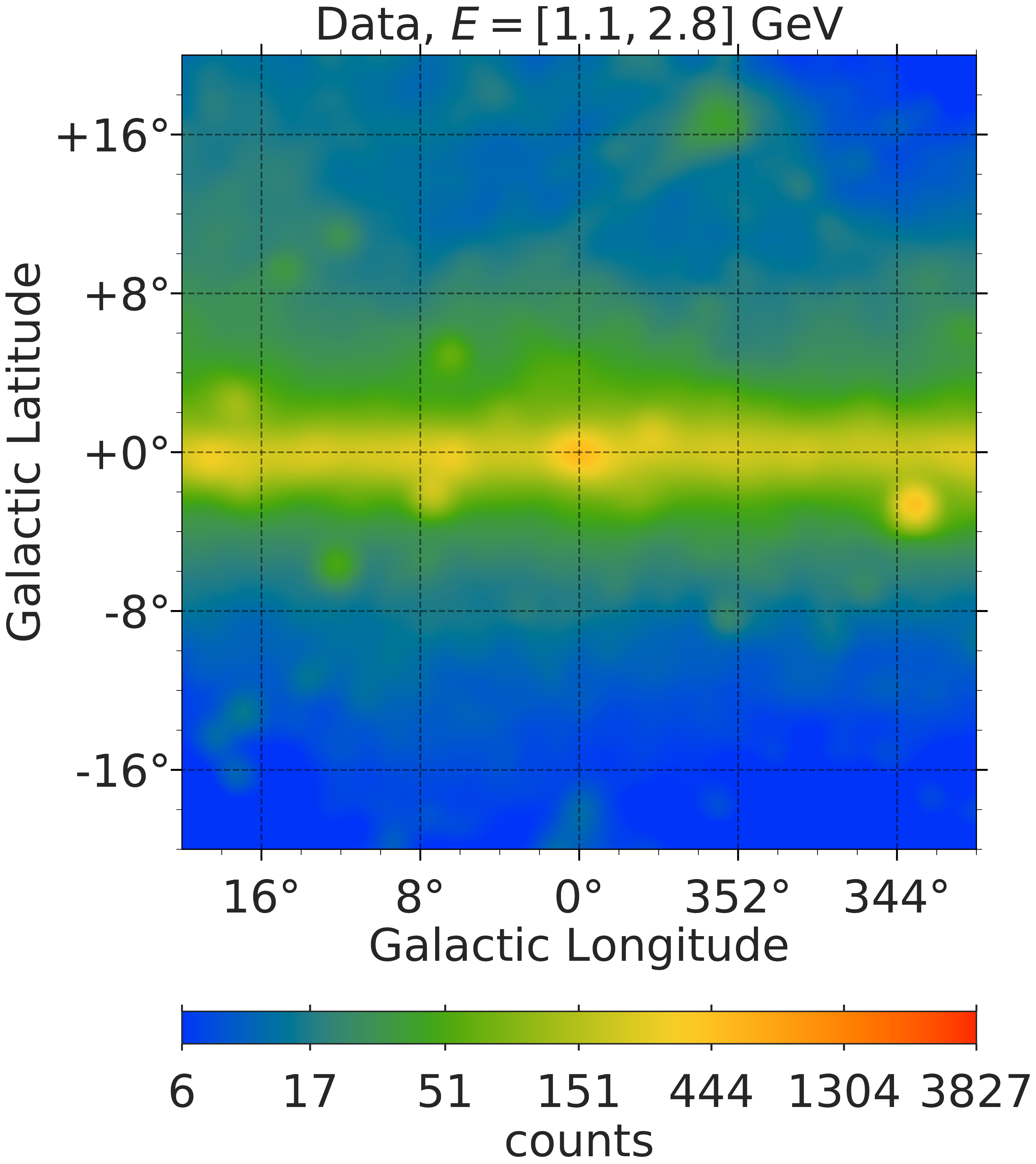} & \includegraphics[width=0.24\textwidth]{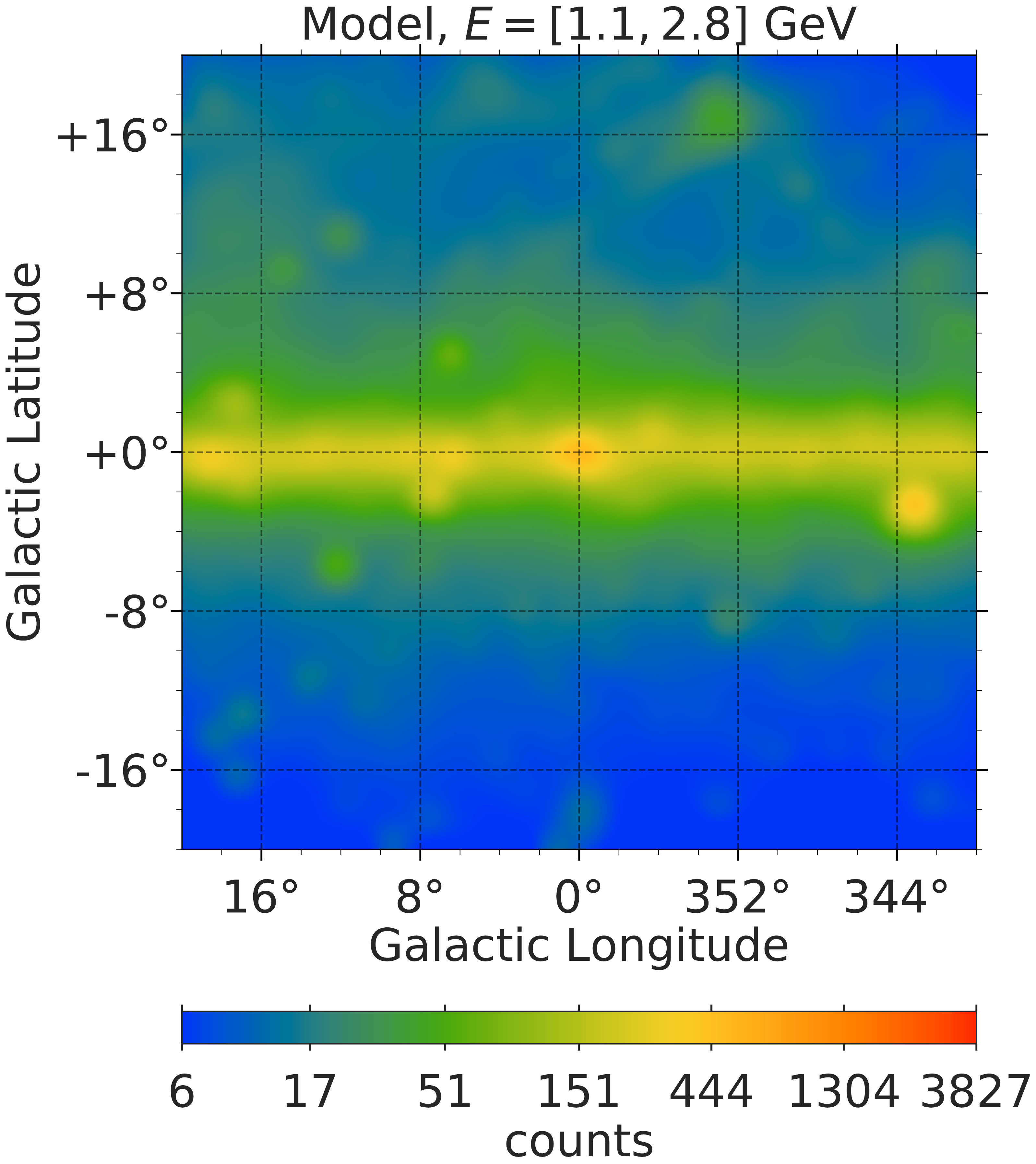}  & \includegraphics[width=0.238\textwidth]{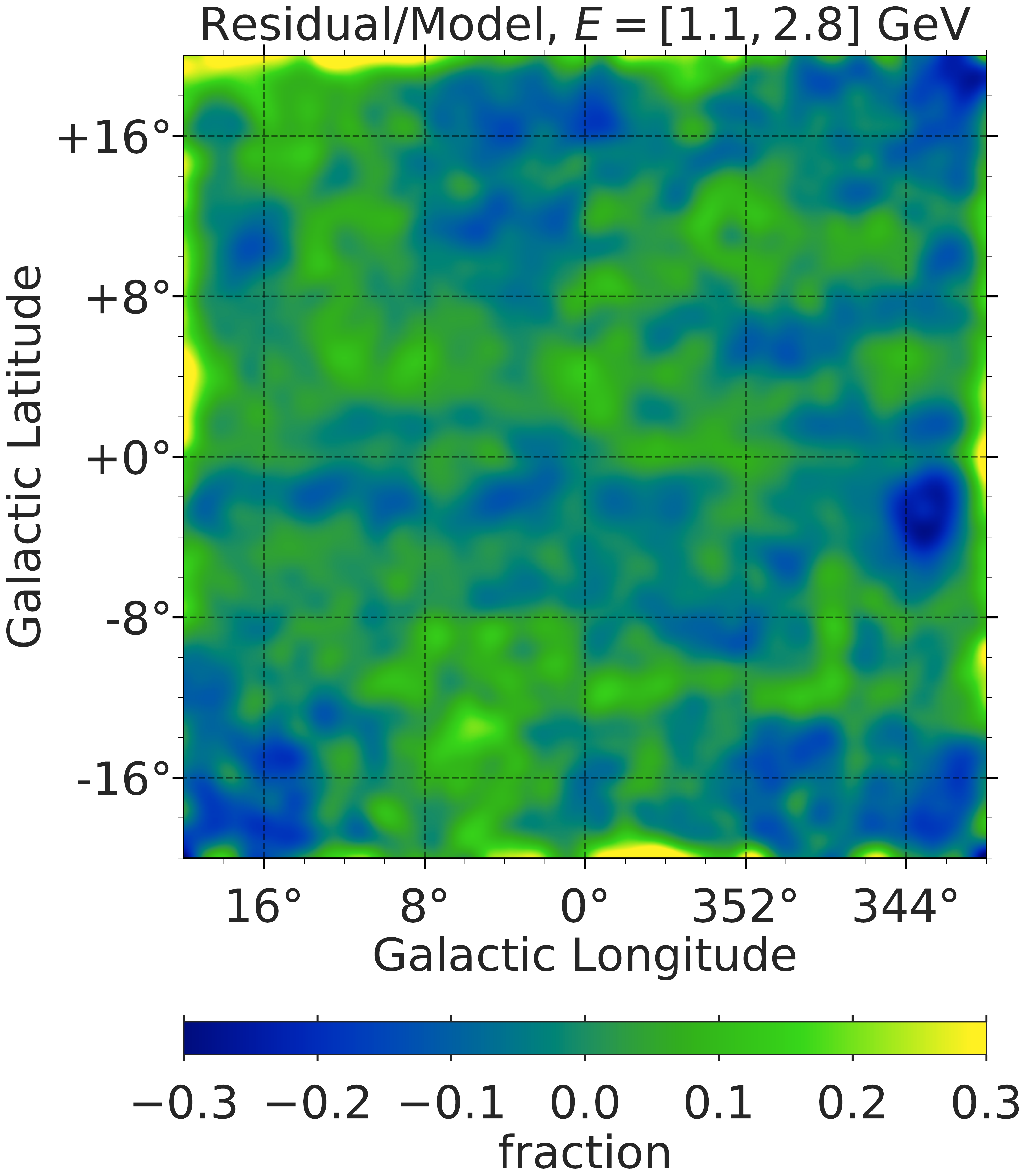}\\
\includegraphics[width=0.24\textwidth]{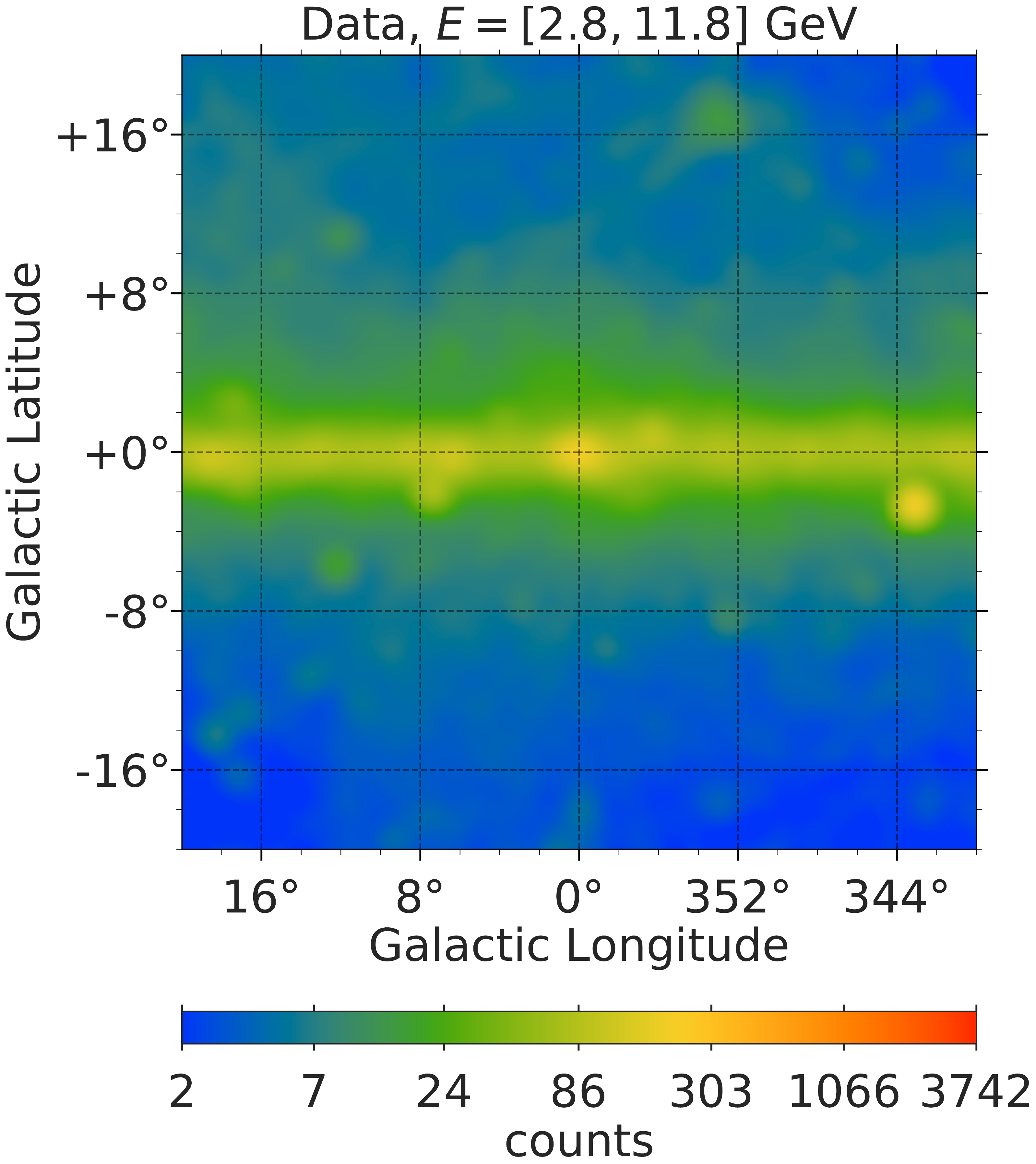} & \includegraphics[width=0.24\textwidth]{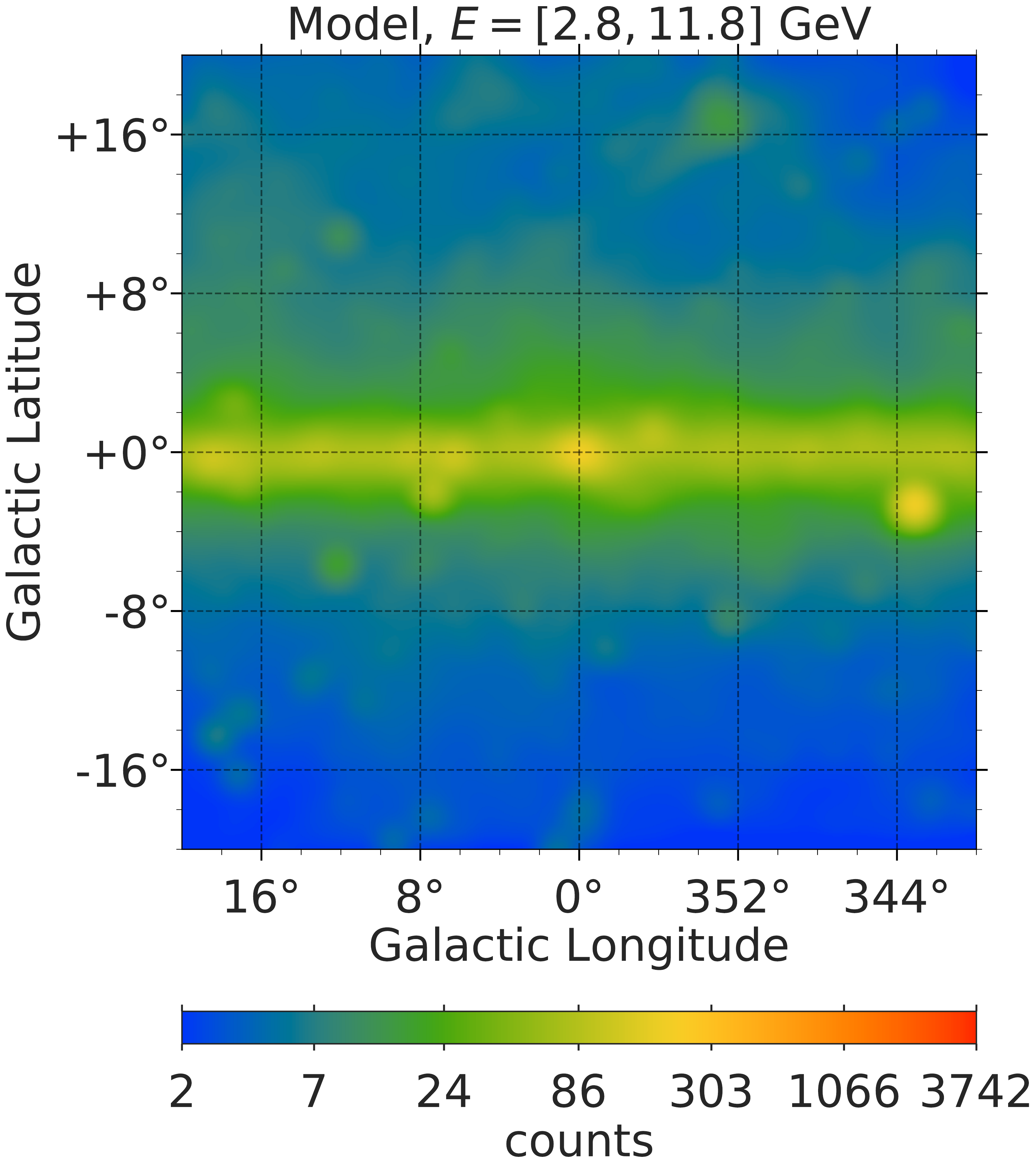}  & \includegraphics[width=0.238\textwidth]{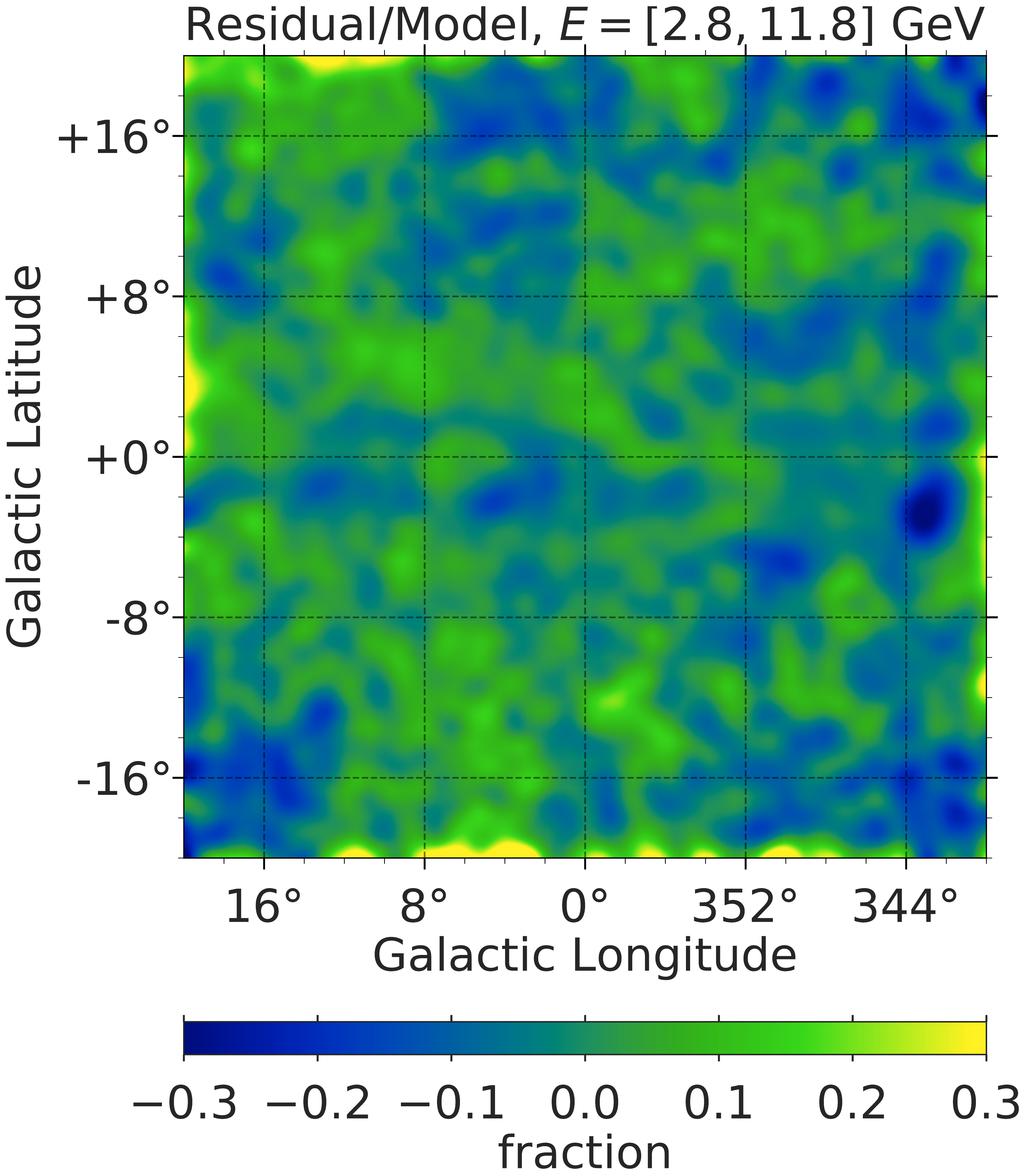}\\
\includegraphics[width=0.24\textwidth]{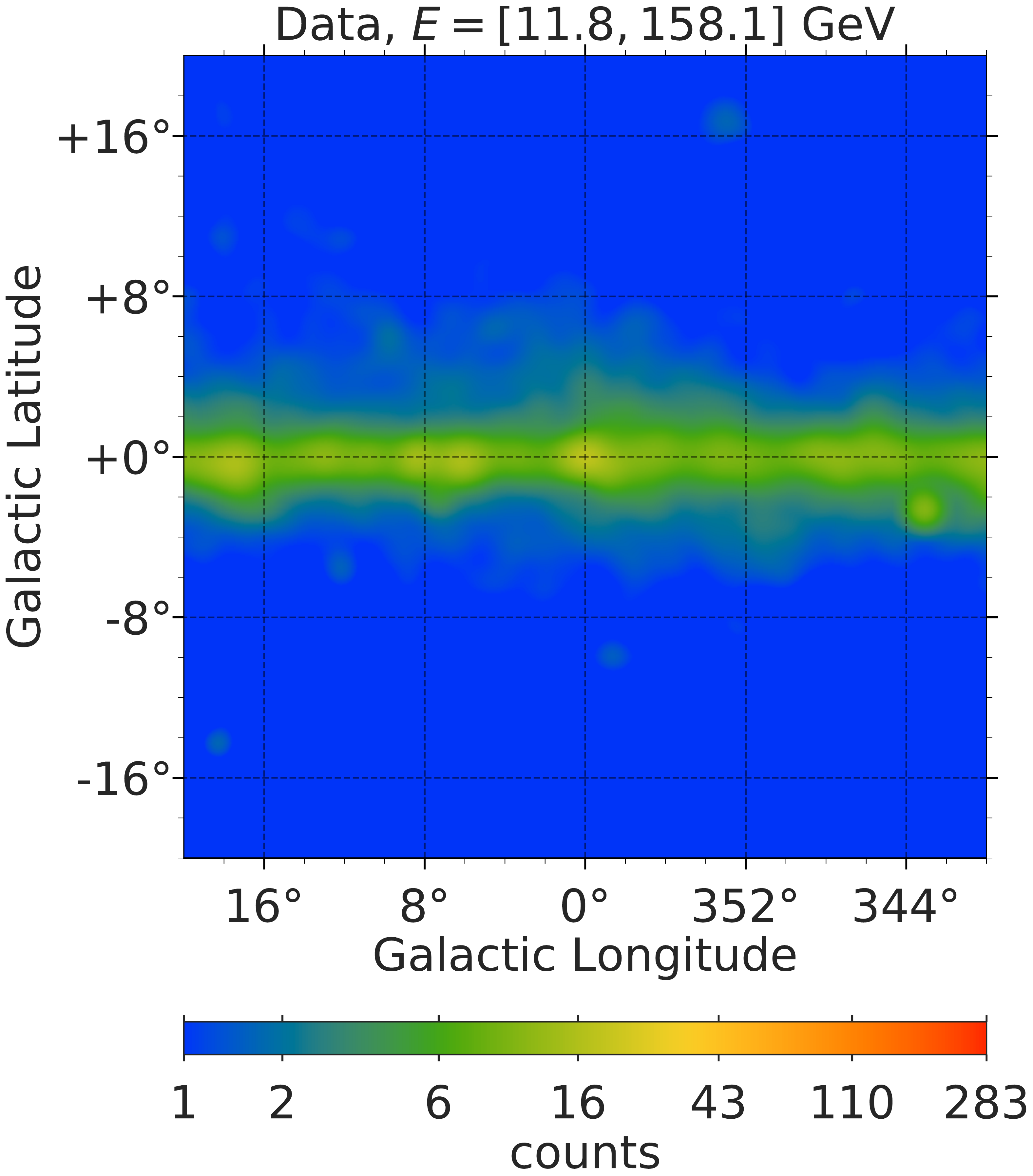} & \includegraphics[width=0.24\textwidth]{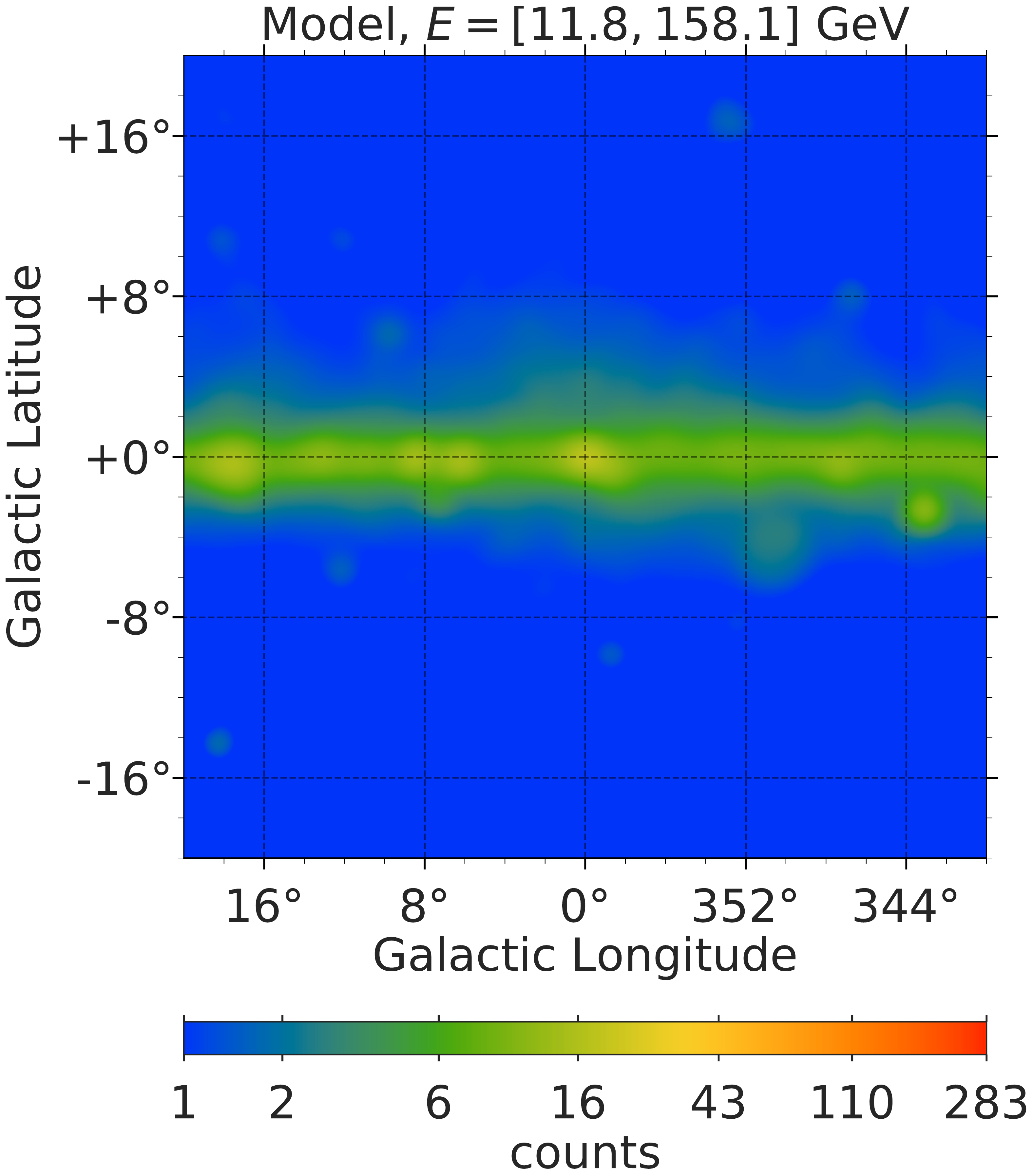}  & \includegraphics[width=0.238\textwidth]{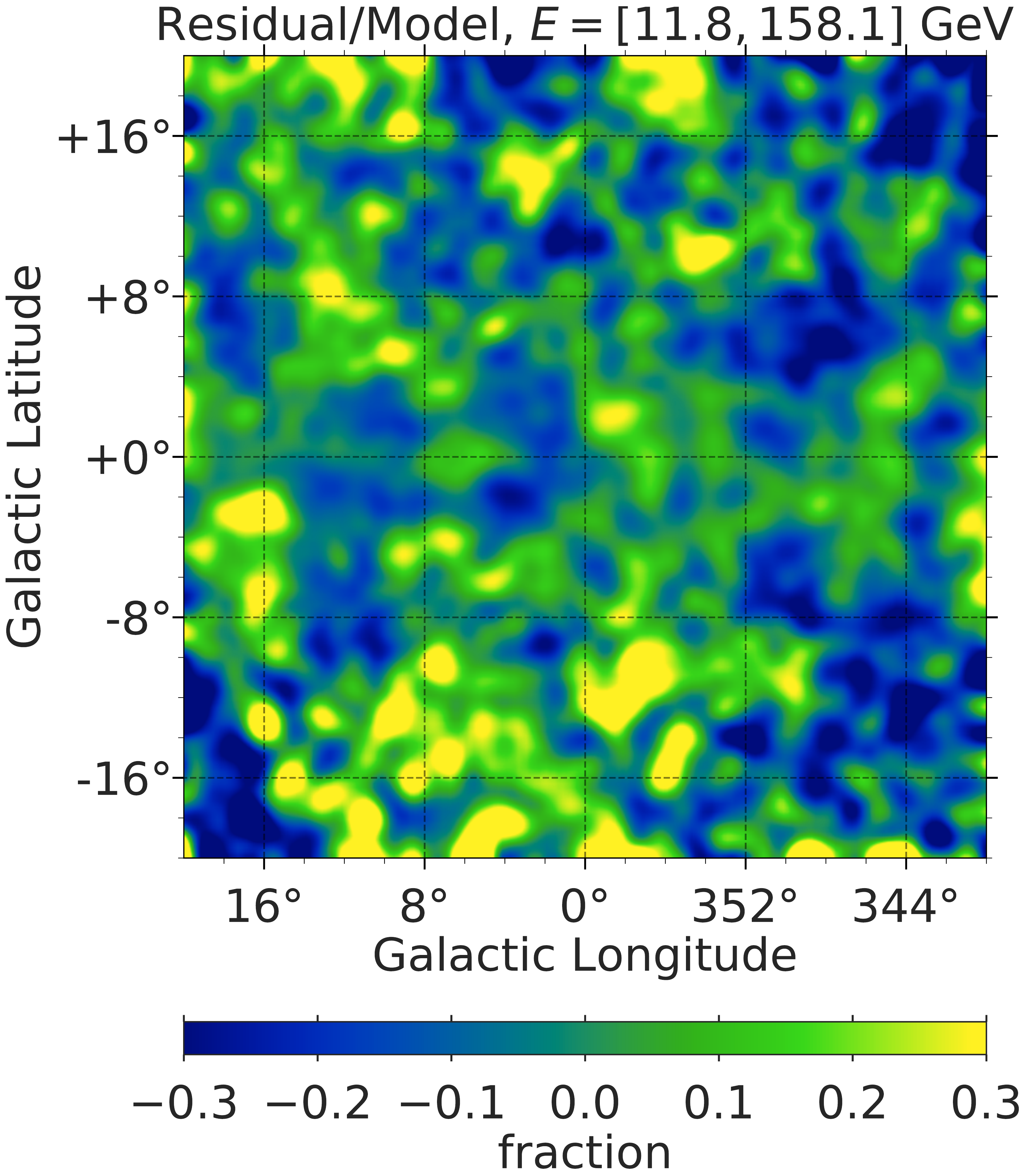}
\end{tabular}
\caption{Measured photon counts (left), best-fit background plus foreground models (middle), and the fractional residuals $(Data-Model)/Model$ (right). For details of the foreground and background model templates, see Table~\ref{Tab:definitions} and Appendix A. The images were constructed by summing the corresponding energy bins over the energy ranges displayed on top of each panel: [0.6,1.1] GeV, [1.1,2.8] GeV, [2.8,11.8] GeV, and [11.8,158.1] GeV, from top to bottom. The maps have been smoothed with a Gaussian filter of radius $0.5^\circ$. The spectrum and flux profiles of the background and foreground model components shown here can be seen in Figs.~\ref{fig:total_spectra} and \ref{fig:fluxprofiles}.}
\label{fig:Residuals}
\end{figure*}

We note that the Sun and the Moon contribute extended gamma-ray emission in our RoI, and not accounting for this emission can bias the spectra of other sources included in our analysis. Templates describing gamma rays originating from the Sun and the Moon need to be independently constructed to match the specific data cuts adopted in the analysis (photon event type, maximum zenith angle cut, energy, and time range). However, constructing newer Sun and Moon templates is bounded by computational costs. As a compromise between computational requirements and photon statistics we used the same data cuts as in the 4FGL~\cite{Fermi-LAT:4FGL} for which there are appropriate Sun and Moon templates readily available. We note that this is another important factor justifying the amount of \textit{Fermi} data included in this analysis.
  
The best-fit spectra of the benchmark background and foreground models are shown in Fig.\ \ref{fig:total_spectra}. As can be seen, our bin-by-bin method produces stable and physically sensible spectra for the model components considered in this work. For display purposes, we have combined the spectra of different sources in groups. It can be observed that the GCE is replaced by the stellar bulge templates. And importantly, the bulge is found to be spectrally distinct to our Fermi bubbles map. Figure~\ref{fig:fluxprofiles} shows the latitudinal and longitudinal flux profiles of the various components included in the fits in comparison with the \textit{Fermi}-LAT data. There are noticeable differences in the shape of the galactic bulge component between the latitudinal and longitudinal flux profiles. This is due to the oblateness of both the boxy bulge and the nuclear bulge templates. It is of importance for this study that the background model components are spatially and spectrally different to the expected galactic bulge emission as this helps preventing possible degeneracies that could impact our log-likelihood scans. 

Figure~\ref{fig:Residuals} (first and second columns) shows a comparison of \textit{Fermi}-LAT data in the $40^\circ\times 40^\circ$ RoI against our best-fitting background/foreground model. Different rows display images combined in four energy windows: $[0.6,1.1]$, $[1.1,2.8]$, $[2.8,11.8]$, and $[11.8,158.1]$ GeV, respectively. All the panels were smoothed with a Gaussian filter of radius $0.5^\circ$. Although the spatial resolution of the LAT is higher than this for energies greater than $10$ GeV, this choice is motivated by limitations in some of our background/foreground model templates. For example, the distribution of the atomic hydrogen column density was derived from the Leiden-Argentine-Bonn 21 cm galactic $H_{\rm I}$ composite survey~\cite{kalberla2005}, which itself has a spatial resolution $0.5^\circ$.

The panels in the third column of of Fig.\ \ref{fig:Residuals} show the fractional residuals, $(\rm{Data}-\rm{Model})/\rm{Model}$, for the benchmark model (see Fig.\ \ref{fig:total_spectra} for the spectrum). It can be seen in the first three energy windows ($[0.6,1.1]$, $[1.1,2.8]$ and $[2.8,11.8]$ GeV) that the model mostly underpredicts the data at the $\lesssim 10\%$ level, with the exception of some more localized negative residuals that reach up to $\sim 20\%$. However, in the last energy window ($[11.8,158.1]$ GeV) the regions of under/overprediction can reach to the $\sim 30\%$ level. Interestingly, these fractional residual images (especially the last energy window) bear some resemblance to the low-latitude Fermi bubbles counterpart (e.g., Fig.8, bottom right of Ref.\ \cite{TheFermi-LAT:2017vmf}). It should be noticed that the Fermi bubbles template used in this analysis is an inpainted version of the original Fermi bubbles template obtained in Ref.\ \cite{TheFermi-LAT:2017vmf}. In that study, a spectral component analysis~\cite{Malyshev:2012mb} was applied to data in the $[1,10]$ GeV energy range in order to reconstruct a morphological template with photons having the same spectrum as that of the Fermi bubbles in the high-latitude region. It is possible that if the same image reconstruction technique is applied to data that include the $[11.8,158.1]$ GeV energy range, regions of under/over-prediction in our last energy window will be ameliorated. A more thorough investigation of the Fermi bubbles template in our last energy window is beyond the scope of our current work, and we leave this interesting possibility for future a analysis. 

We note that when a DM-like template is included as a model for the positive residual, this is unable to account for all of the residual emission. In this sense, our DM limits should be seen as conservative. Even though the residual emission does not appear spherically symmetric distributed, our fitting procedure allows sufficient freedom to the DM template to try account for most of the residual photons. 

Our main concern in this section was to investigate the extent at which the computed DM constraints depend on the specific fore-/background model assumed. It was not our aim to perform an exhaustive search for an alternative foreground model that matches the LAT data best in the GC region. Indeed, in Ref.~\cite{Macias:2019omb}, we have shown that GDE models that assume the hydrodynamical gas and the new 3D IC maps are better fits to the \texttt{Fermi} data. However, here, we used the different variations in the fore-/background model for the purpose of testing the impact they had in our limits and estimating their expected variance. 

\section{Dark Matter Injection and Recovery tests}\label{appx:DM_injections}

Given that our upper limit procedure allows for all the sources to vary in the fits~\footnote{We varied the normalization of the 146 model components included in the fits.}, it is crucial to verify that our foreground/background model would not absorb a DM signal if one were present in the data. For this, we have artificially injected DM signals of different characteristics into the real data and consecutively applied our upper limits procedure to each augmented dataset.

Our tests are similar to those carried out in Refs.~\cite{Chang:2018bpt,Leane:2019xiy}; we have simulated DM injections by taking a random Poisson draw of DM maps generated for a range of DM masses and annihilation cross sections. In particular, we considered self-annihilating DM in the $\bar{b}b$ channel; DM masses of 10, 25, 100, and 500 GeV; annihilation cross sections in the range $[10^{-27},3\times10^{-25}]$ cm$^{3}$/s; and two different DM spatial morphologies (gNFW $\gamma=1.2$ and cored profiles). For a given realization, we obtained the 95\% C.L. flux upper limits by requiring a change in the log likelihood of 2.71/2 from the best-fitting point.

\begin{figure*}[ht!]
\centering
\includegraphics[width=1.0\textwidth]{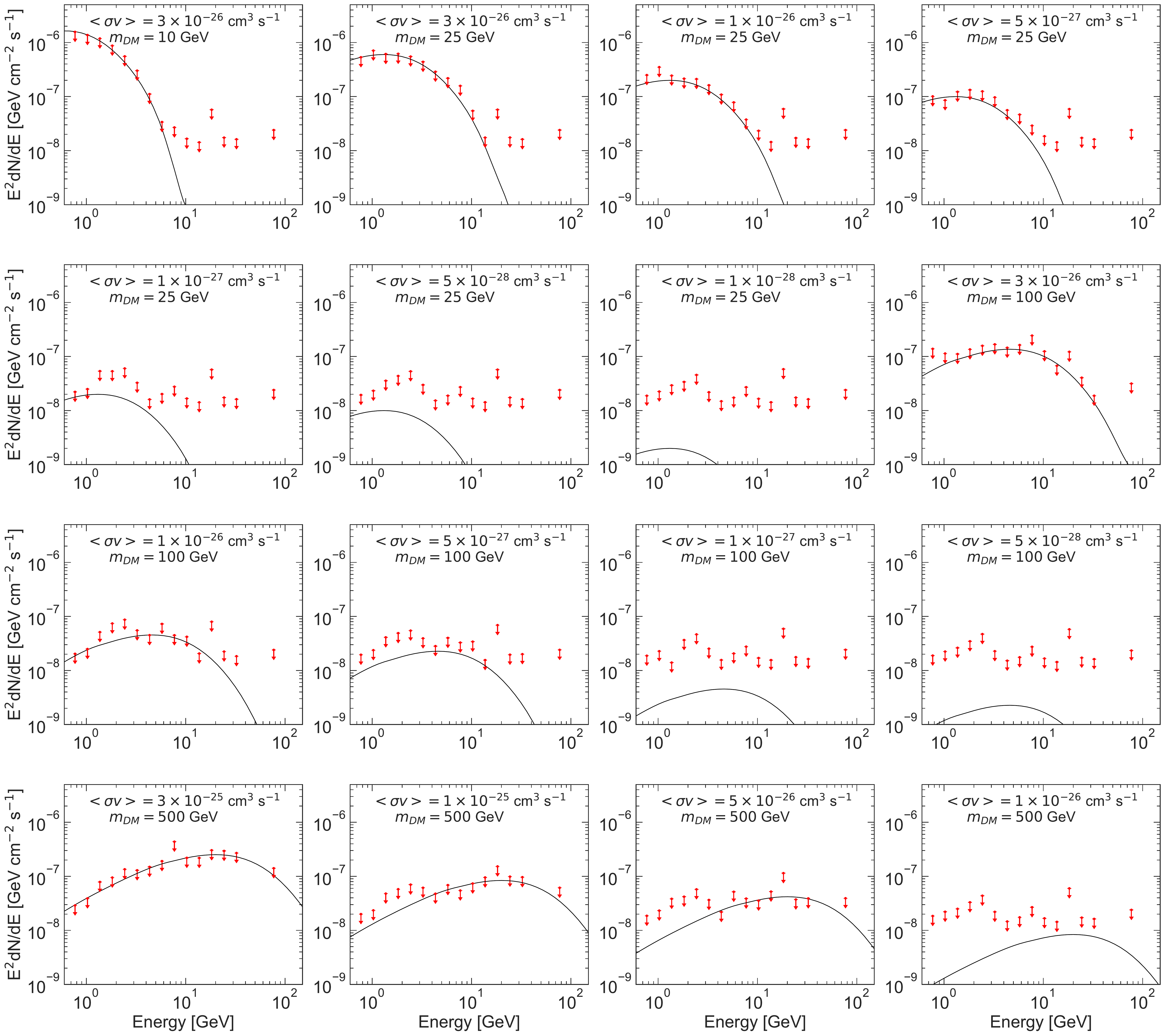}
\caption{Each panel shows the comparison between an artificial DM signal injected into the real data and corresponding bin-by-bin 95\% C.L. flux upper limits  recovered after passing those through our upper limits procedure. All realizations assume DM particles self-annihilating into the $\bar{b}b$ channel. The spatial morphology assumed corresponds to a NFW profile with $\gamma=1.2$. The assumed foreground/background model is benchmark model shown in Fig.~\ref{fig:total_spectra}. The bin-by-bin upper limits method allows all the 4FGL and GDE components to float in the fit (see text). } 
\label{fig:DM_InjectionNFW}
\end{figure*}

\begin{figure*}[t!]
\centering
\includegraphics[width=1.0\textwidth]{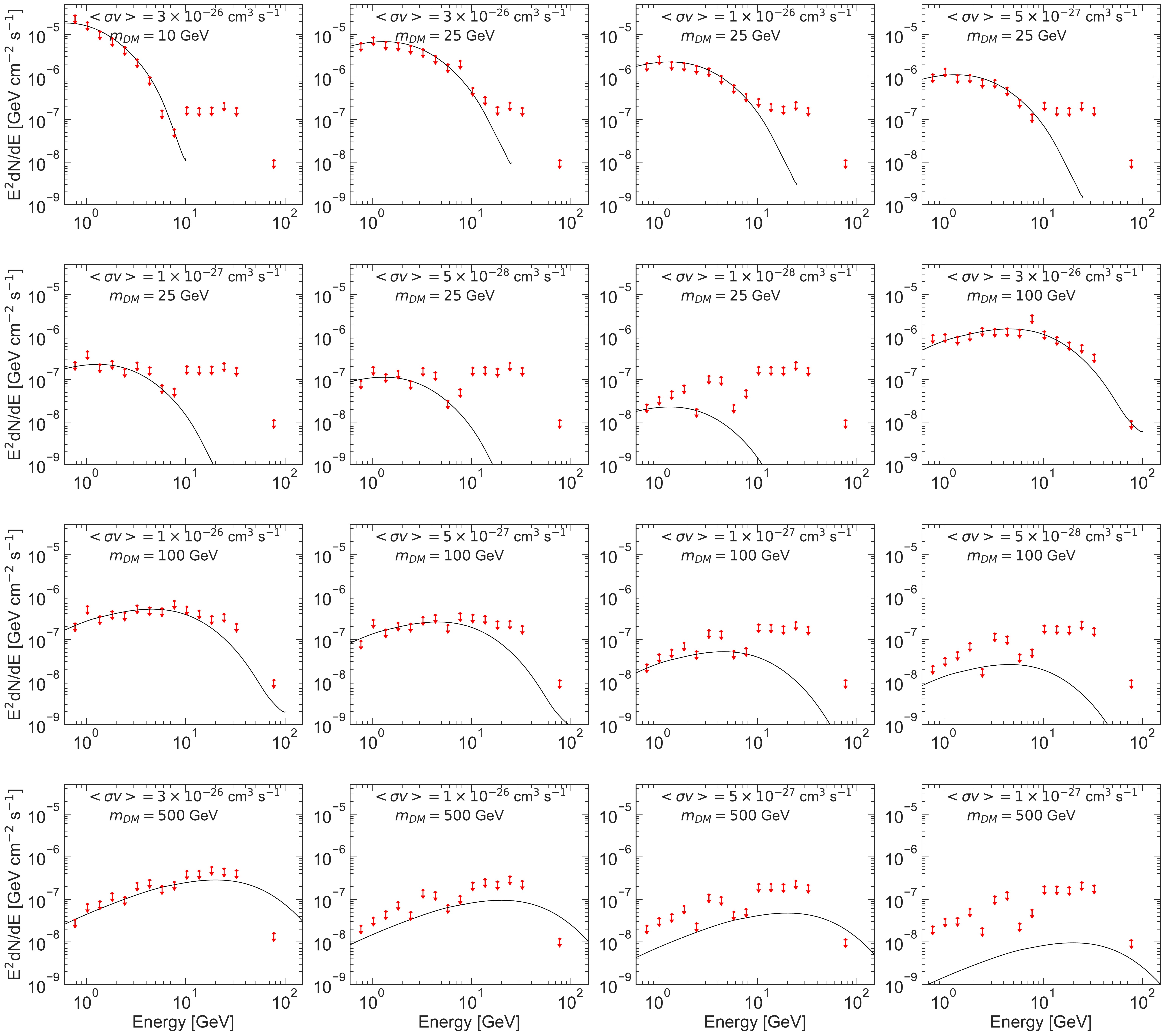}
\caption{Same as Fig.~\ref{fig:DM_InjectionNFW}, except that here the spatial morphology of the DM source is given by a cored profile. The flux upper limit in the last energy bin was found not to have the correct statistical coverage. Conservatively we have removed the log-likelihood data corresponding to this bin from our limits setting procedure. See also Fig.~\ref{fig:UL_spectra2}.} 
\label{fig:DMInjection_Read}
\end{figure*}

The results of our DM injection tests are presented in Figs.~\ref{fig:DM_InjectionNFW} (gNFW $\gamma=1.2$) and~\ref{fig:DMInjection_Read} (cored profile). In each panel, the black line shows the DM signal that was injected into the data, and the red arrows display the 95\% C.L. flux upper limits recovered with our log-likelihood profile scan method.  As can be seen from these figures, for a large majority of our realizations, the recovered bin-by-bin flux upper limits have the correct statistical coverage. There are a few cases in which our upper limits are below the injected DM signal; for most of those we nonetheless obtain that the upper limits weaken in a way that is consistent with the strength of the injected signal. The only exception to this pattern was observed in the highest energy bin ($37.49-158.11$ GeV) for the  DM injections corresponding to the cored profile. In this case, it was found that the flux upper limits did not have the correct statistical coverage for all our high DM mass injection trials. It is possible that this is due to a combination of complicating factors. First, the cored profile is much flatter than the gNFW profile. Second, in the highest energy bin the statics are low. Degeneracies between the injected DM signal and the GDE model components appear to be difficult to resolve under these conditions.

\begin{figure*}
    \centering
    \includegraphics[width=0.49\textwidth]{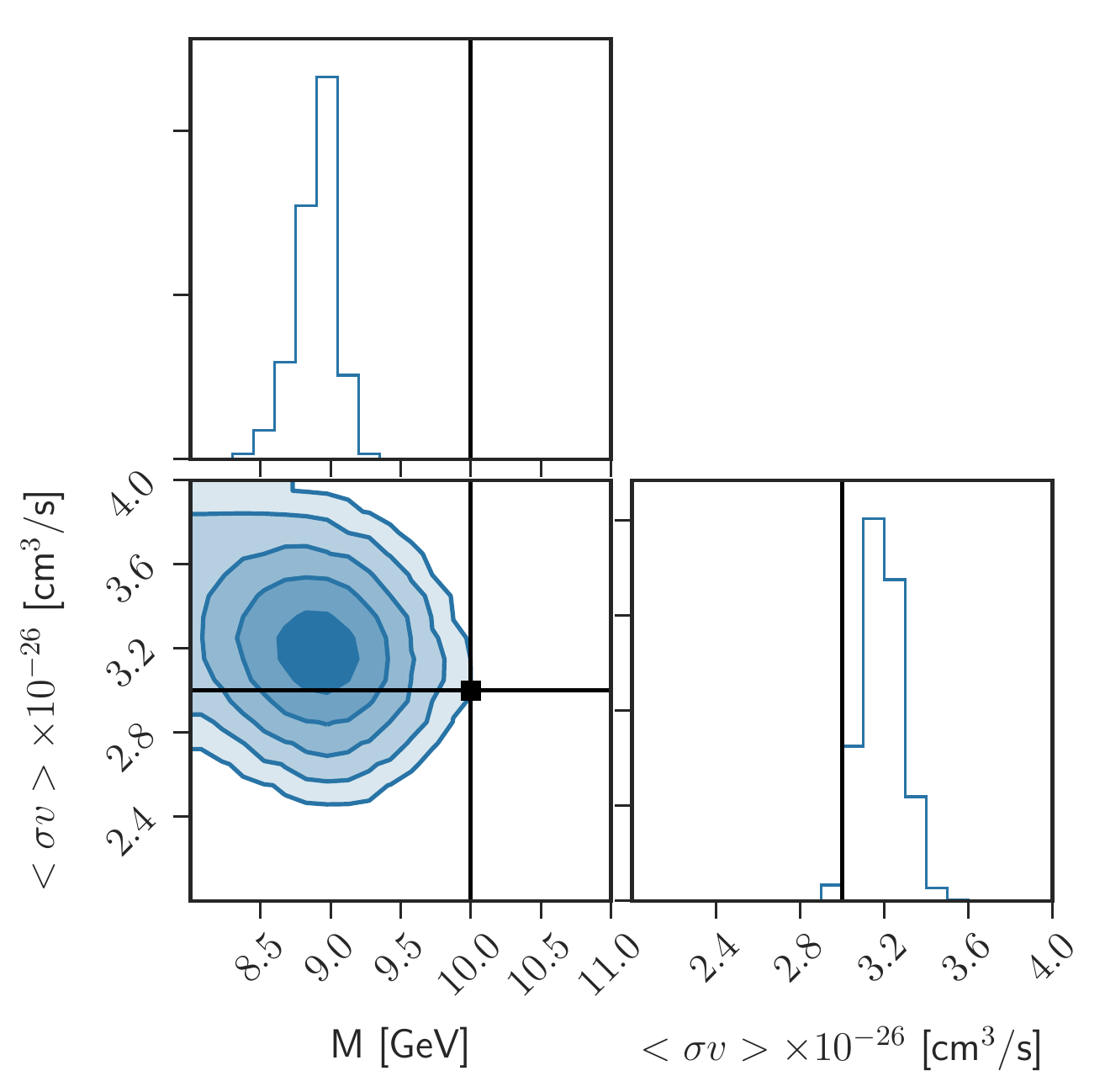}
    \includegraphics[width=0.49\textwidth]{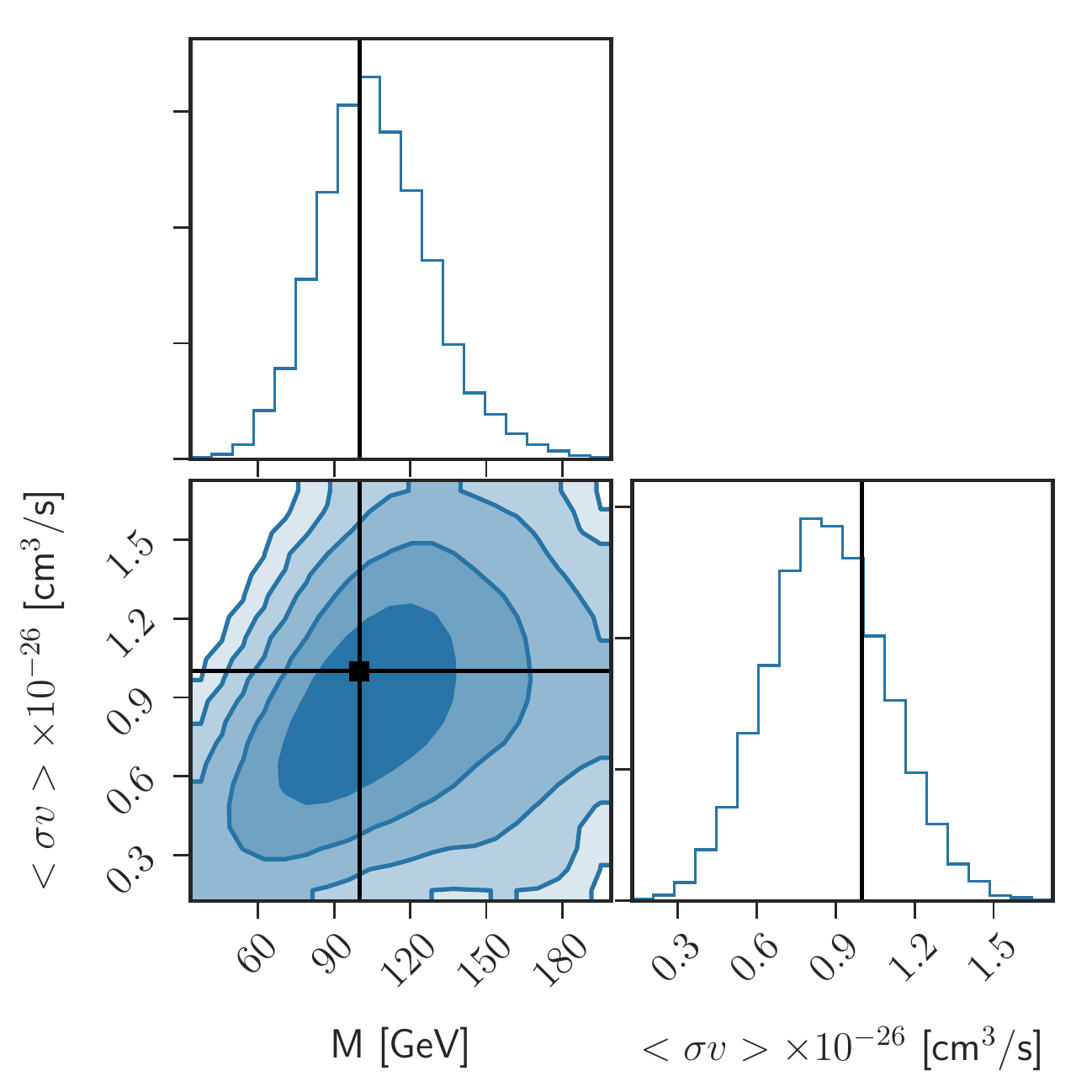}\\
    \includegraphics[width=0.49\textwidth]{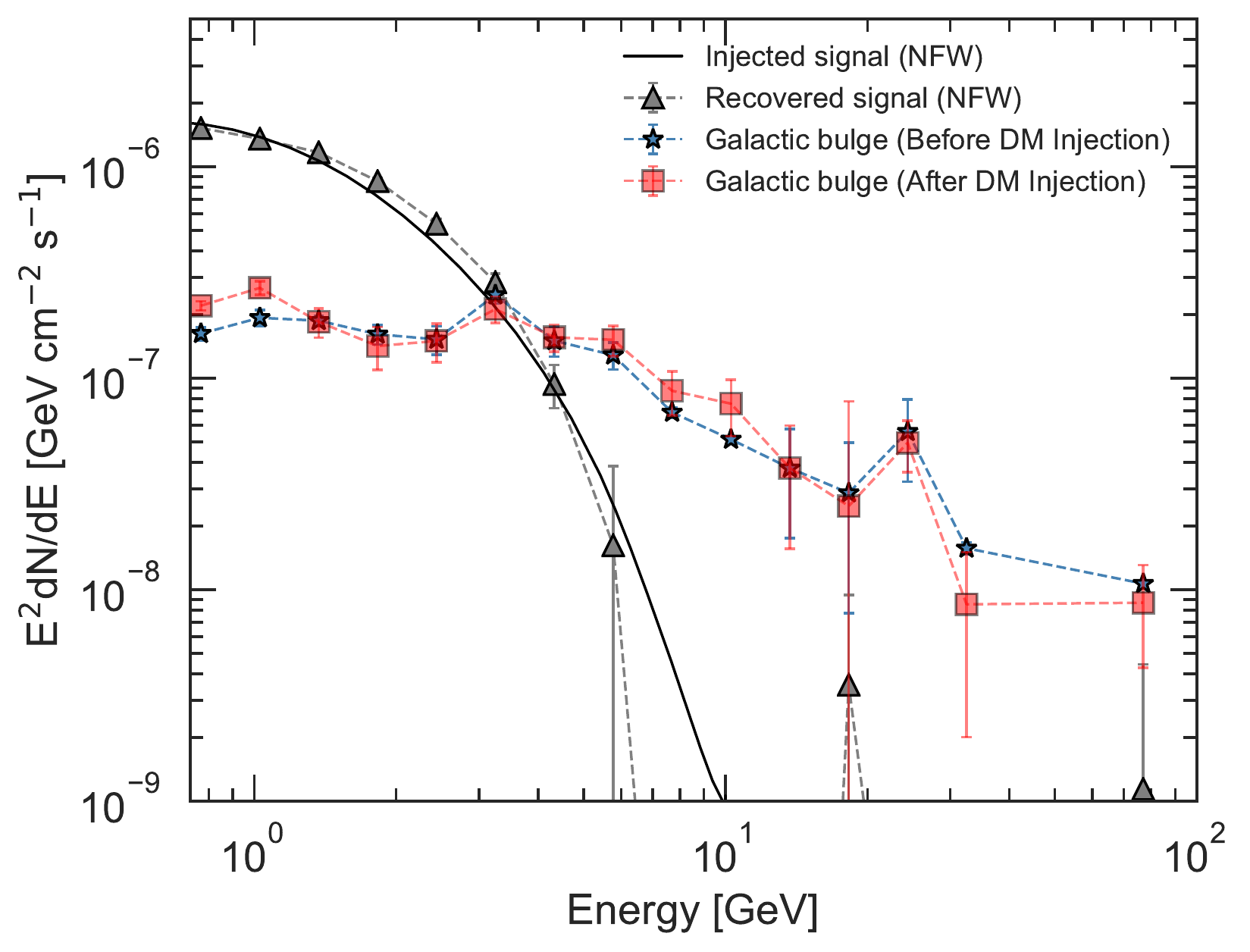}
    \includegraphics[width=0.49\textwidth]{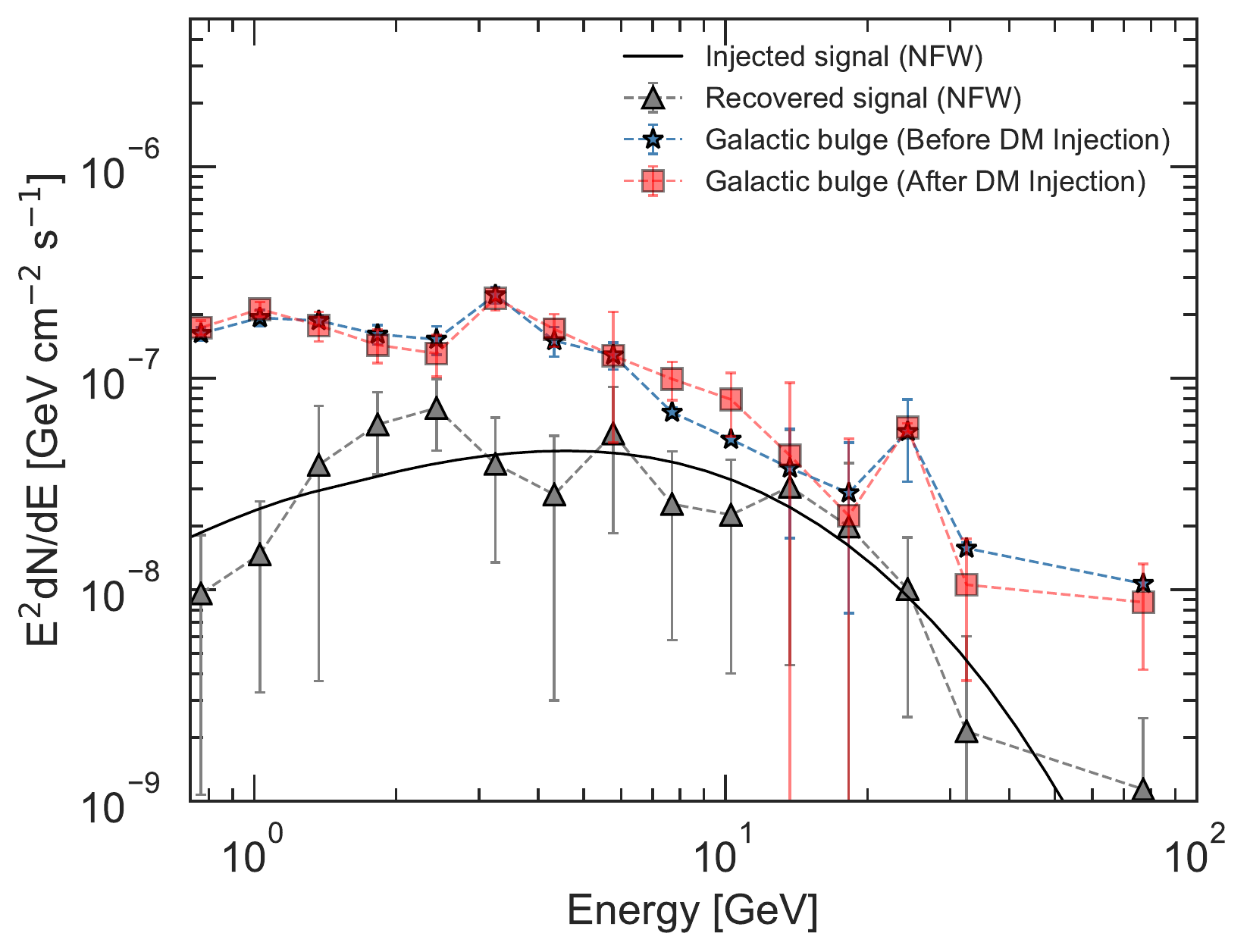}\\
    \caption{Results of our bin-by-bin analysis procedure applied to \textit{Fermi} data containing injected DM signals. The left column corresponds to the injection of a NFW signal with $m_{\rm DM}=10$ GeV and $\langle \sigma v \rangle=3\times 10^{-26}$ cm$^{3}/$s, and the right column corresponds to $m_{\rm DM}=100$ GeV and $\langle \sigma v \rangle=1\times 10^{-26}$ cm$^{3}/$s. The bottom panels show the injected and recovered DM signal, as well as the Galactic bulge spectra obtained before and after the DM injections. In the top panels, we show the results of a DM parameter scan using MCMC methods. The black crosses represent the injected signals, and the contours represent $1\sigma,2\sigma,$...$,5\sigma$ (statistical-only) confidence regions. }
    \label{fig:DMInjection_contours}
\end{figure*}

It is interesting to inspect in more detail the characteristics of the recovered DM spectra for some of our injection tests. For this, we present two example injection points in Fig.~\ref{fig:DMInjection_contours}. The left column corresponds to the injection of a NFW signal with $m_{\rm DM}=10$ GeV and $\langle \sigma v \rangle=3\times 10^{-26}$ cm$^{3}/$s, and the right column corresponds to an injection with the same spatial morphology, $m_{\rm DM}=100$ GeV and $\langle \sigma v \rangle=1\times 10^{-26}$ cm$^{3}/$s. The bottom panels show the injected DM spectra in comparison to the recovered DM spectra for each case, respectively. We also show the Galactic bulge spectra obtained before and after the signal injections. We note that the level of degeneracy between the injected DM signal and the Galactic bulge is small. This is evident from the fact that the Galactic bulge spectra remain largely unchanged after our bin-by-bin analysis has been applied to the data containing the injected signal. 

The triangle plots in the top panels of Fig.~\ref{fig:DMInjection_contours} show the results of a  DM parameter scan that we performed using the recovered bin-by-bin DM spectra. In particular, we ran a Markov chain Monte Carlo
(MCMC) routine using the \texttt{emcee}\footnote{\url{https://emcee.readthedocs.io/en/stable/}} package to scan the 2D parameter space given by ($m_{\rm DM}$, $\langle \sigma v \rangle$). We report the probability distributions for these two parameters, and their respective confidence contours ($1\sigma,2\sigma$,...,$5\sigma$). The true injected values are represented by black dots. For the injection point displayed in the top-left panel of Fig.~\ref{fig:DMInjection_contours}, we recovered a signal with $m_{\rm DM}=9.0\pm 0.2$ GeV and $\langle \sigma v \rangle=(3.2\pm0.2)\times 10^{-26}$ cm$^{3}/$s. This point is $\sim 5\sigma$ away from the true injected signal. However, we note that the contours displayed in this figure account for statistical errors only. If the systematic uncertainties associated with the background model were included, the level of agreement between the recovered and injected points may be better. Note that the bias in the recovered parameters for this example is roughly 10\% and it arises from similar level of differences in the injected and recovered fluxes in some energy bins. In the case of the injection point shown in the top-right panel of Fig.~\ref{fig:DMInjection_contours} we observe a very good agreement (true point lies within the $1\sigma$ contour) between the injected and recovered signal. 

\section{Selection of the region of interest}\label{appx:RoI_selection}

Using a very similar fitting procedure to the one employed in this work, the \textit{Fermi} team made a careful analysis of the impact that the choice of RoI size has in their fits~\cite{TheFermi-LAT:2017vmf}. It was shown that relatively small RoIs allowed more freedom for the interstellar gas templates to reproduce the features in the data and reduced the effects of several modeling assumptions. Importantly, they noted that although relatively small RoIs (\textit{e.g.,} $|b|$, $|l|< 10^{\circ}$) are sufficient to resolve the gas-correlated templates, the IC templates---being smoother and broader than the gas maps---are generally more challenging to pin down in such small RoIs. In addition, Ref.~\cite{TheFermi-LAT:2017vmf} demonstrated that the intensity of the GCE is reduced in fits performed in small RoIs.

One of the major improvements in our GDE modeling for this work is the generation of more sophisticated IC templates that are divided in different galactocentric rings so that the uncertainties in the CR energetics and radiation fields can be more rigorously accounted for in the fits. We have tested our pipeline using a smaller $15^\circ\times15^\circ$ RoI  and the IC maps (divided in four rings),  and we could not
get stable and physically plausible spectra for the annular IC templates. This 
was our main motivation to choose a larger RoI ($40^\circ\times40^\circ$) for the main results in this analysis. 
 
Bearing in mind the caveats above, we explored a smaller RoI ($15^\circ\times15^\circ$) with GDE models where the IC was not split into independent rings. However, in this case they do not pass our injection tests: namely, our flux upper limits did not have the correct statistical coverage in the first two and five energy bins for the gNFW ($\gamma=1.2$) and cored profiles cases, respectively. It is possible that this issue is due to flux oversubtractions in energy bins where the point spread function of the \textit{Fermi} instrument is comparatively worse and therefore model template degeneracies are more acute.

\section{Dark Matter limits for other annihilation channels}\label{appx:DM_limits_otherchannels}

\begin{figure*}
    \centering
    \includegraphics[width=0.49\textwidth]{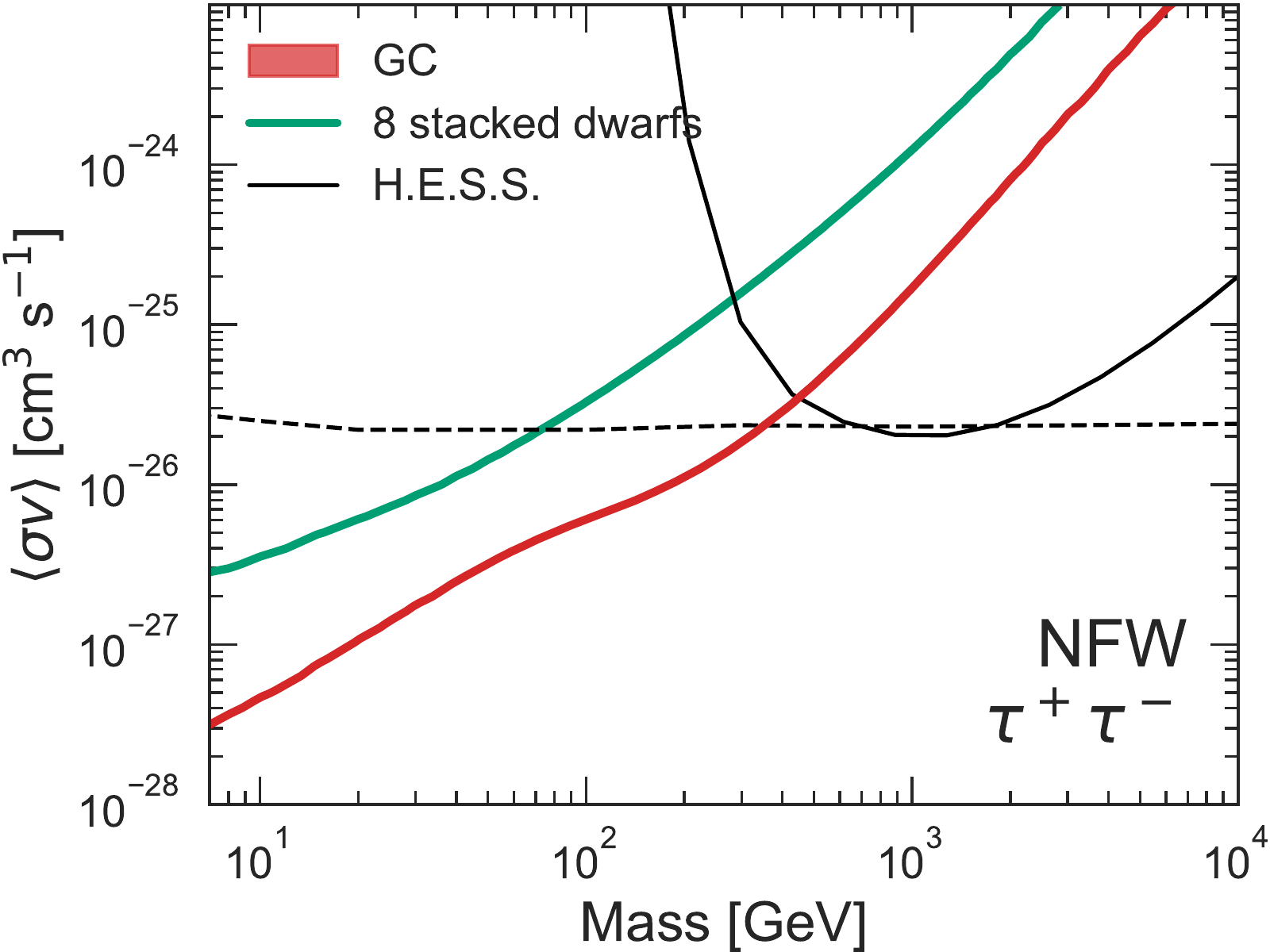}
    \includegraphics[width=0.49\textwidth]{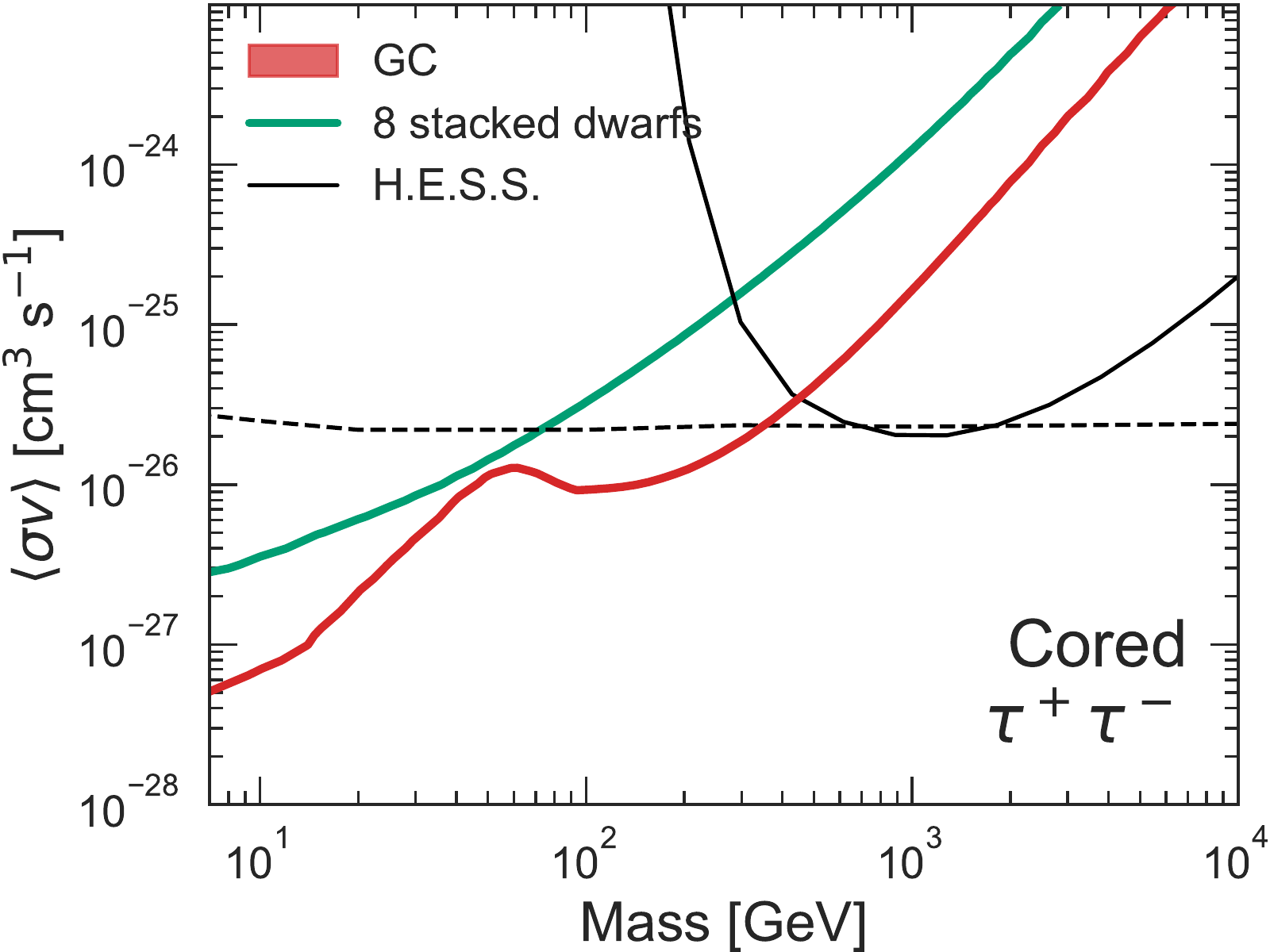}\\
    \includegraphics[width=0.49\textwidth]{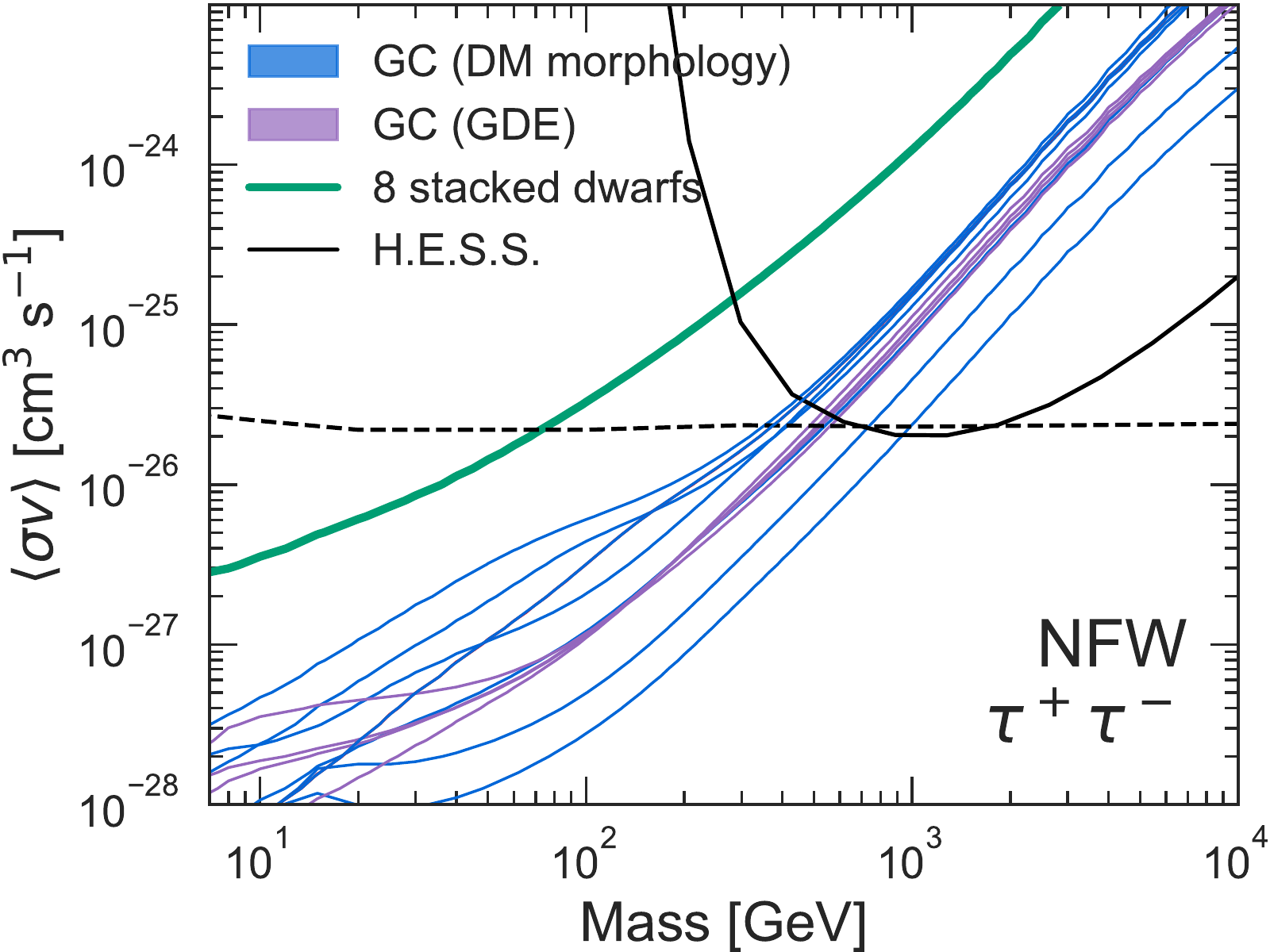}
    \includegraphics[width=0.49\textwidth]{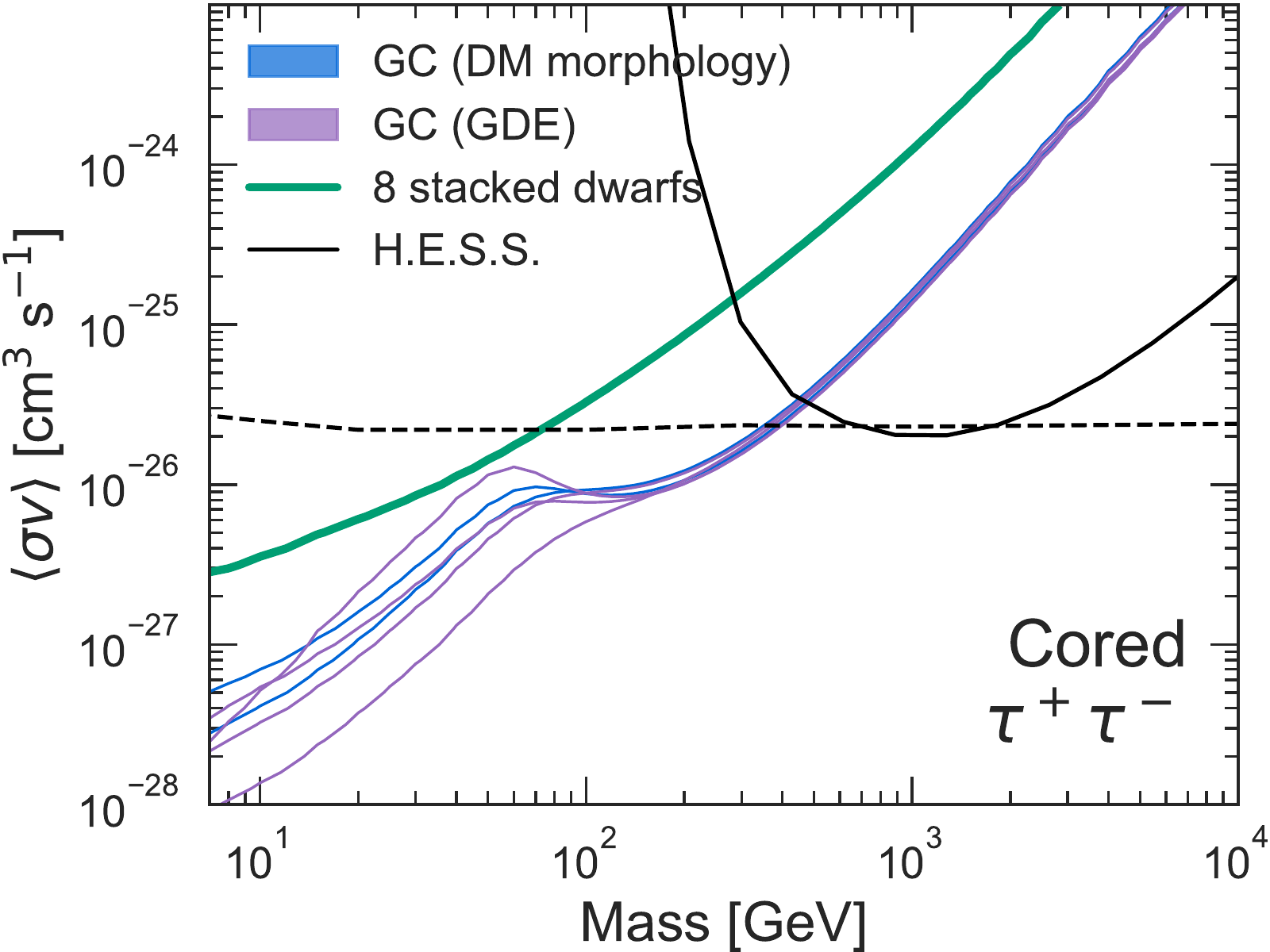}\\
    \caption{The least constraining upper limits for annihilation through $\tau$ leptons among all of the considered DM morphology and GDE variations around NFW-$\gamma$ (top left), and around the cored profile (top right). The bottom panels show the upper limits for each individual variation of the DM morphology and GDE around the NFW profile (bottom left) and the cored profile (bottom right). }
    \label{fig:tauchannel}
\end{figure*}

\begin{figure*}
    \centering
    \includegraphics[width=0.49\textwidth]{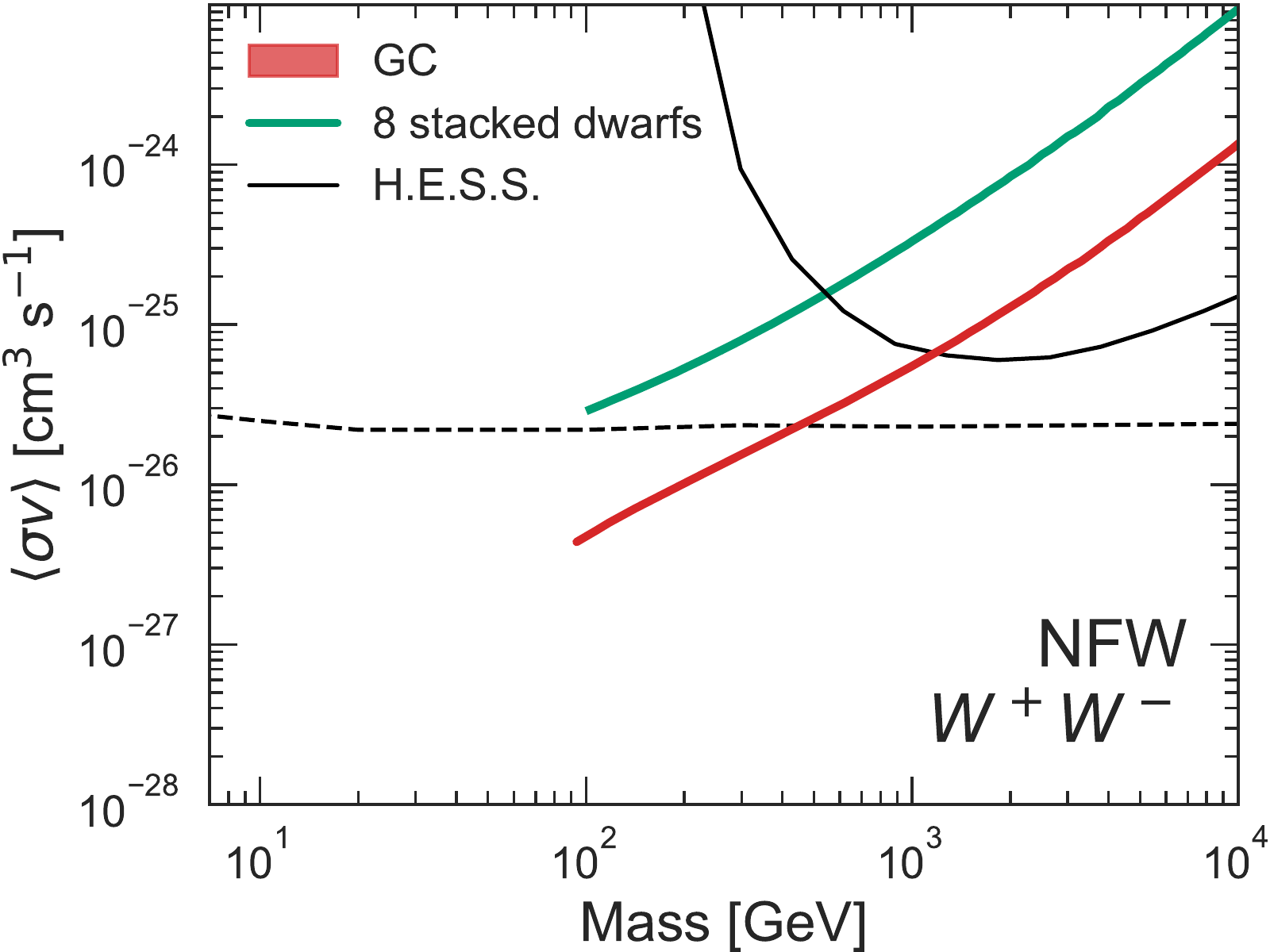}
    \includegraphics[width=0.49\textwidth]{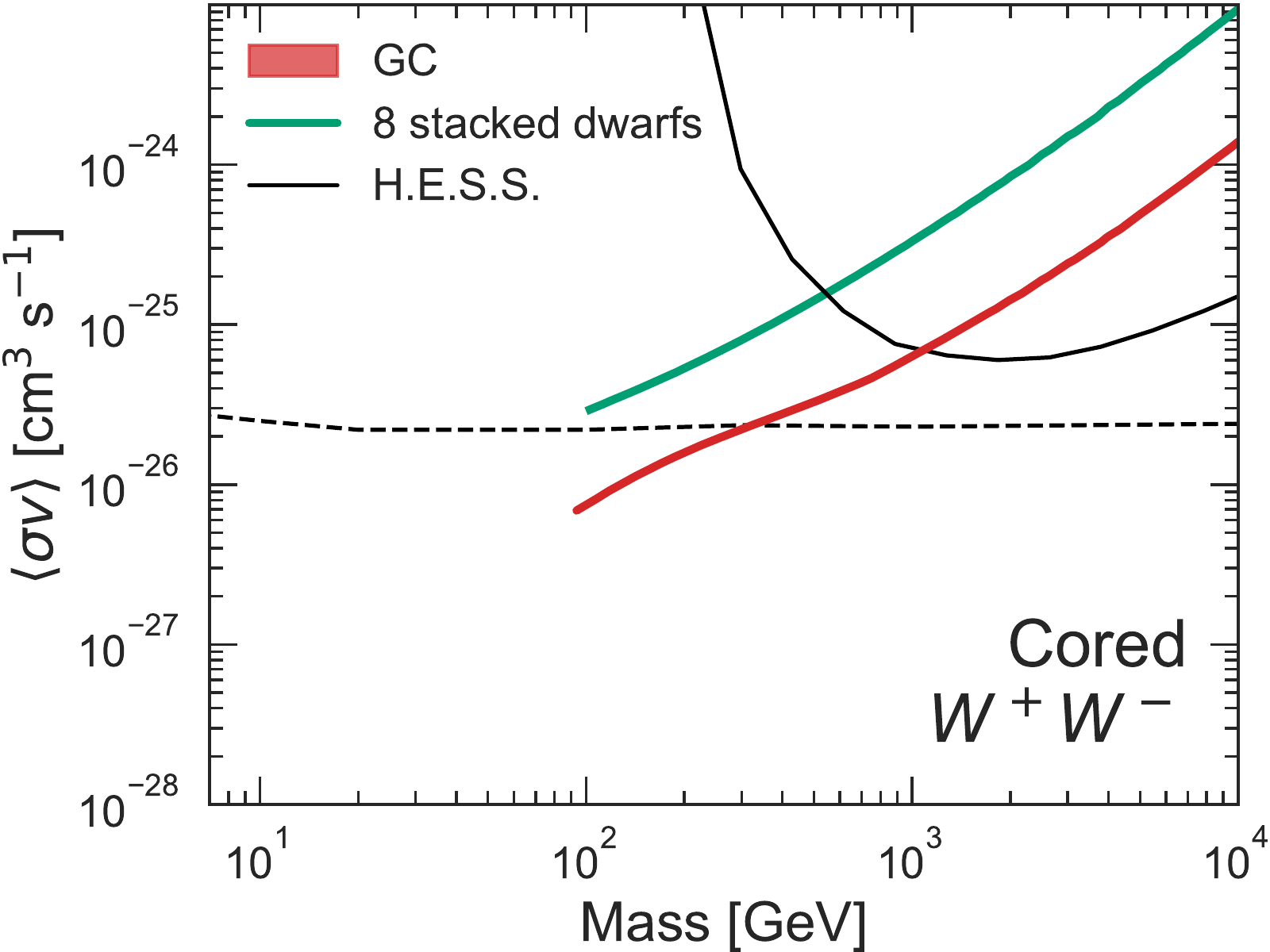}\\
     \includegraphics[width=0.49\textwidth]{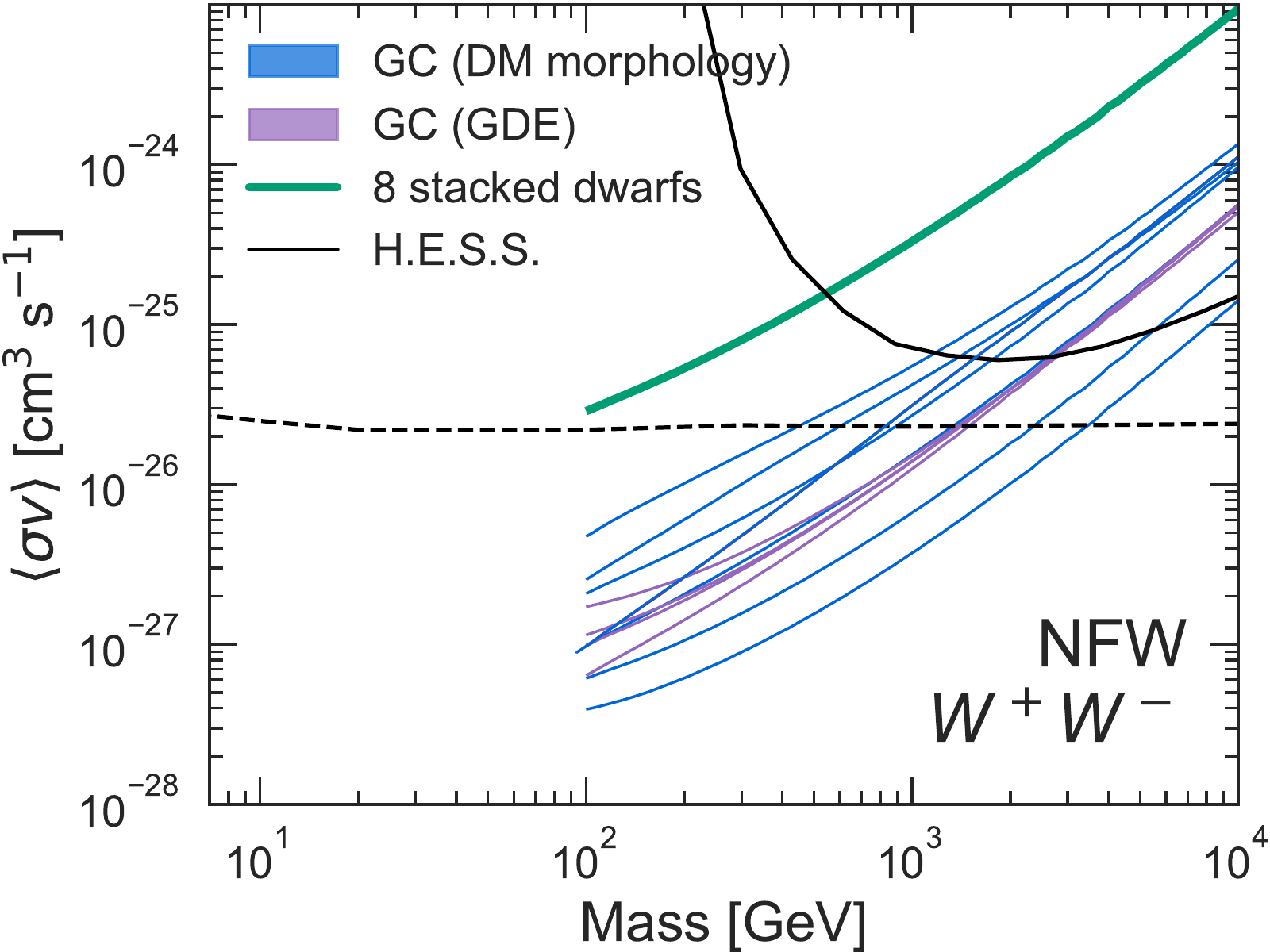}
    \includegraphics[width=0.49\textwidth]{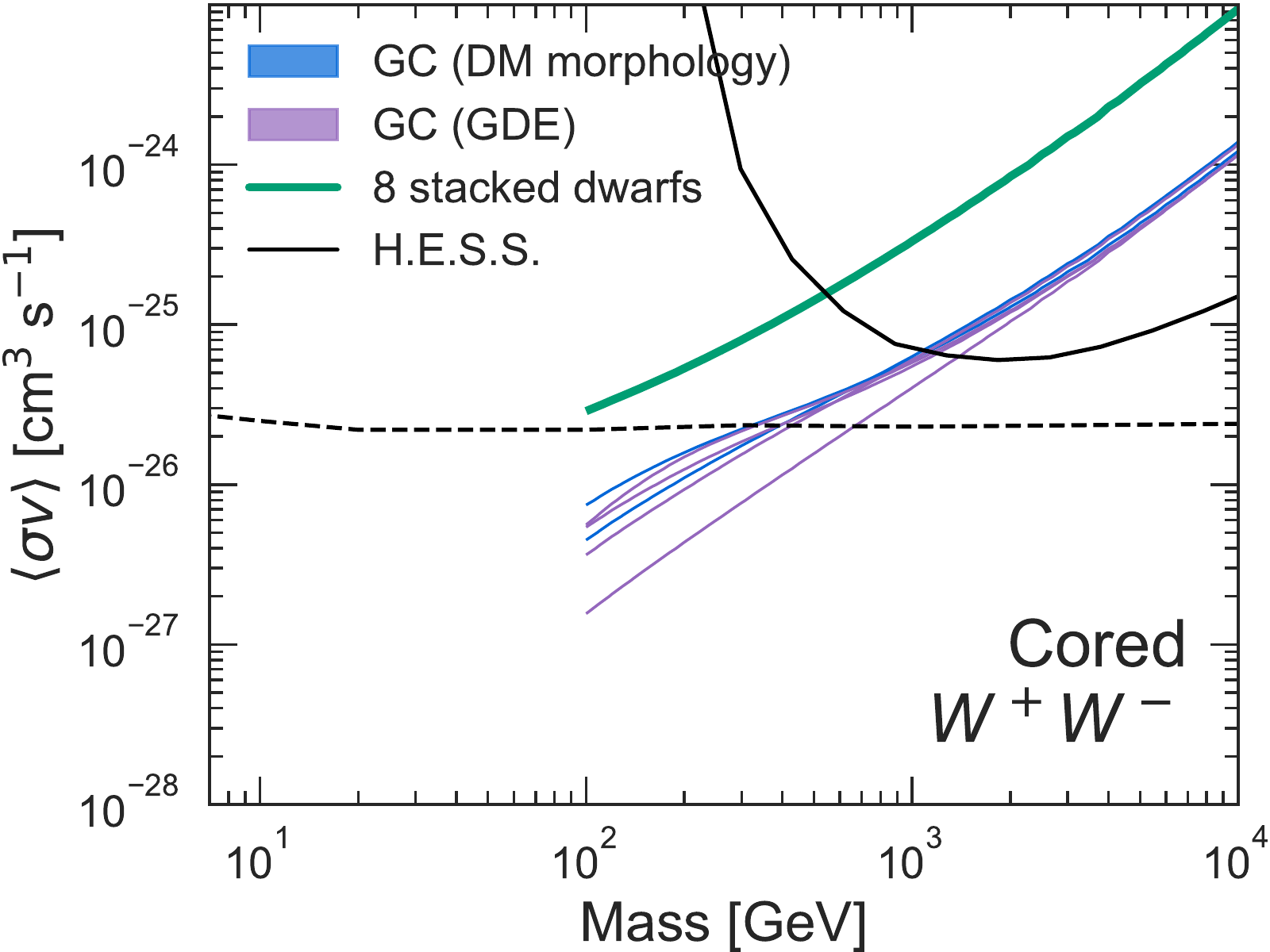}
    \caption{The least constraining upper limits for annihilation through W bosons among all of the considered variations around NFW-$\gamma$ (top left),  and around the cored profile (top right). The bottom panels show the upper limits for each individual variation of the DM morphology and GDE around the NFW profile (bottom left) and the cored profile (bottom right). }
    \label{fig:Wchannel}
\end{figure*}

\begin{figure*}
    \centering
    \includegraphics[width=0.49\textwidth]{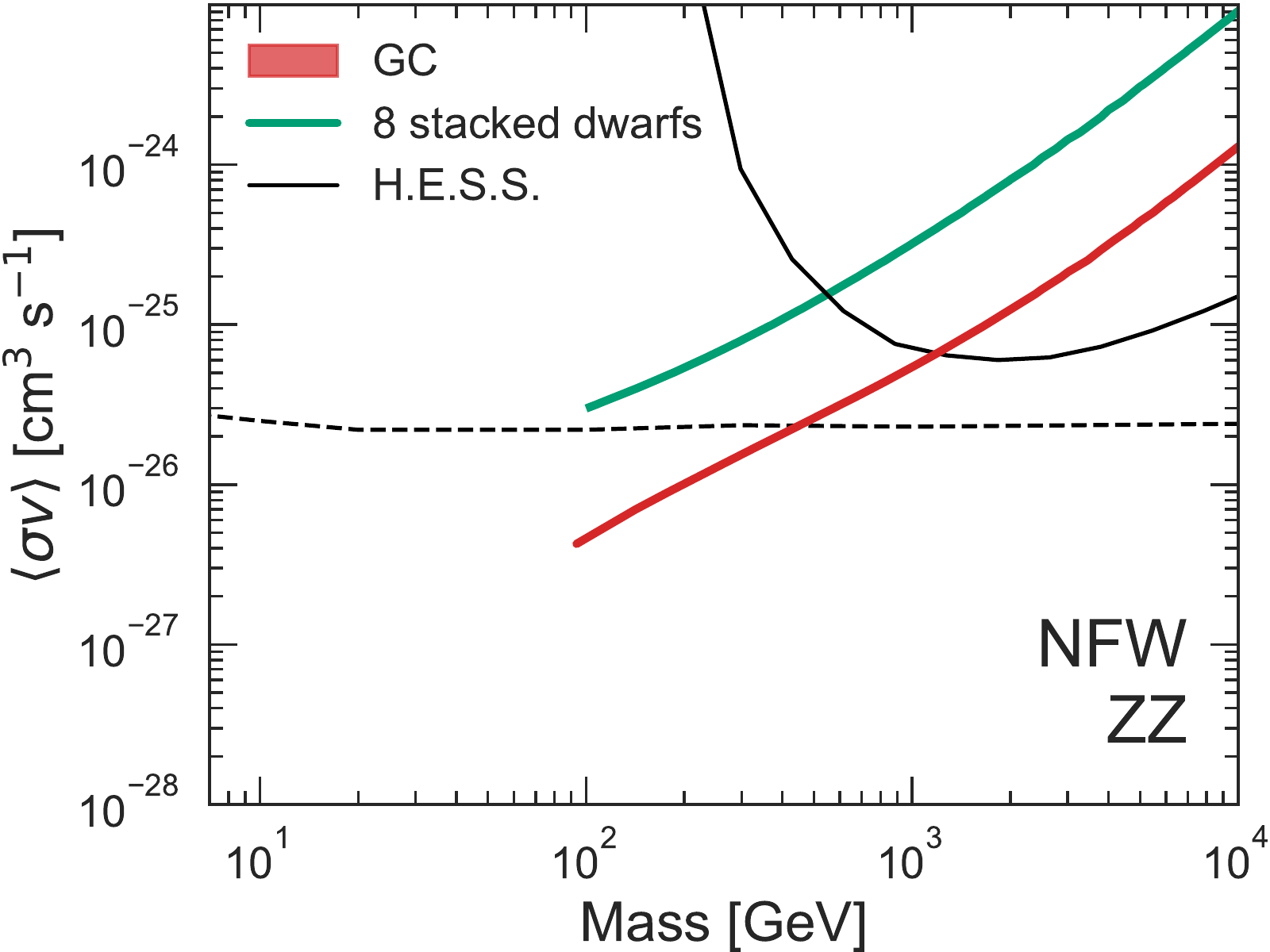}
    \includegraphics[width=0.49\textwidth]{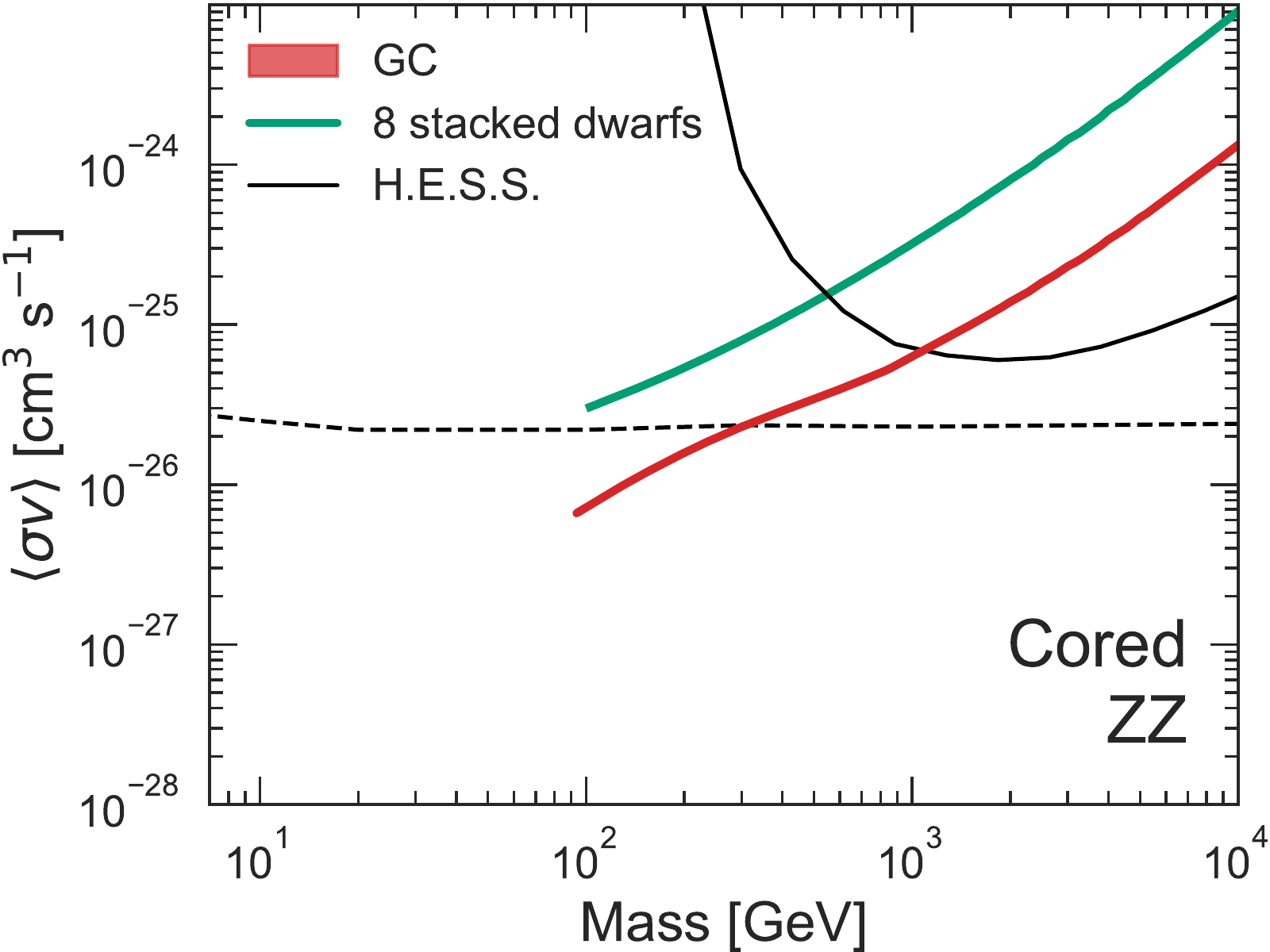}\\
    \includegraphics[width=0.49\textwidth]{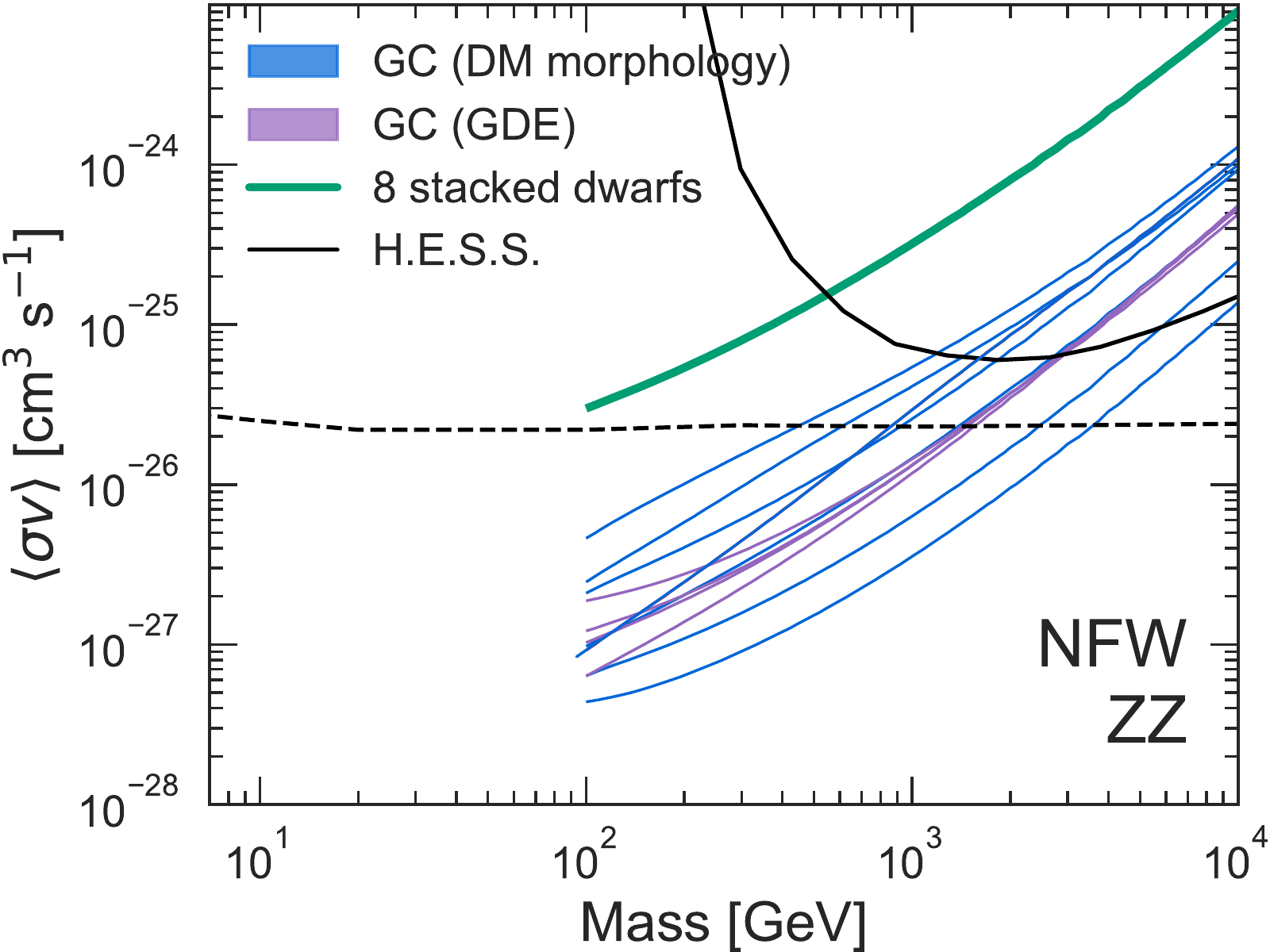}
    \includegraphics[width=0.49\textwidth]{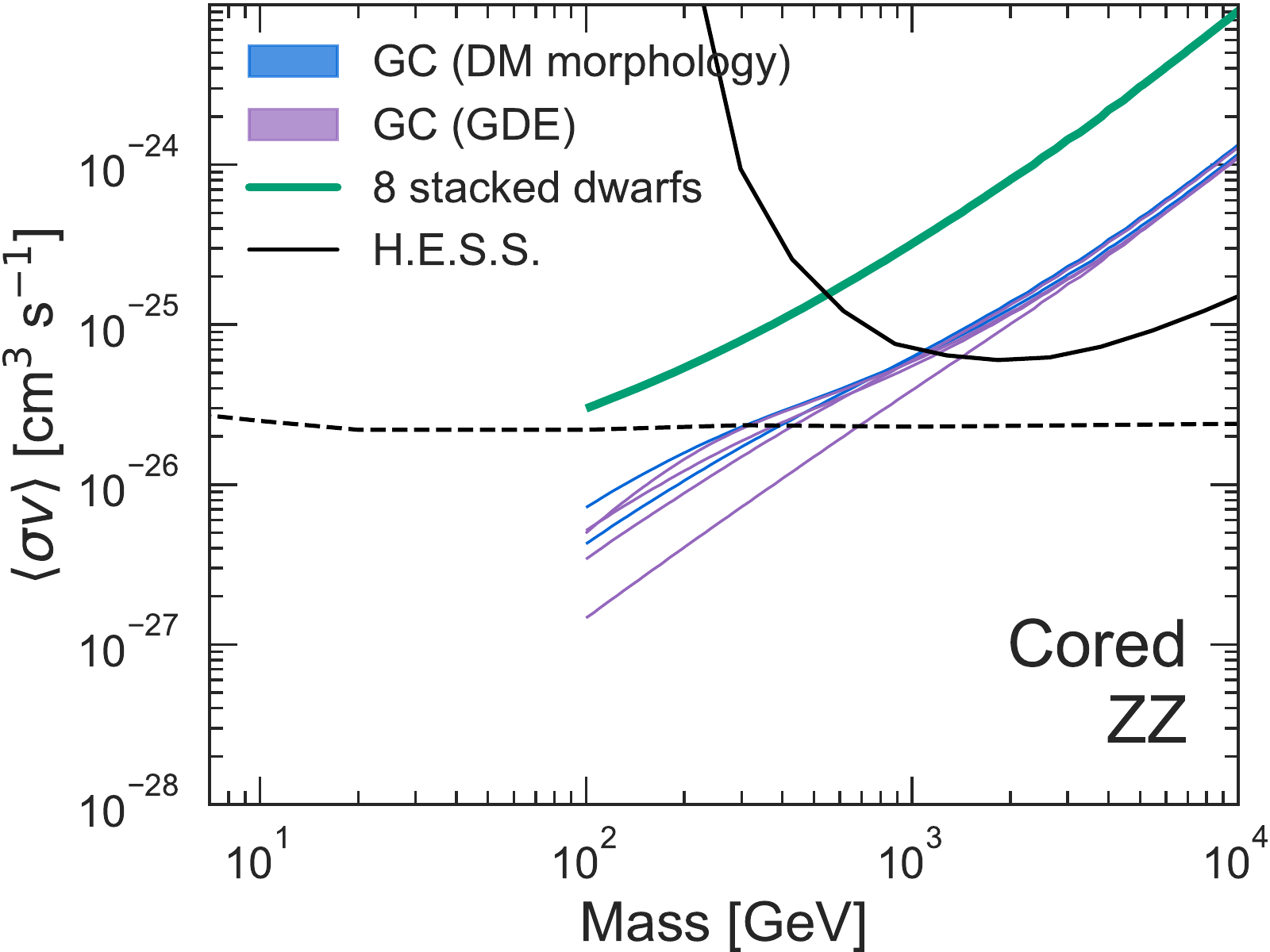}
    \caption{The least constraining upper limits for annihilation through Z bosons among all of the considered variations around NFW-$\gamma$ (top left),  and around the cored profile (top right). The bottom panels show the upper limits for each individual variation of the DM morphology and GDE around the NFW profile (bottom left) and the cored profile (bottom right). }
    \label{fig:Zchannel}
\end{figure*}

\begin{figure*}
    \centering
    \includegraphics[width=0.49\textwidth]{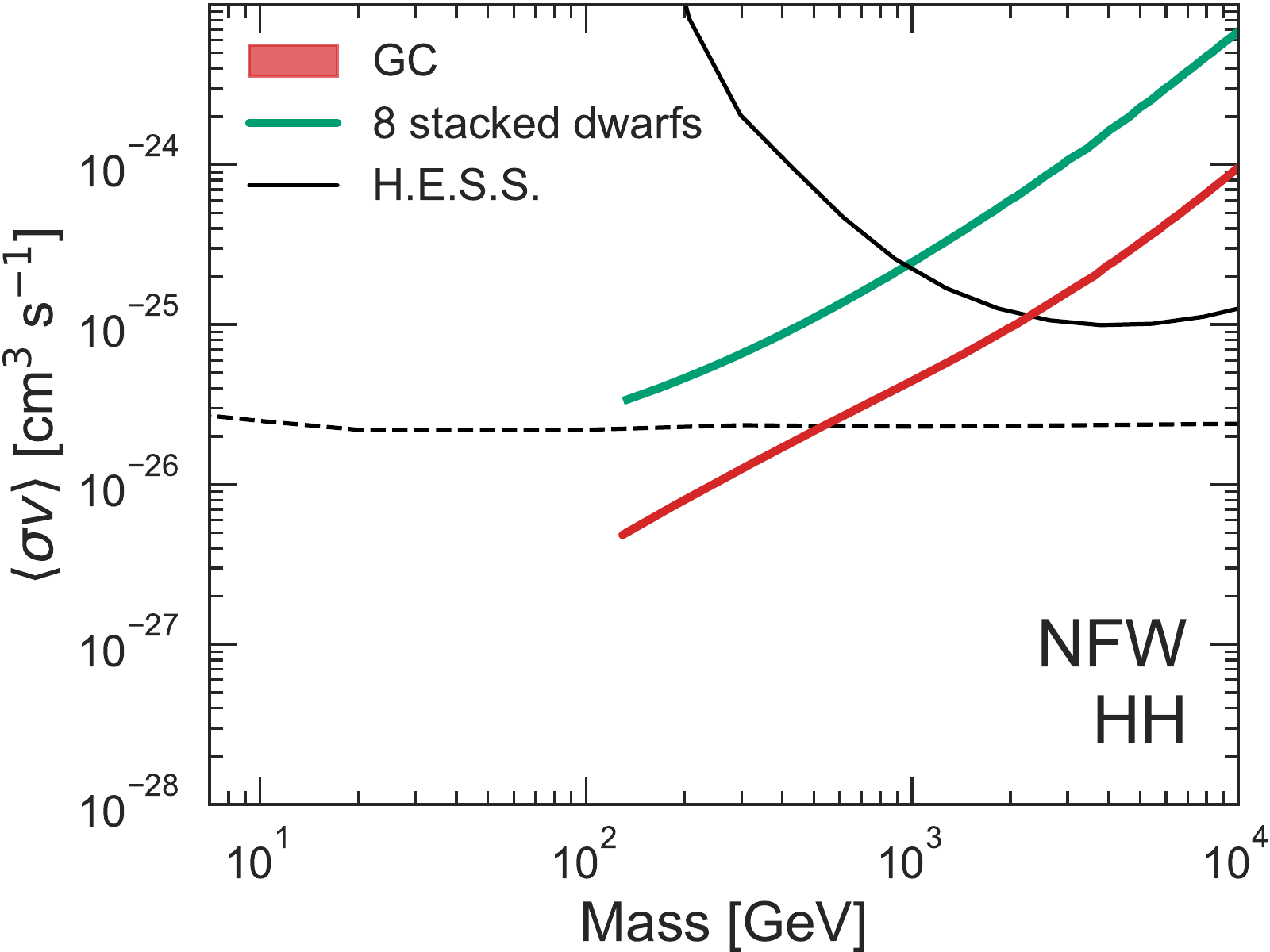}
    \includegraphics[width=0.49\textwidth]{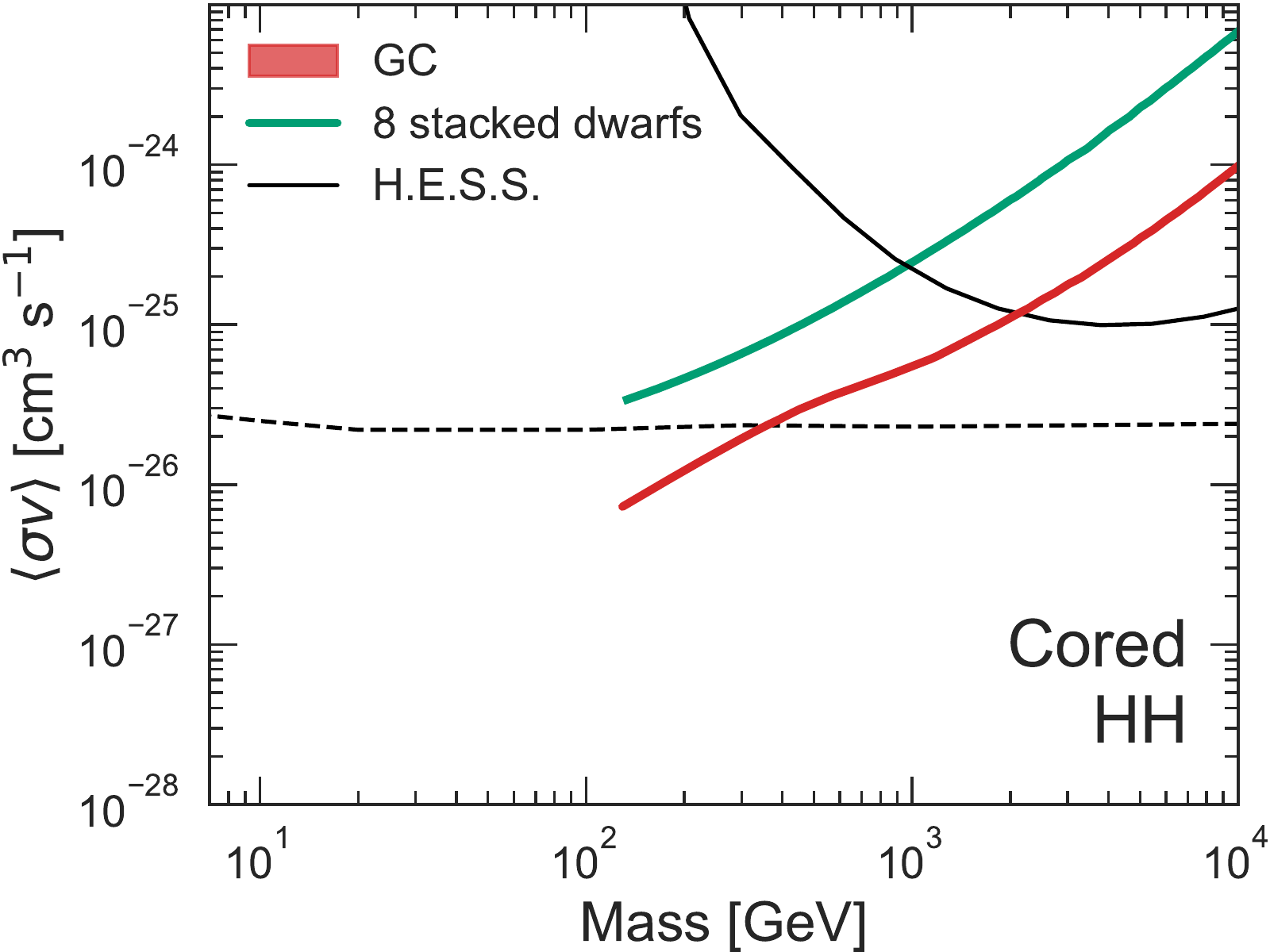}\\
        \includegraphics[width=0.49\textwidth]{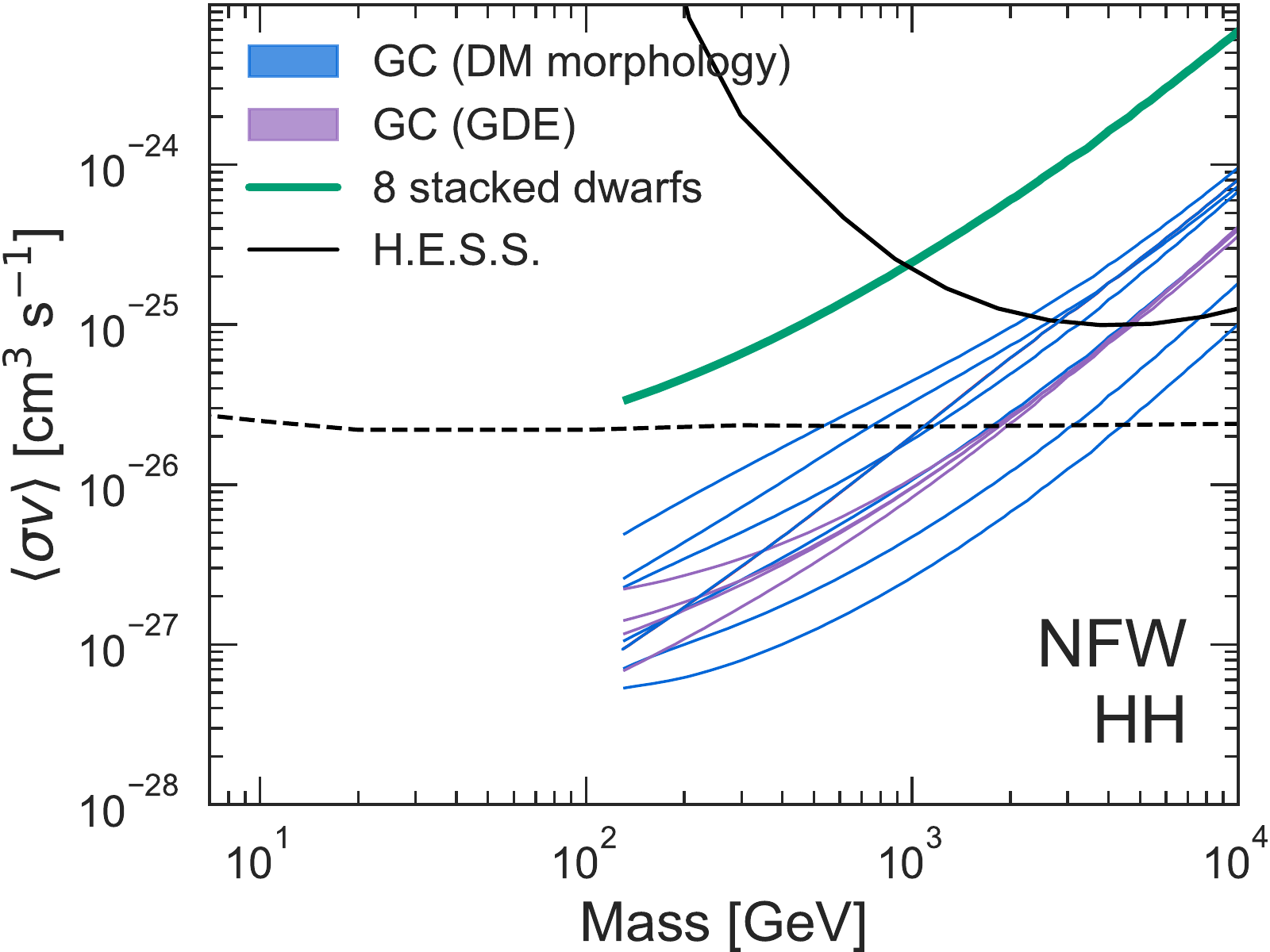}
    \includegraphics[width=0.49\textwidth]{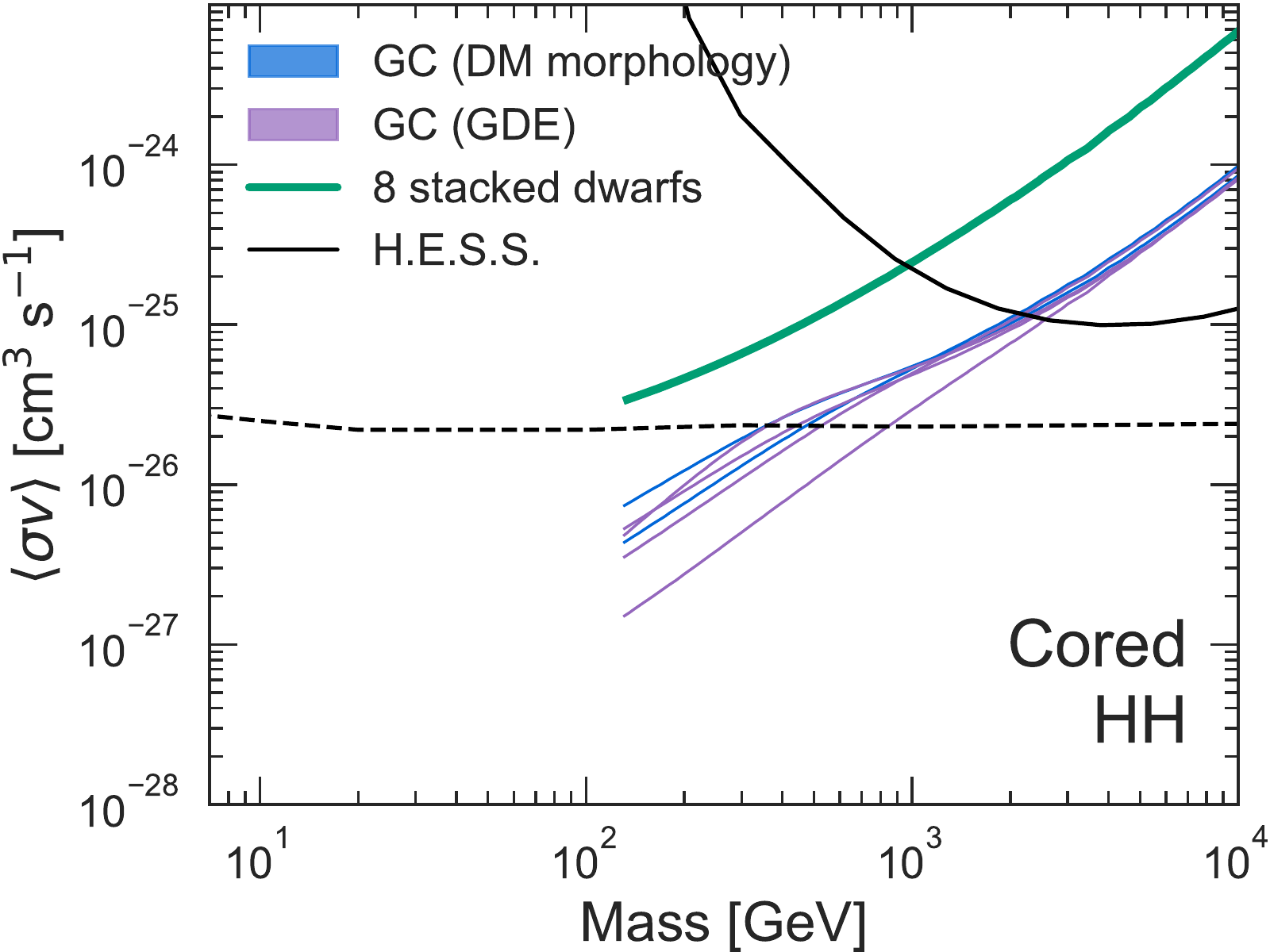}
    \caption{The least constraining upper limits for annihilation through Higgs boson among all of the considered variations around NFW-$\gamma$ (top left),  and around the cored profile (top right). The bottom panels show the upper limits for each individual variation of the DM morphology and GDE around the NFW profile (bottom left) and the cored profile (bottom right).}
    \label{fig:Hchannel}
\end{figure*}

In this section we investigate the ability of \textit{Fermi}-LAT GC observations to constrain the predicted DM emission when other possible DM annihilation channels are considered. Figures~\ref{fig:tauchannel}--\ref{fig:Hchannel} show the 95\% C.L.\ upper limits for final states producing a hard gamma-ray spectrum such as $\tau^+\tau^-$, $W^+W^-$, $ZZ$ and $HH$.

Similar to Figs.~\ref{fig:limitsNFW} and \ref{fig:limitsDiffuse} in the main text, Figs.~\ref{fig:tauchannel}--\ref{fig:Hchannel} illustrate how the upper limits on the DM annihilation cross section change when different spatial morphologies for the DM source and GDE models are assumed. For comparison purposes, we also display the limits obtained from dwarfs~\cite{Fermi-LAT:2016uux}. We omit the ultrafaints whose J-factors are more uncertain, e.g., Ref.~\citep{Ando:2020yyk}. Bottom-left panels show the limits obtained when the DM source is a NFW with various values of $\gamma$ and various GDE models (i.e., different IC models and interstellar gas and dust maps). Bottom-right panels show the same, except this time the DM source is modeled with a cored profile. The top panels show the weakest constraint from the set of variations shown in the bottom panels. 
 
 The results shown in these figures bracket realistic DM halo shapes that depart from the traditional NFW morphology. These illustrate how even by making conservative model assumptions, \textit{Fermi}-LAT observations of the GC provide very stringent constraints on thermal dark matter.

\end{document}